\documentclass[a4paper,twocolumn,pra]{revtex4}
\usepackage[english]{babel}

\usepackage{amssymb,amsthm,mathrsfs,amscd,amsmath}
\usepackage{multirow}
\usepackage{array}

\usepackage{graphicx,color}

\begin{document}

\title{Propagating edge states in strained  honeycomb lattices}

\author{Grazia Salerno, Tomoki Ozawa, Hannah M. Price and Iacopo Carusotto}
\affiliation{INO-CNR BEC Center and Dipartimento di Fisica, Universit\`a di Trento, I-38123 Povo, Italy}

\date{\today}

\begin{abstract}
We investigate the helically-propagating edge states associated with pseudo-Landau levels in strained honeycomb lattices. 
We exploit chiral symmetry to derive a general criterion for the existence of these propagating edge states in the presence of only nearest-neighbour hoppings and we verify our criterion using numerical simulations of both uni-axially and trigonally strained honeycomb lattices. We show that the propagation of the helical edge state can be controlled by engineering the shape of the edges. 
Sensitivity to chiral-symmetry-breaking next-nearest-neighbour hoppings is assessed.
Our result opens up an avenue toward the precise control of edge modes through manipulation of the edge shape.
\end{abstract}

\maketitle

\section{Introduction}

The existence of unidirectionally propagating edge states is one of the most important features of two-dimensional systems in the presence of a magnetic field~\cite{Girvin, Yoshioka}. 
In particular, such \textit{chiral} edge states are responsible for the quantized conductance in the quantum Hall effect, and their existence is guaranteed by a topological property of the bulk system, which does not depend on the specific type of edge termination.

In honeycomb lattices, such as graphene, a strong artificial pseudo-magnetic field can be implemented through strain engineering \cite{Kane,Vozmediano, CastroNeto, Goerbig,Guinea,GuineaPRB,deJuan}, and associated relativistic pseudo-Landau levels have been experimentally observed in the density of states \cite{Levy}.
It is, {\it prima facie}, natural to expect propagating edge states also in the presence of such a pseudo-magnetic field.
However, an important difference between the real magnetic field and the pseudo-magnetic field induced by strain is that, since strain does not break time reversal symmetry, the pseudo-magnetic field has opposite signs in the two Dirac valleys. 
Consequently, unlike for a real magnetic field, chiral edge states cannot exist in the presence of a pseudo-magnetic field only.
Instead, \textit{helical} edge states can exist where states from different valleys experience opposite magnetic fields and propagate in opposite directions. 
It has been noted by several authors that such helical edge states do not always exist \cite{Low,Chang,Ghaemi,Brendel,Atteia} depending on the edge shape and the type of strain.
However, a general condition specifying when propagating edge states exist is still lacking.

In this paper, we show that the existence of the helical edge states in strained honeycomb lattices strongly depends on the type of edge termination as well as on the type of strain. We also compare the strained system with a pristine lattice in the presence of a real magnetic field.
We give a general criterion that explains the termination-dependence and the strain-dependence of the propagating edge states. The criterion is obtained for a strained honeycomb lattice in the presence of only nearest-neighbour hoppings, and it is based on the chiral symmetry of the tight-binding Hamiltonian, as well as on the particular form of Landau levels. Introducing next-nearest-neighbour hoppings, which break the chiral symmetry, limits the validity of our criterion and introduces new features.

In the presence of nearest-neighbour hoppings only, the criterion is confirmed through numerical simulations of both uni-axial and trigonal strain. 
Our finding suggests a powerful and simple avenue to build valley filters and control currents in solid-state graphene using strain and the engineering of the edge termination. This approach can also be implemented in artificial graphene for various platforms, such as photonic, optomechanical or phononic systems \cite{Brendel, Bellec2014, Jacqmin, Bellec2013a, Bellec2013b, Rechtsman}.

The paper is organized as follows.
In Section \ref{sec:model} we introduce the system and review the main properties of the pseudo-Landau levels in strained honeycomb lattices.
In Section \ref{sec:criterion} we present our criterion for the existence of propagating edge states of the $0$-th pseudo-Landau level. 
In Section \ref{sec:uniaxial_dispersion} we apply our criterion to two cases of uni-axial strain. 
In Section \ref{sec:uniaxial_dispersionA} we consider uni-axial strain along the $x$ direction to study the energy dispersion for the zigzag and bearded terminations.
In Section \ref{sec:uniaxial_dispersionB} a uni-axial strain along the $y$ direction is considered to study the armchair termination. 
In Section \ref{sec:magnetic_dispersion}, we show that our criterion can also be applied to the case of a real magnetic field.
The steady-state of artificial graphene under a coherent driving is studied in Section~\ref{sec:steadystate} for both the uni-axial and the trigonal strains. We also show that, by a controlled edge engineering, the propagating edge states can be valley filtered.
In Section \ref{sec:NNN}, we discuss the effect of a next-nearest-neighbour hopping on the propagating edge states.
Finally, we conclude in Section \ref{sec:conclusions}.

\section{The model} 
\label{sec:model}

\begin{figure}[t]
\centering
\includegraphics[width=0.47\textwidth]{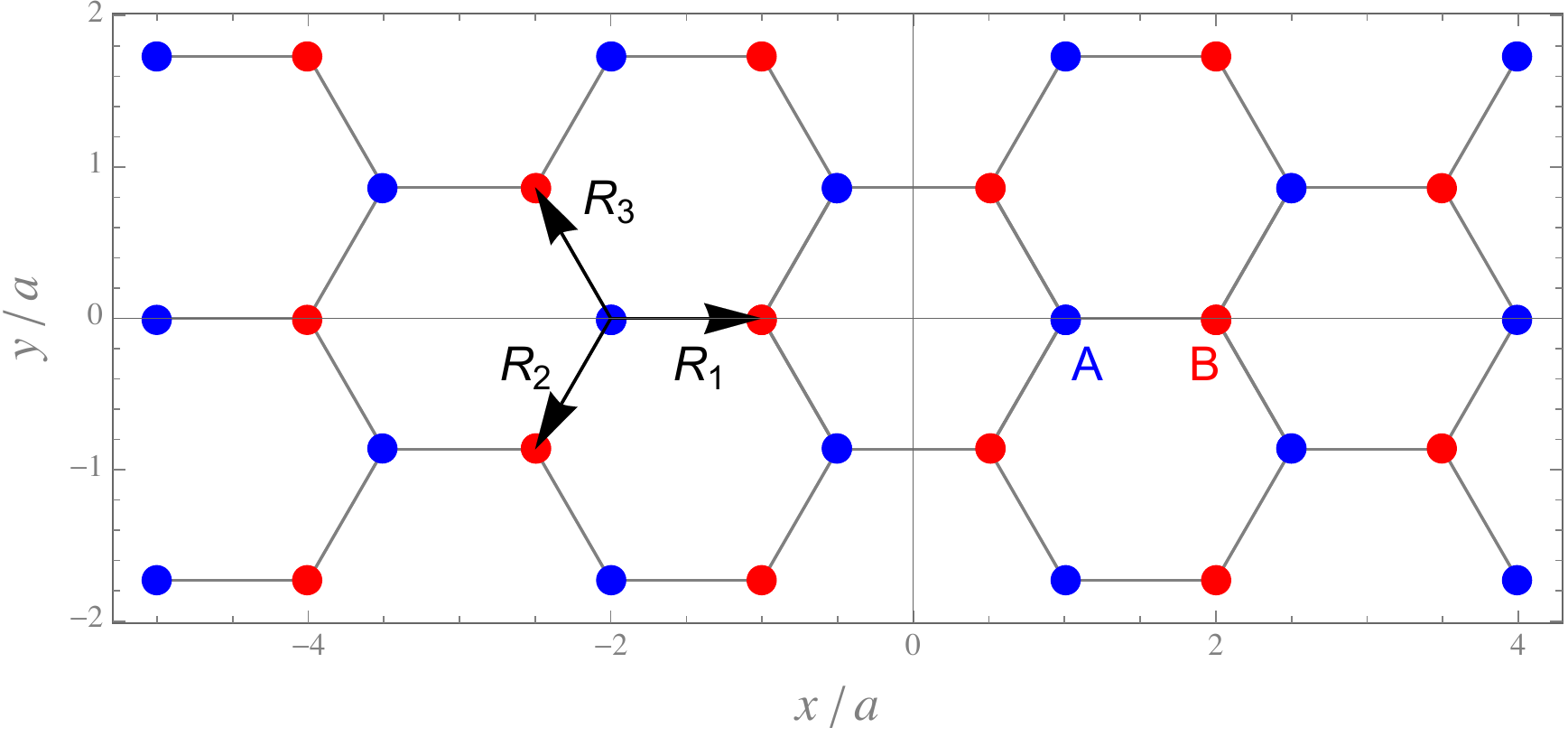}
\caption{A honeycomb lattice. The different sites $A$ and $B$, that form two different sublattices, are coloured in blue and red respectively. The vectors $\mathbf{R}_j$ connect nearest neighbour sites. In the main text, we always refer to this orientation of the lattice. On the left edge, we show the bearded termination, while on the right edge we show the zigzag termination. The bottom and top edges have armchair terminations.
This ribbon has $N_x=7$ unit cells counted along the armchair direction (although the last unit cell is not complete, as it is missing a $B$ site), and $N_y=5$ unit cells counted along the zigzag/bearded direction.}
\label{fig:system}
\end{figure}

We start by reviewing the main properties of relativistic pseudo-Landau levels stemming from strain in a honeycomb lattice.
The tight-binding Hamiltonian of a strained honeycomb lattice with only nearest neighbour hopping takes, in real space, the following form:
\begin{align}
\mathcal{H} = -\sum_{\mathbf{r},j} &\left( t_j(\mathbf{r}) \hat{a}^{\dagger}_{\mathbf{r} - \mathbf{R}_j} \hat{b}_{\mathbf{r}} +\text{H.c.}\right)
\label{strained_ham}
\end{align}
where $\hat{a}_\mathbf{r}$ and $\hat{b}_\mathbf{r}$ are annihilation operators of a particle at position $\mathbf{r} = (x,y)$ in $A$ and $B$-sublattices, respectively.
The vectors $\mathbf{R}_j$, with $j=1,2,3$ connect nearest neighbour sites, as shown in Fig.~\ref{fig:system}, and $t_j$ is the nearest-neighbour hopping along $\mathbf{R}_j$. 
In particular, we have $\mathbf{R}_1=(a,0)$, $\mathbf{R}_2=(-a/2,-\sqrt{3}a/2)$ and $\mathbf{R}_3=(-a/2,\sqrt{3}a/2)$, where $a$ is the lattice spacing.
For almost all the manuscript, we will use the form in \eqref{strained_ham} which contains only nearest neighbour hoppings. This restriction allows us to exploit chiral symmetry to provide a simple criterion for the existence of propagating edge states. In section~\ref{sec:NNN} we will relax this constraint and add also next-nearest-neighbour hoppings, showing that the qualitative features of our predictions are still valid.

The first Brillouin zone of the unstrained honeycomb lattice can be taken in the form of a hexagon, where the Dirac points $K$ and $K'$ are located at its corners. Around such Dirac points, when $t_{1,2,3}=t$, the system has a linear energy dispersion, whose slope defines the Dirac velocity $v_D=3at/(2\hbar)$. 
In our model, the strain is implemented as a spatial dependence of the hopping parameters $t_j$ and our results are independent of the actual deformation of the underlying honeycomb lattice one needs to realize the spatial variation of hopping in a specific physical system. A relation between the inter-site length and the hopping amplitude, in fact, depends on the particular system such as solid-state graphene, microwave cavity arrays or exciton-polariton micropillars.
Equation~\eqref{strained_ham} allows us for a clear description of the strain effects from a theoretical tight-binding perspective.

The effect of a spatially homogeneous strain can be described in momentum space around these Dirac points $K$ and $K'$ as a synthetic vector potential  $\boldsymbol{\mathcal{A}}$ for the low energy modes \cite{Vozmediano, CastroNeto, Goerbig}:
\begin{equation} 
v_D^x e \mathcal{A}_x = \frac{\sqrt{3}\xi}{2} \left(t_2-t_3\right),\,
v_D^y e \mathcal{A}_y= \frac{\xi}{2} \left(2t_1-t_2-t_3\right)
\label{A}
\end{equation}
where $\xi= \pm1$ is the valley index which distinguishes between $K$ and $K'$, and the Dirac velocity is, in general, no longer isotropic $v_D^x \neq v_D^y \neq v_D$ \cite{deJuan,Salerno}.
From Eq.~\eqref{A} we see that a non-uniform strain with position-dependent hoppings $t_j(\mathbf{r})$ induces a non-zero synthetic pseudo-magnetic field $\boldsymbol{\mathcal{B}}=\nabla \times \boldsymbol{\mathcal{A}}$. 
Since the vector potential in Eq.~\eqref{A} is opposite for the two valleys $K$ and $K'$, also the pseudo-magnetic field has opposite signs in the two valleys. 
We now review the properties of pseudo-Landau levels for three different choices of the strain, highlighting the key features which will be important when presenting the criterion for the existence of propagating edge states.

\subsection{Uni-axial strain along x}
We first consider the following uni-axial strain along the $x$ direction:
\begin{equation}
	t_1(\mathbf{r}) = t \left(1 +  \frac{\boldsymbol{x}\cdot\mathbf{R}_1}{3a^2} \,\tau \right), \qquad
	t_{2,3}(\mathbf{r}) = t
\label{uniaxialX}
\end{equation}
where $\boldsymbol{x} \equiv (x,0)$. The dimensionless parameter $\tau$ controls the amount of strain. 
When $\tau$ is positive (negative), the strain increases (decreases) linearly towards the right edge. 
We choose the strain so that we have $t_1=t$ in the middle of the system, and such that on the edges of the system $t_1>0$ and $t_1<2t$.
The $x$-dependent form of the strain in Eq.~\eqref{uniaxialX} is convenient for studying the zigzag and bearded edges for the system oriented as in Fig.~\ref{fig:system}, since the system is translationally invariant along the $y$ direction under periodic boundary conditions along $y$, and so the quasi-momentum $k_y$ is a good quantum number.
The uni-axial strain considered in Eq.~\eqref{uniaxialX} is straightforwardly implementable in artificial graphene while is difficult to realize in solid-state graphene by applying mechanical forces to the sample.

We shall now concentrate on the case of positive $\tau$, namely where the hopping along the $x$ direction is minimal at the left edge.
The artificial pseudo-magnetic vector potential is  $e\boldsymbol{\mathcal{A}} = \xi 2\hbar \tau/(9a^2) (0, x)$, which is the Landau gauge with the vector potential oriented along $y$.
The artificial magnetic field is $ e\mathcal{B} = \xi 2 \hbar \tau /(9 a^2)$, where the corresponding magnetic length is $l_B = \sqrt{\hbar/|e\mathcal{B}|}$.
Relativistic pseudo-Landau levels form near Dirac points in momentum space and their energy levels at the Dirac points are given by $E_n = \mathrm{sign} (n)t\sqrt{\tau |n|}$, where $n \in \mathbb{Z}$.
For a given value of the ``guiding center" $x_0=-\xi l_B^2 k_y$, the wavefunction of $n$-th pseudo-Landau level in the $A-B$ sublattice basis is given, for $n\neq 0$, by
\begin{equation}
	\psi_n (x,y) = e^{i k_y y} e^{-\frac{(x-x_0)^2}{2l_B}}
	\begin{pmatrix}
	H_{|n|-1} \left( \frac{x-x_0}{l_B}\right) \\ \mathrm{sign} (n) H_{|n|} \left(\frac{x-x_0}{l_B}\right)
	\end{pmatrix},
	\label{LLn}
\end{equation}
and where, for any $m \geq 0$, $H_m(x)$ is a Hermite polynomial of degree $m$.
For the $0$-th pseudo-Landau level, the wavefunction is a Gaussian completely localized on the $B$-sublattice \cite{Goerbig}.
Each level is almost degenerate, since wavefunctions localized around different $x$-coordinates $x_0$ share the same energy up to a shift due to the position dependence of the Dirac velocity \cite{deJuan,Salerno}. 

As we shall see better in Section \ref{sec:criterion}, depending on the details of the termination, when the guiding center $x_0$ hits the physical edge of the system, the energy of the pseudo-Landau level can shift significantly from around its bulk value and the states are localized at the edge. Since the energy dispersion acquires a non-zero slope as a function of $k_y$, that is opposite around the two valleys, these states are helically-propagating edge states.

\subsection{Uni-axial strain along y}

We now consider the following uni-axial strain along the $y$-direction:
\begin{equation}
t_1(\mathbf{r})= t, \quad t_{2,3}(\mathbf{r}) = t \left(1 +  2\frac{\boldsymbol{y}\cdot\mathbf{R}_{2,3}}{3a^2} \,\tau \right),
\label{uniaxialY}
\end{equation}
where $\boldsymbol{y} \equiv (0,y)$. 
This $y$-dependent form of the strain is convenient in studying the armchair edge in Fig.~\ref{fig:system}, since the system is translationally invariant along the $x$ direction when periodic boundary conditions are applied along $x$, and so the quasi-momentum $k_x$ is a good quantum number.
The strain in Eq.~\eqref{uniaxialY} creates an artificial pseudo-magnetic vector potential oriented along $x$ in the Landau gauge: $e\boldsymbol{\mathcal{A}} = \xi 2\hbar \tau /(3a^2) (- y,0)$.
The artificial pseudo-magnetic field is $e \mathcal{B}=\xi 2\hbar \tau/(3a^2)$, which corresponds to pseudo-Landau levels at energies  $E_n = \mathrm{sign} (n)t\sqrt{3 \tau |n|}$, where $n \in \mathbb{Z}$.
For a given value of the guiding center $y_0=-\xi l_B^2 k_x$, the wavefunction of $n$-th pseudo-Landau level is given, when $n\neq 0$, by
\begin{equation}
	\psi_n (x,y) = e^{i k_x x} e^{-\frac{(y-y_0)^2}{2l_B}}
	\begin{pmatrix}
	H_{|n|-1} \left( \frac{y-y_0}{l_B}\right) \\ \mathrm{sign} (n) H_{|n|} \left(\frac{y-y_0}{l_B}\right)
	\end{pmatrix}
	\label{LLny}
\end{equation}
For the $0$-th pseudo-Landau level, the wavefunction is again a Gaussian completely localized on the $B$-sublattice.
As we will see in Section \ref{sec:criterion}, this localization on one sublattice plays an important role in determining when propagating edge states may exist.

\subsection{Trigonal strain}

We now consider the case of trigonal strain:
\begin{equation}
t_j(\mathbf{r}) = t \left(1 + \frac{\mathbf{r}\cdot\mathbf{R}_j}{3a^2} \,\tau \right),
\label{trigonal}
\end{equation}
for $j = 1, 2,$ and $3$.
This type of strain can be applied to solid-state graphene by engineering the distribution of the forces applied to the perimeter of the graphene flake \cite{Guinea}. The trigonal strain has been widely used in experiments on solid-state graphene and in various types of artificial graphene, such as photonic graphene \cite{Poli, Rechtsman}.

The trigonal strain implements the artificial pseudo-magnetic vector potential $e \boldsymbol{\mathcal{A}}= \xi \hbar \tau/(3a^2) (-y,x)$ in the symmetric gauge and the artificial pseudo-magnetic field $e\mathcal{B} = \xi 2 \hbar \tau  /(3a^2)$.
While analytical expressions for Landau levels in the symmetric gauge are available, for the following we will only need that the $0$-th pseudo-Landau level wavefuntion is again localized only on $B$-sublattice for a positive $\tau>0$ for both valleys.
Since the system is not translationally invariant in any direction, periodic boundary conditions can not be applied for studying the energy dispersion in momentum space, and we discuss only the numerical results later in Section \ref{sec:steadystate}.

\section{Criterion for the existence of propagating edge states}
\label{sec:criterion}

We now give an intuitive physical criterion for the existence of propagating edge states for strained honeycomb lattices.
The criterion is based on the chiral symmetry of the tight-binding Hamiltonian, on the structure of the relativistic Landau level wavefunction, and on the existence of non-propagating zero-energy edge states in the absence of a space-dependent strain.

The Hamiltonian in Eq.~\eqref{strained_ham} has a chiral symmetry; that is, under the transformation $a_{\mathbf{r}} \to a_{\mathbf{r}}$ and $b_{\mathbf{r}} \to -b_{\mathbf{r}}$, we have $\mathcal{H} \to - \mathcal{H}$.
By applying this transformation to an eigenstate with energy $E \neq 0$, we obtain a different orthogonal eigenstate with energy $-E$ where the sign of the wavefunction on the $B$-sublattice is flipped.
As an important consequence, if an eigenstate is localized only on one sublattice then, by the chiral symmetry, its energy must be zero.

We now apply this argument to the Landau levels when a guiding center is close to the edge of the system.
We first consider the case of a level with $n\neq 0$, for which the Landau level wavefunction lives on both $A$- and $B$-sublattices. When the guiding center of the wavefunction is far from the edge of the system, the energy of the state is given by the (almost) degenerate Landau level energy, as one can see for example in Fig.~\ref{fig:Spectra_all_terminations}. However, when the guiding center is close to the edge, the energy of the states lifts significantly from the degenerate value, and the states have a large non-zero group velocity along the edge direction. 
These edge states are helically-propagating and always exist at both ends of the ribbon for Landau levels with $n\neq 0$.

The situation is drastically different for the pseudo-Landau level with $n=0$, where we need to consider the specific form of the $0$-th pseudo-Landau level wavefunction, which is nonzero only on the $B$-sublattice for positive $\tau>0$ for both valleys.
Even when the guiding center is close to the edge, the energy of the $n=0$ pseudo-Landau level remains zero due to the chiral symmetry, as long as the wavefunction is always localized on $B$-sublattice.
This state can never propagate because it has a zero group velocity.
In order to have a propagating edge state, the $0$-th pseudo-Landau level wavefunction localized on the $B$-sublattice needs to mix with zero-energy edge states which have a non-zero amplitude on the other $A$-sublattice \footnote{This criterion supersedes the argument proposed in \cite{Ghaemi}. In \cite{Ghaemi}, it is erroneously stated that the zeroth Landau level localized on a A-sublattice can only mix with the zero-energy edge states localized on the same A-sublattice. Their statement is in disagreement with the chiral symmetry argument we give in this paper, which we have fully numerically verified. In \cite{Ghaemi}, it was also stated that armchair edges do not support counterpropagating edge modes, which is not generally true as we show in Section~\ref{sec:uniaxial_dispersionB}.}.

\begin{figure}[t]
\centering
\includegraphics[width=0.23\textwidth]{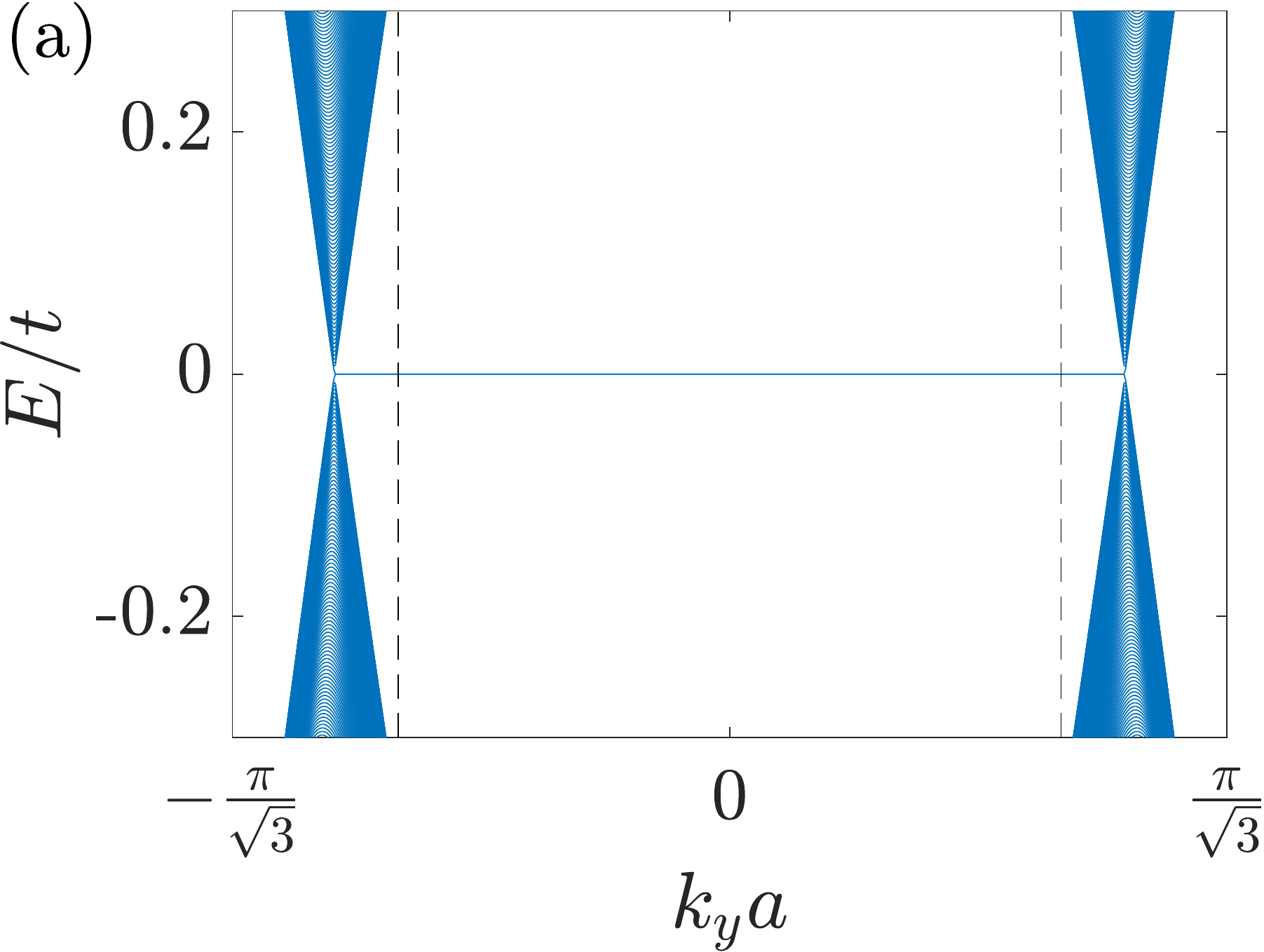}
\includegraphics[width=0.23\textwidth]{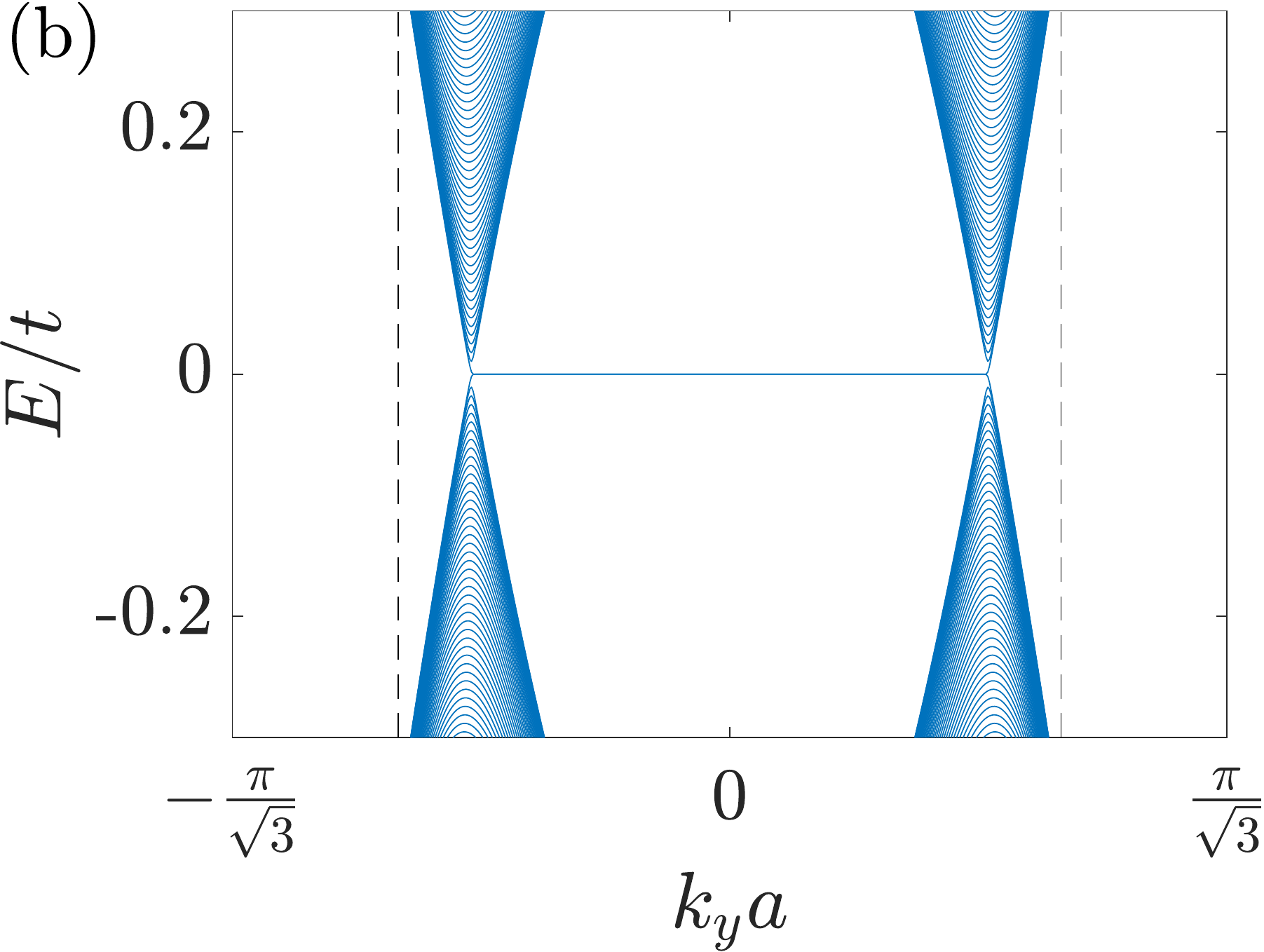}
\includegraphics[width=0.23\textwidth]{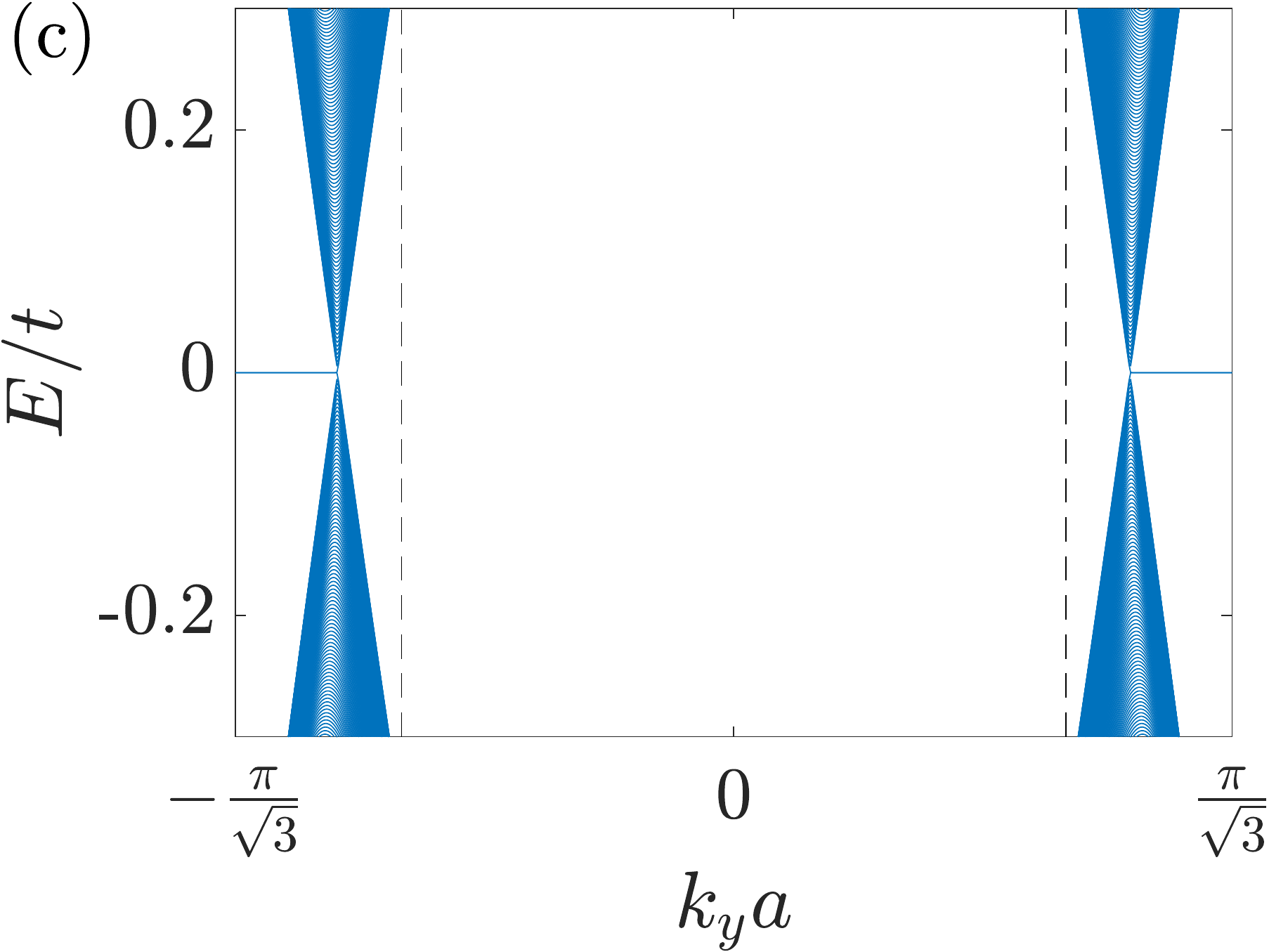}
\includegraphics[width=0.23\textwidth]{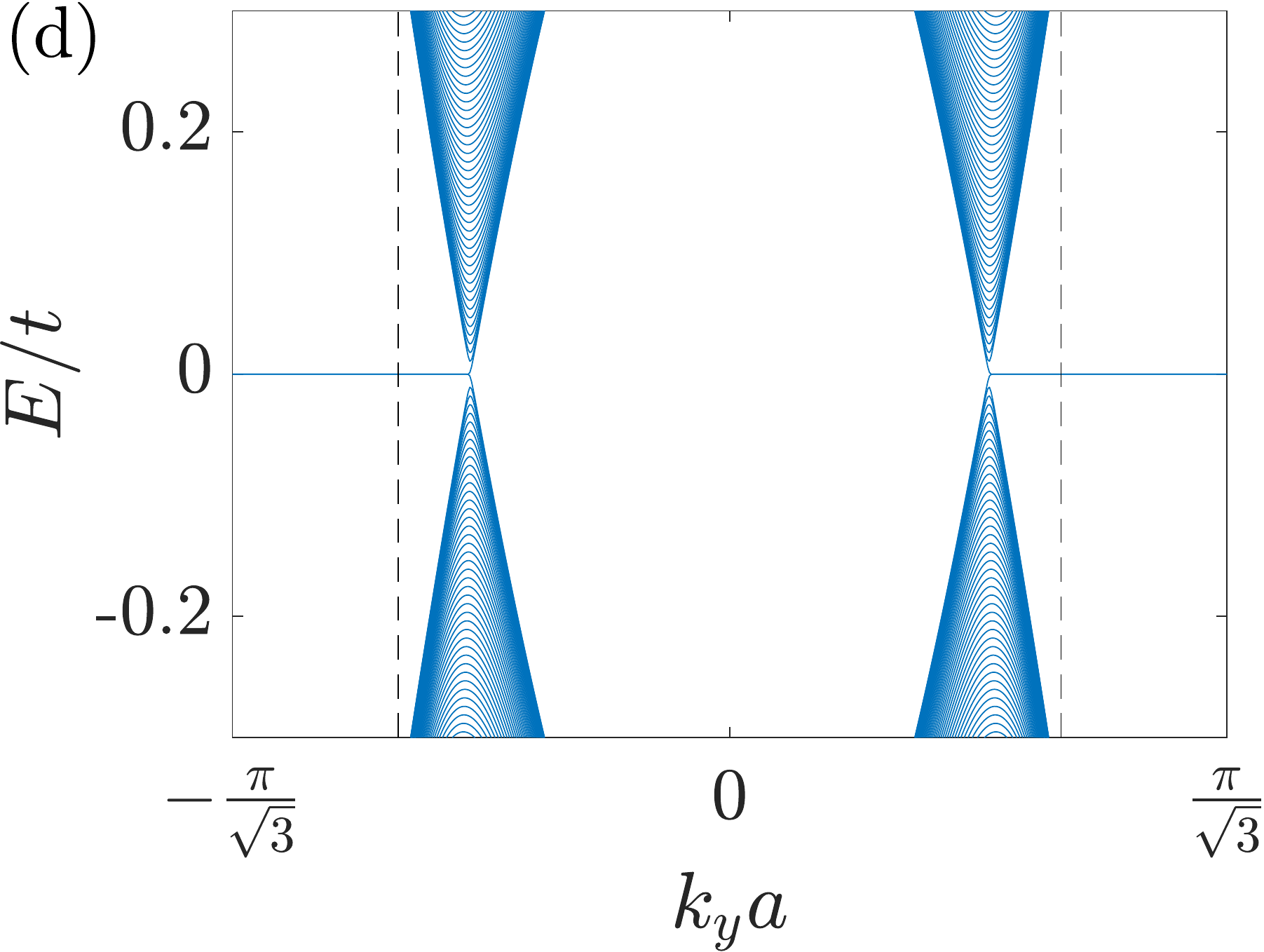}
\includegraphics[width=0.23\textwidth]{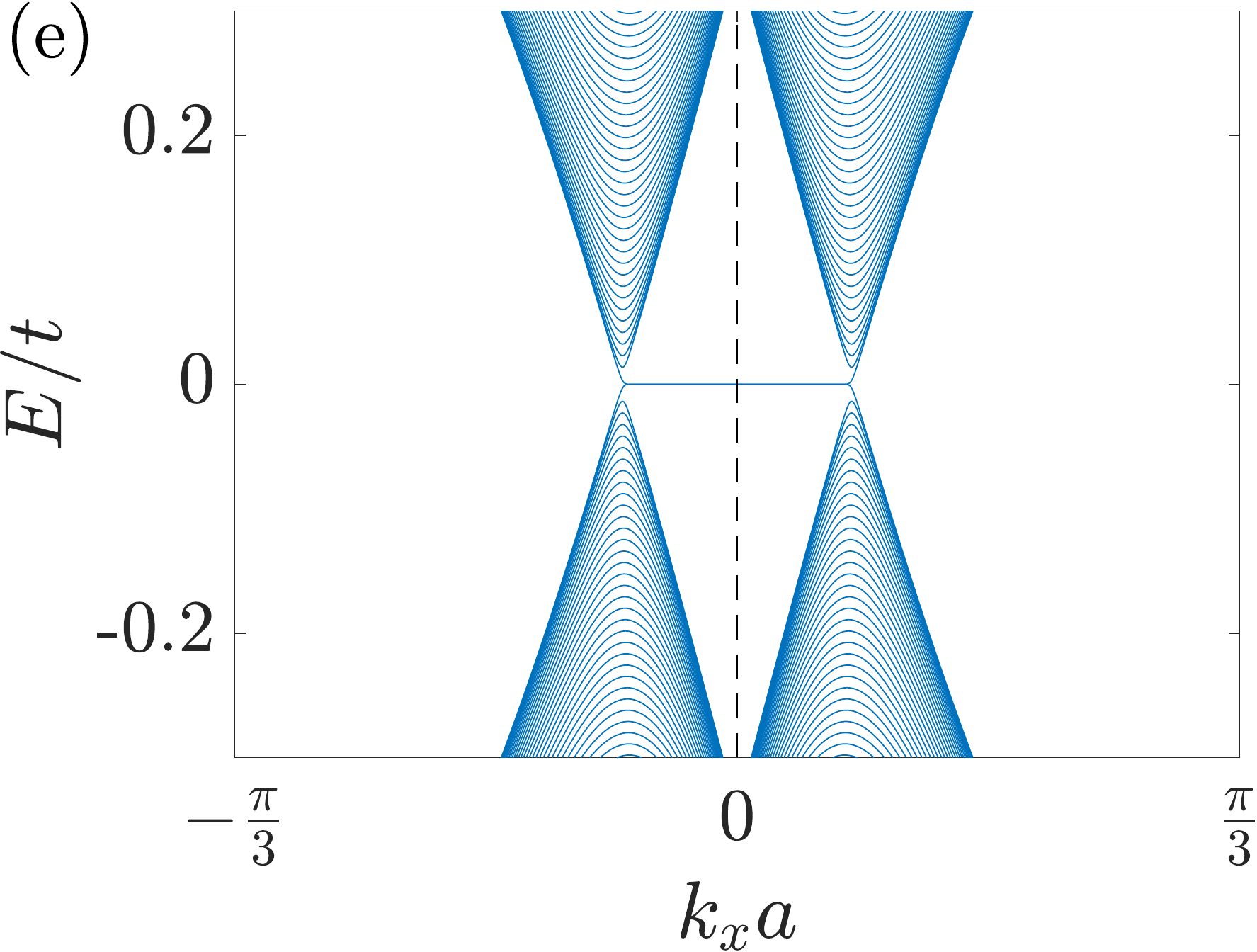}
\caption{
Energy dispersion of a uniformly strained ribbon, \textit{i.e.} when hoppings are spatially constant but not equal. Dirac cones are shifted in momentum space because of the synthetic pseudo-magnetic vector potential, and these Dirac cones are connected by flat lines corresponding to the non-propagating edge states.
Panels~ (a)-(d) are for different configurations with $t_1 \neq t$ and $t_{2,3}=t$ and periodic boundary conditions along $y$.
Specifically, panel~(a) is for bearded terminations on both ends and $t_1< t$, panel~(b) is for bearded terminations on both ends and $t_1> t$, panel~(c) is for zigzag terminations on both ends and $t_1< t$, and panel~(d) is for zigzag terminations on both ends and $t_1> t$. 
Panel~(e) is for armchair terminations and $t_2<t_1=t<t_3$ with periodic boundary conditions along $x$. The vertical dashed lines indicate the position of the Dirac points $K$ and $K'$ of the pristine lattice. In all cases, the non-propagating edge states are doubly degenerate and lives on both open edges.}
\label{fig:localpicture}
\end{figure}

We can understand when these zero-energy edge states can exist in a non-uniformly strained system from a \textit{local picture}.
In this local picture, we assume that the hoppings of the whole system have constant values that are determined by their values at the edge. 
In the uniform case $t_{1,2,3}=t$, the honeycomb lattice has zero-energy edge states localized on one sublattice for the zigzag and the bearded termination, but not for the armchair termination. 
The existence of these non-propagating edge states can be related to a topological quantity called the winding number \cite{Bellec2014, Delplace}.
In the case of a uniform strain, when the hoppings are different $t_{1,2,3}\neq t$ but constant in space, all three types of edges have these zero-energy states, as long as the gap-opening Lifshitz transition is not reached \cite{Kohmoto, Bellec2014}.
In Figs.~\ref{fig:localpicture}(a)-(d), we show the energy dispersion of a uniformly strained ribbon for $t_1 \neq t$ and $t_{2,3}=t$, for bearded and zigzag terminations with periodic boundary conditions along $y$.  Figure~\ref{fig:localpicture}(e) shows the energy dispersion of a uniformly strained ribbon for $t_{2,3} \neq t$ and $t_1=t$ for the armchair termination with periodic boundary conditions along $x$. 
In all cases, we see that non-propagating edge states exist at $E/t = 0$, but are restricted to a limited window of $k_{x,y}$.

As discussed before, due to the chiral symmetry, only the zero-energy edge states that are localized on $A$-sublattice in a local picture can mix with the $0$-th pseudo-Landau level (that for $\tau>0$  is on the $B$-sublattice) and disperse. Due to momentum conservation, this mechanism can happen only between modes at the same $k_{x,y}$ along the edge direction.
These zero-energy edge states are localized on the $A$-sublattice for the bearded termination at the left edge and for the zigzag termination at the right edge. Vice versa, a zigzag termination at the left edge and a bearded termination at the right edge would have zero-energy edge states localized on the $B$-sublattice.
Therefore, the propagating edge states of the $0$-th pseudo-Landau level appear on the bearded edge on the left and on the zigzag edge on the right.

A similar argument can be applied also to the armchair termination. 
In fact, the armchair edge possesses zero-energy edge states when $t_{2} \neq t_3$, as visible in Fig.~\ref{fig:localpicture}(e).
In particular, when $t_2<t_3$ the edge state is localized on the $A$-sublattice on the bottom edge and on the $B$-sublattice at the top edge. 
Vice versa, when $t_2>t_3$ the edge state is localized on the $B$-sublattice on the bottom edge and on the $A$-sublattice at the top edge \cite{Bellec2014}.
Within the local picture, for the particular non-uniform strain given in Eq.~\eqref{uniaxialY} where the hoppings are equal in the center of the ribbon, we have that $t_2<t_3$ at the bottom and $t_2>t_3$ at the top, such that the zero-energy edge state is  localized on the $A$-sublattice at both edges and it can mix with the $n=0$ pseudo-Landau level wavefunction and give rise to a propagating edge state.
 
For the sake of completeness, we now consider the case of a negative $\tau<0$.
In this case, the role of $A$ and $B$-sublattices in Eqs.~\eqref{LLn} and \eqref{LLny} is flipped, and the $0$-th pseudo-Landau level wavefunction is non-zero only on the $A$-sublattice. 
The propagating edge states exist when the pseudo-Landau level wavefunctions mix with states which have a non-zero amplitude on the $B$-sublattice.
This happens on the vertical edges that terminate with a $B$ site, that are the zigzag on the left part and the bearded on the right part, and on both horizontal armchair edges for the strain in Eq.~\eqref{uniaxialY}, with $\tau<0$.

It is worth noticing that the uni-axial strain in Eq.~\eqref{uniaxialX} and~\eqref{uniaxialY} corresponds to a pseudo-magnetic vector potential expressed in two different gauges, hence the existence of propagating edge states is gauge-dependent, as well as termination-dependent. 

As a key remark, we recall that, in the presence of a real magnetic field, the $0$-th Landau level wavefunctions around different valleys are localized on different sublattices. 
A very important consequence of this is that the $0$-th Landau level wavefuntions from different Dirac points can always give rise to propagating edge states, regardless of the shape of the edge or the underlying existence of zero-energy modes.
More details on the case of a honeycomb lattice in the presence of a real magnetic field are given in Section~\ref{sec:magnetic_dispersion}.

We now summarize our criterion. \textit{A particular termination can host a propagating edge state associated with the $0$-th Landau level if, in the previously defined local picture, there is a zero-energy edge state localized on a different sublattice than the $0$-th Landau level wavefunction.}

\section{Propagating edge states of a uni-axially strained system} 
\label{sec:uniaxial_dispersion}

We now numerically validate our criterion by calculating the energy dispersion of a uni-axially strained system with a ribbon geometry and periodic boundary conditions along one direction. 

\subsection{Uni-axial strain along x}
\label{sec:uniaxial_dispersionA}

We first consider a ribbon with the uni-axial strain along $x$ given in Eq.~\eqref{uniaxialX}.
The ribbon is oriented as in Fig.~\ref{fig:system}, with $N_x$ unit cells along the armchair direction and periodic boundary conditions along the $y$ direction.
The ribbon can be terminated on the left and right edges with either a bearded or a zigzag type of edge.
Since the strain in Eq.~\eqref{uniaxialX} breaks translational invariance only along the $x$-direction, we can diagonalize the tight-binding Hamiltonian in the quasi-momentum space $k_y$ and obtain the energy dispersion.

\begin{figure}[t]
\centering
\includegraphics[width=0.47\textwidth]{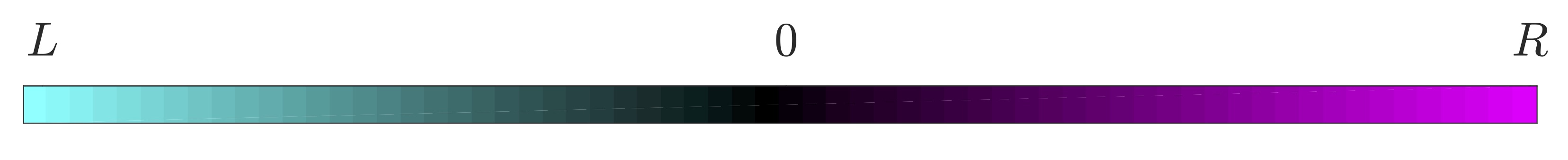}
\includegraphics[width=0.23\textwidth]{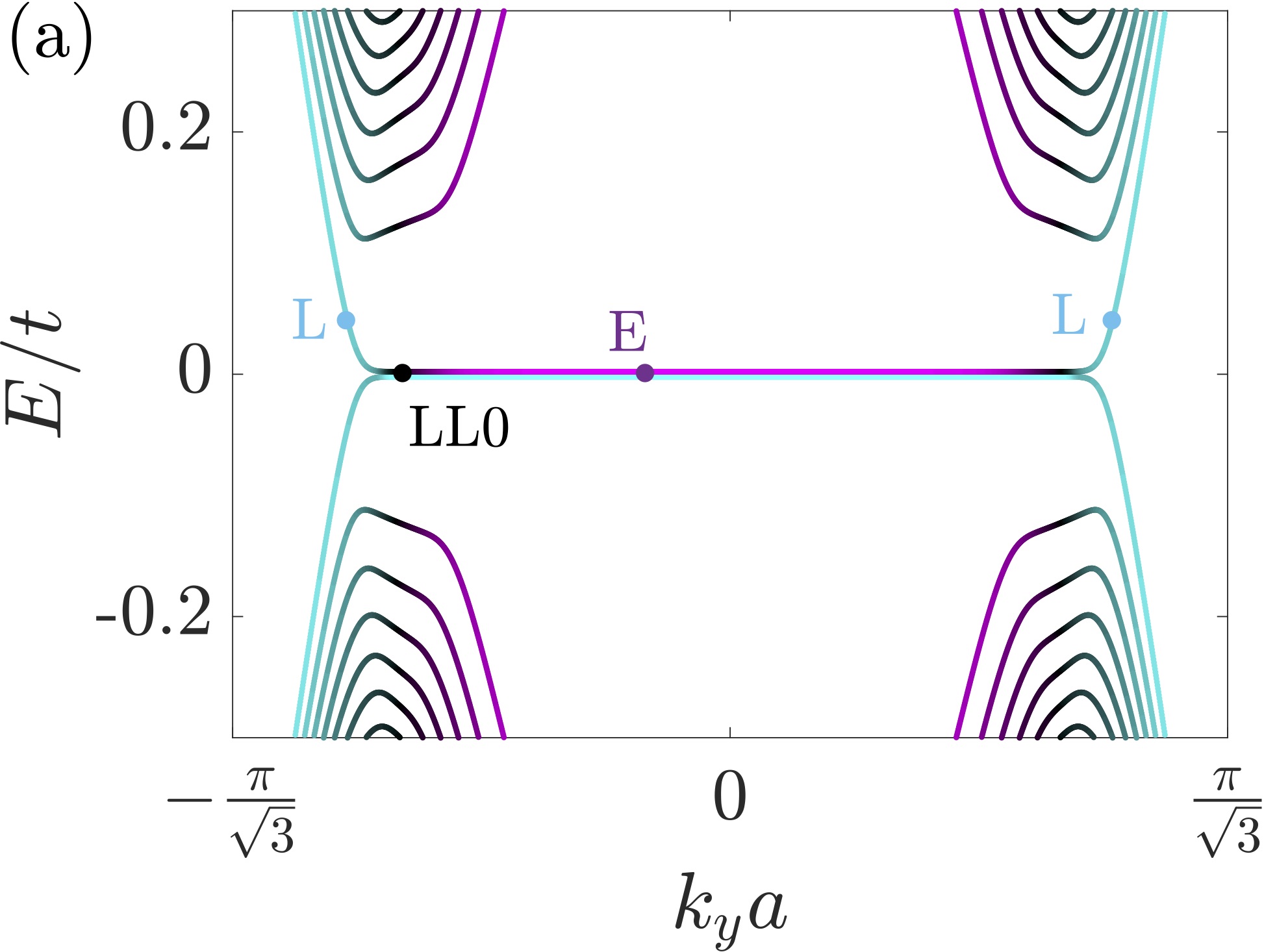}
\includegraphics[width=0.23\textwidth]{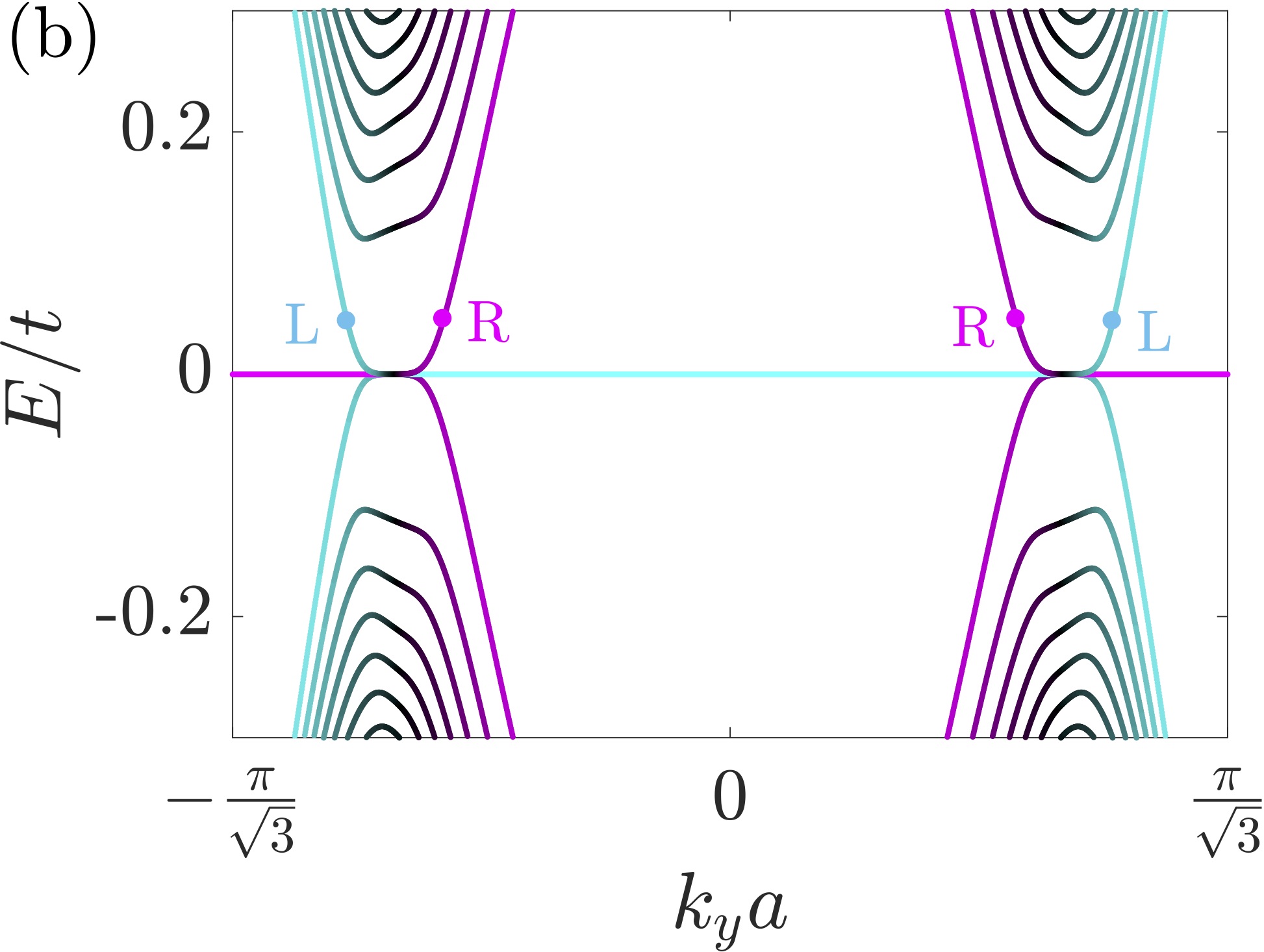}
\includegraphics[width=0.23\textwidth]{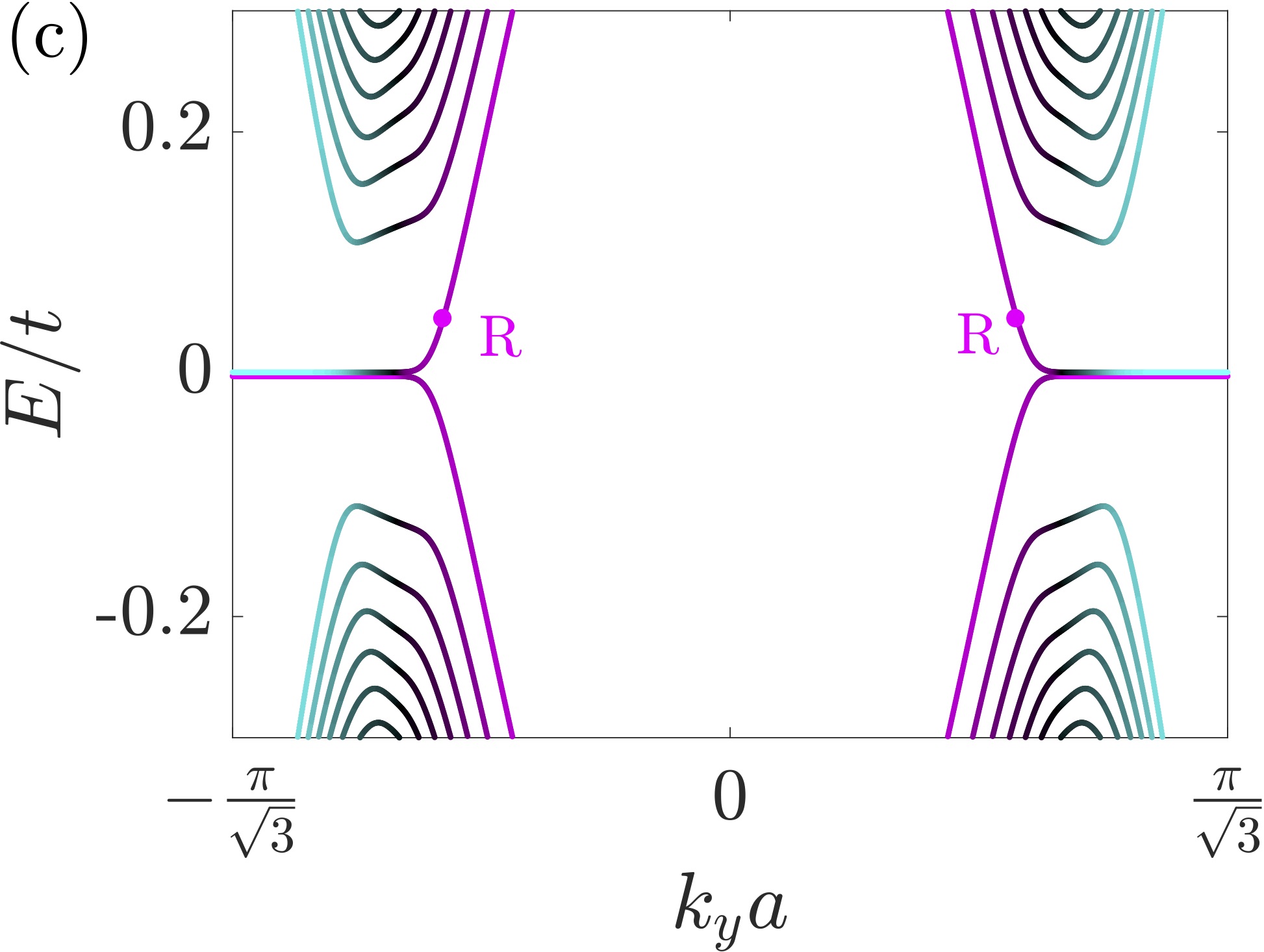}
\includegraphics[width=0.23\textwidth]{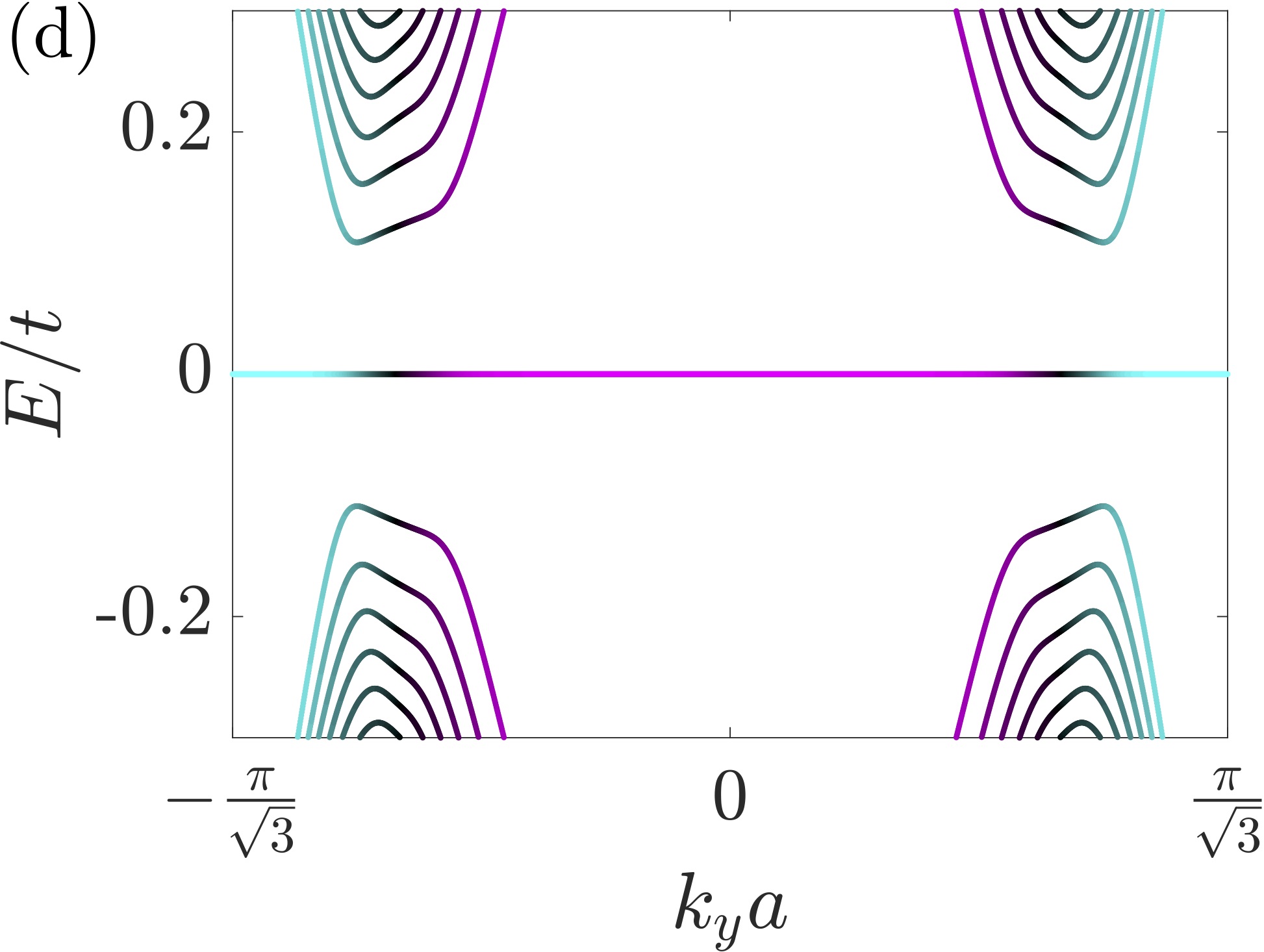}
\caption{Energy dispersion for a uni-axially strained ribbon along $x$, oriented as in Fig.~\ref{fig:system}, with $N_x=99$ along the armchair direction and periodic boundary conditions along $y$, for different terminations. The strain strength is $\tau=0.015$, corresponding for parameters of solid-state graphene, to a pseudo-magnetic field of $\mathcal{B}=30$~T. 
Panel~(a) is for bearded terminations on both ends, panel (b) is for a bearded termination on the left and a zigzag termination on the right, panel~(c) is for zigzag terminations on both ends and panel~(d) is for a zigzag termination on the left and a bearded termination on the right.
Each state is colored according to the mean $\langle x \rangle$ position of its spatial wavefunction, where cyan stands for states localized on the left edge, magenta stands for states localized on the right edge and black is for bulk states.
The spatial wavefunction of the states associated to the cyan, magenta, violet and black dots are plotted in Fig.~\ref{fig:wavefunction_LR}. Labels L and R indicate whether the eigenstate is localized on the left or on the right edge. The label E indicates the non-propagating edge state, while the label LL0 is used to indicate an example of a $n=0$ Landau level state.}
\label{fig:Spectra_all_terminations}
\end{figure}

In Fig.~\ref{fig:Spectra_all_terminations} we show the low-energy dispersion of a ribbon with $N_x=99$ and $\tau=0.015$ for various edge terminations on the left and right edges. 
Figure~\ref{fig:Spectra_all_terminations}(a) is for bearded terminations on both ends, Fig.~\ref{fig:Spectra_all_terminations}(b) is for a bearded termination on the left and a zigzag termination on the right, Fig.~\ref{fig:Spectra_all_terminations}(c) is for zigzag terminations on both ends and Fig.~\ref{fig:Spectra_all_terminations}(d) is for a zigzag termination on the left and a bearded termination on the right.
For each state associated to the energy dispersion in Fig.~\ref{fig:Spectra_all_terminations}, we have calculated the mean position of the corresponding spatial wavefunction. This mean position is reported as a color scale in the energy dispersion. States that are localized on the left (right) edge are colored in cyan (magenta), while bulk states are represented with darker colors.

In all panels we see the appearance of quantized relativistic pseudo-Landau levels in the vicinity of the two $K$ and $K'$ points, located at $k_y a= 2\pi/(3\sqrt{3})$ and $k_y a= -2\pi/(3\sqrt{3})$ respectively.  
The level energies follow a square-root law and their tilting is due to the slight spatial non-uniformity of the Dirac velocity \cite{Salerno}. 

Near the Dirac points, states with different $k_y$ correspond to pseudo-Landau level wavefunctions with different guiding centers $x_0$. 
When the guiding center reaches the physical edge of the system there is a sudden increase in the energy dispersion of the $n \neq 0$ pseudo-Landau levels, as we discussed earlier.
These states are localized at the edge and have a non-zero group velocity along the $y$-direction. 
We indeed see, from Fig.~\ref{fig:Spectra_all_terminations}, that for the higher pseudo-Landau levels ($|n|>0$) each valley has propagating edge states with both positive and negative velocities. This is in agreement with our criterion since these higher pseudo-Landau levels have a non-zero component on both sublattices.
The four panels differ in the behaviour of the edge states of the $0$-th pseudo-Landau level, which can be understood fully through our criterion, as explained in Sec.~\ref{sec:criterion}. For sake of simplicity, we focus only on the propagating edge states with positive energies, as the ones with negative energies can be obtained, due to the chiral symmetry, by flipping the sign of the $B$-sublattice component of the wavefunction.

\begin{figure}[t]
\centering
\includegraphics[width=0.23\textwidth]{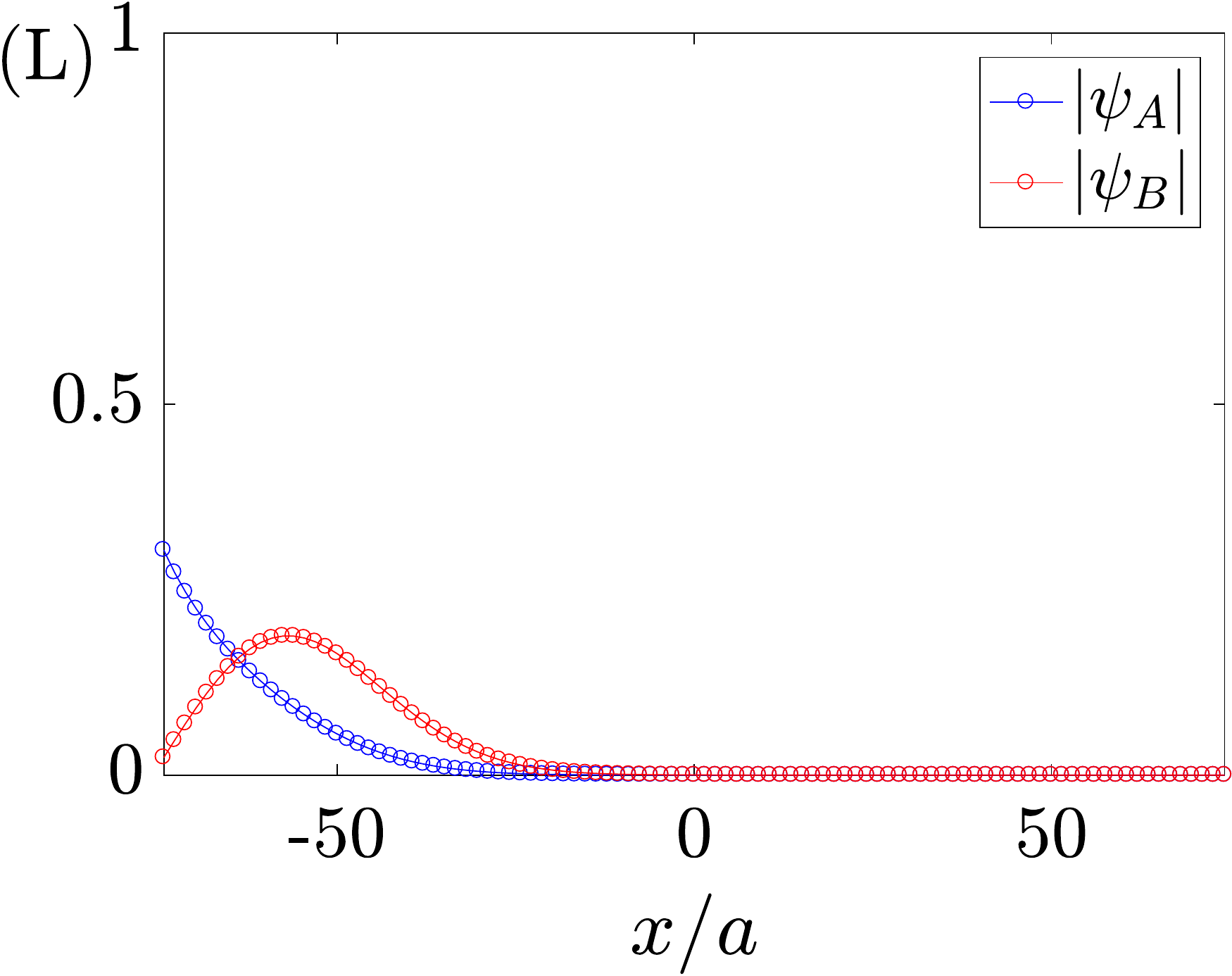}
\includegraphics[width=0.23\textwidth]{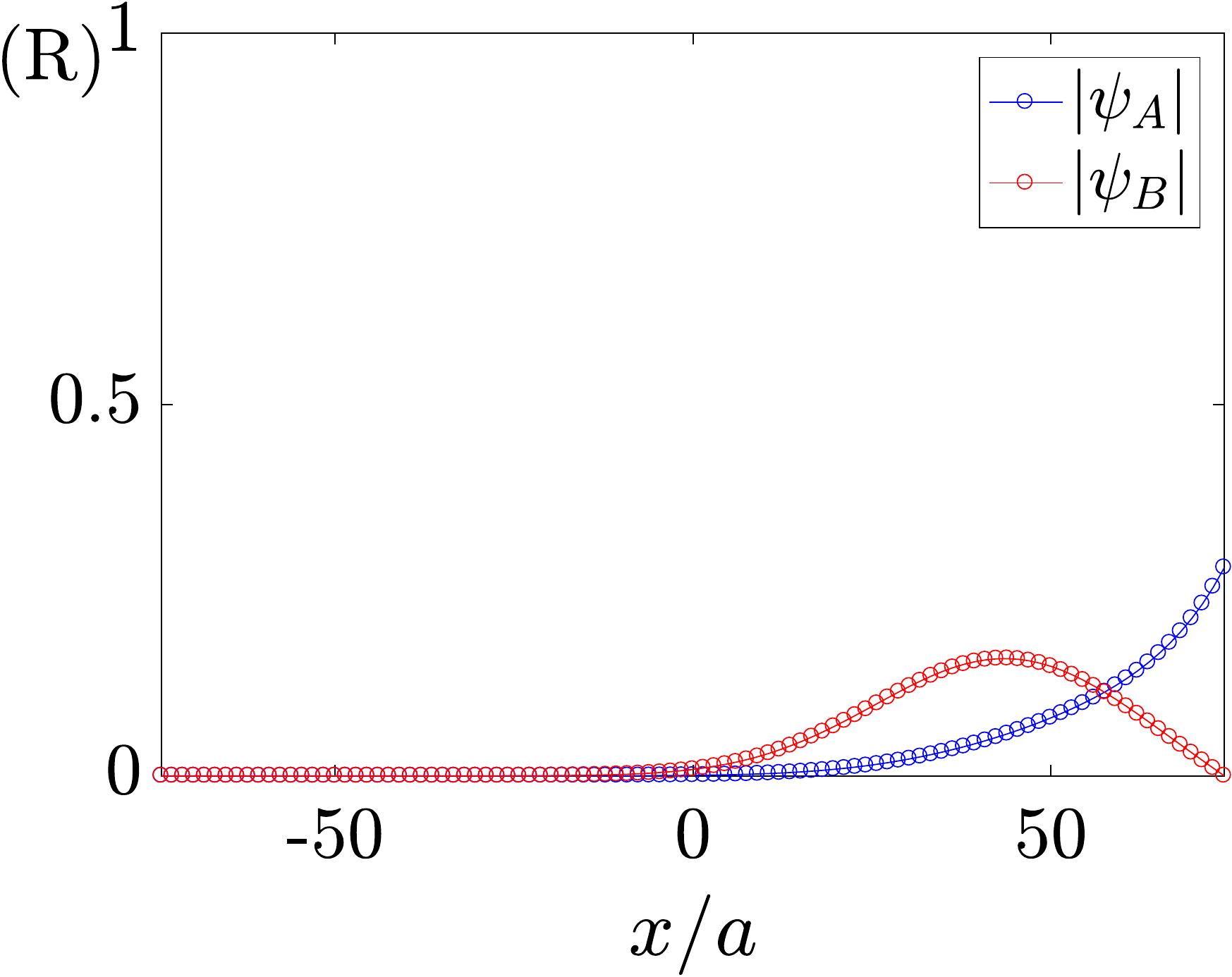}
\includegraphics[width=0.23\textwidth]{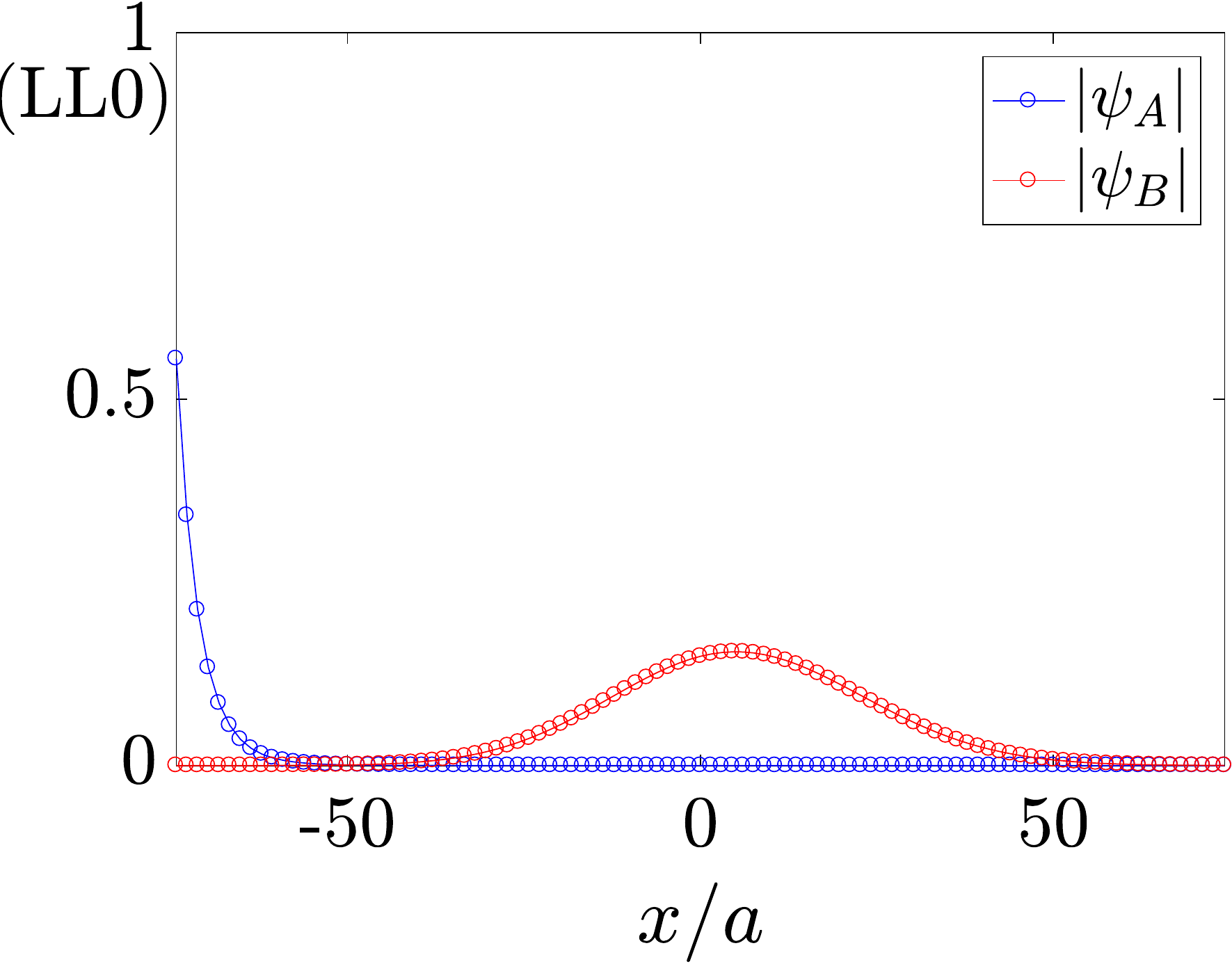}
\includegraphics[width=0.23\textwidth]{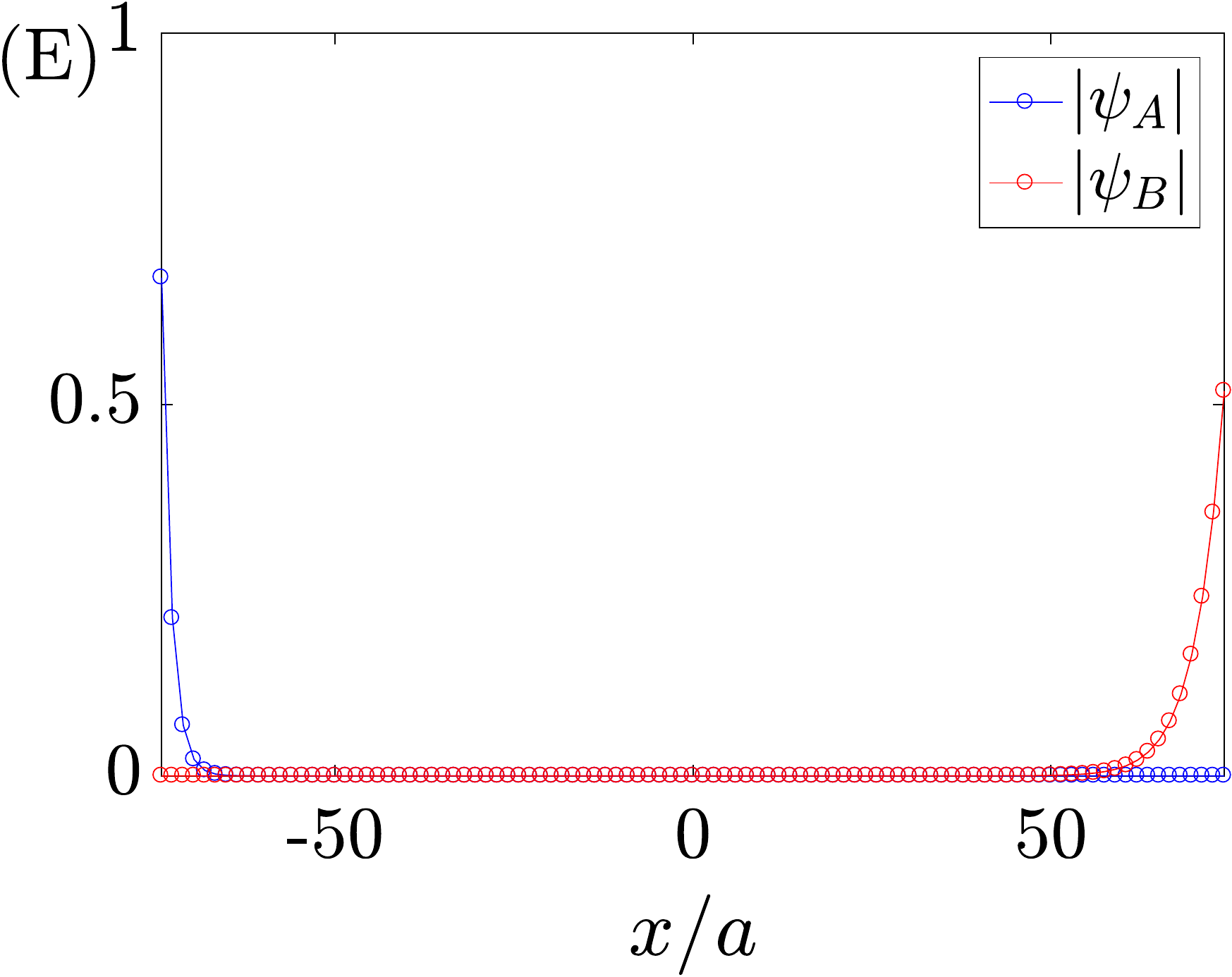}
\caption{Modulus of the numerical eigenfunctions for parameters in Fig.~\ref{fig:Spectra_all_terminations}. Panel (L) corresponds to the cyan dots in Fig.~\ref{fig:Spectra_all_terminations}, panel (R) corresponds to the magenta dots in Fig.~\ref{fig:Spectra_all_terminations}. Panel (LL0) corresponds to the black dot in Fig.~\ref{fig:Spectra_all_terminations}(a), where we see the wavefunction of the pseudo-Landau level with $n=0$ together with the non-propagating edge state of the bearded left end. Panel (E) corresponds to the violet dot in Fig.~\ref{fig:Spectra_all_terminations}(a), showing the superposition of the doubly degenerate non-propagating edge states.}
\label{fig:wavefunction_LR}
\end{figure}

In Figs.~\ref{fig:Spectra_all_terminations}(a) and \ref{fig:Spectra_all_terminations}(c), we see that each valley has only one propagating edge state, while in Fig.~\ref{fig:Spectra_all_terminations}(b), the two valleys have both a pair of propagating edge states. 
The states are labelled in Fig.~\ref{fig:Spectra_all_terminations} with L or R and colored in cyan or magenta according to their spatial wavefunction being localized at the left or at the right edge, as we see in Fig.~\ref{fig:wavefunction_LR}. 
We also notice from Fig.~\ref{fig:Spectra_all_terminations} that the group velocity of the lowest propagating edge states for the two $K$ and $K'$ valleys is opposite, therefore these states are helically-propagating.
Finally, in Fig.~\ref{fig:Spectra_all_terminations}(d), we show the situation in which there are no propagating edge states of the $n=0$ pseudo-Landau level. 

In Fig.~\ref{fig:wavefunction_LR} we show the spatial structure of the wavefunction corresponding to the propagating edge states identified by the dots in Fig.~\ref{fig:Spectra_all_terminations}. 
The contribution to the wavefunction from the two sublattices $A$ and $B$ has been separated and plotted in respectively blue and red. 
Figure~\ref{fig:wavefunction_LR}(L) represents the wavefunction of the states identified with the cyan dots and labelled with L in Fig.~\ref{fig:Spectra_all_terminations}, while Fig.~\ref{fig:wavefunction_LR}(R) represents the wavefunction of the states identified with the magenta dots and labelled with R in Fig.~\ref{fig:Spectra_all_terminations}.
Figure~\ref{fig:wavefunction_LR}(LL0) shows the wavefunction of a superposition of the doubly degenerate states highlighted by the black dot in Fig.~\ref{fig:Spectra_all_terminations}(a). Such states are a pseudo-Landau level with $n=0$ with a guiding center in the middle of the system and the non-propagating edge state of the left (bearded) end.
Figure~\ref{fig:wavefunction_LR}(E) corresponds to the violet dot in Fig.~\ref{fig:Spectra_all_terminations}(a), and shows the wavefunction of a superposition of the doubly degenerate non-propagating edge states on each bearded edge. We notice that these non-propagating edge states are localized on a much shorter distance than the propagating states of Figs.~\ref{fig:wavefunction_LR}(L) and \ref{fig:wavefunction_LR}(R). In fact, the propagating edge states are mixed with the pseudo-Landau levels, which are localized on a longer length-scale set by the magnetic length.

All these results are in perfect agreement with our criterion for the existence of the propagating edge states of the $0$-th pseudo-Landau level discussed in Section~\ref{sec:criterion}.

\subsection{Uni-axial strain along y}
\label{sec:uniaxial_dispersionB}

We now consider a ribbon with the uni-axial strain along $y$ given in Eq.~\eqref{uniaxialY}. 
The ribbon is still oriented as in Fig.~\ref{fig:system}, but now has $N_y$ unit cells along the vertical direction and periodic boundary conditions along the $x$ direction. We diagonalize the tight-binding Hamiltonian in the quasi-momentum space $k_x$ and obtain the energy dispersion.

\begin{figure}[t]
\centering
\includegraphics[width=0.23\textwidth]{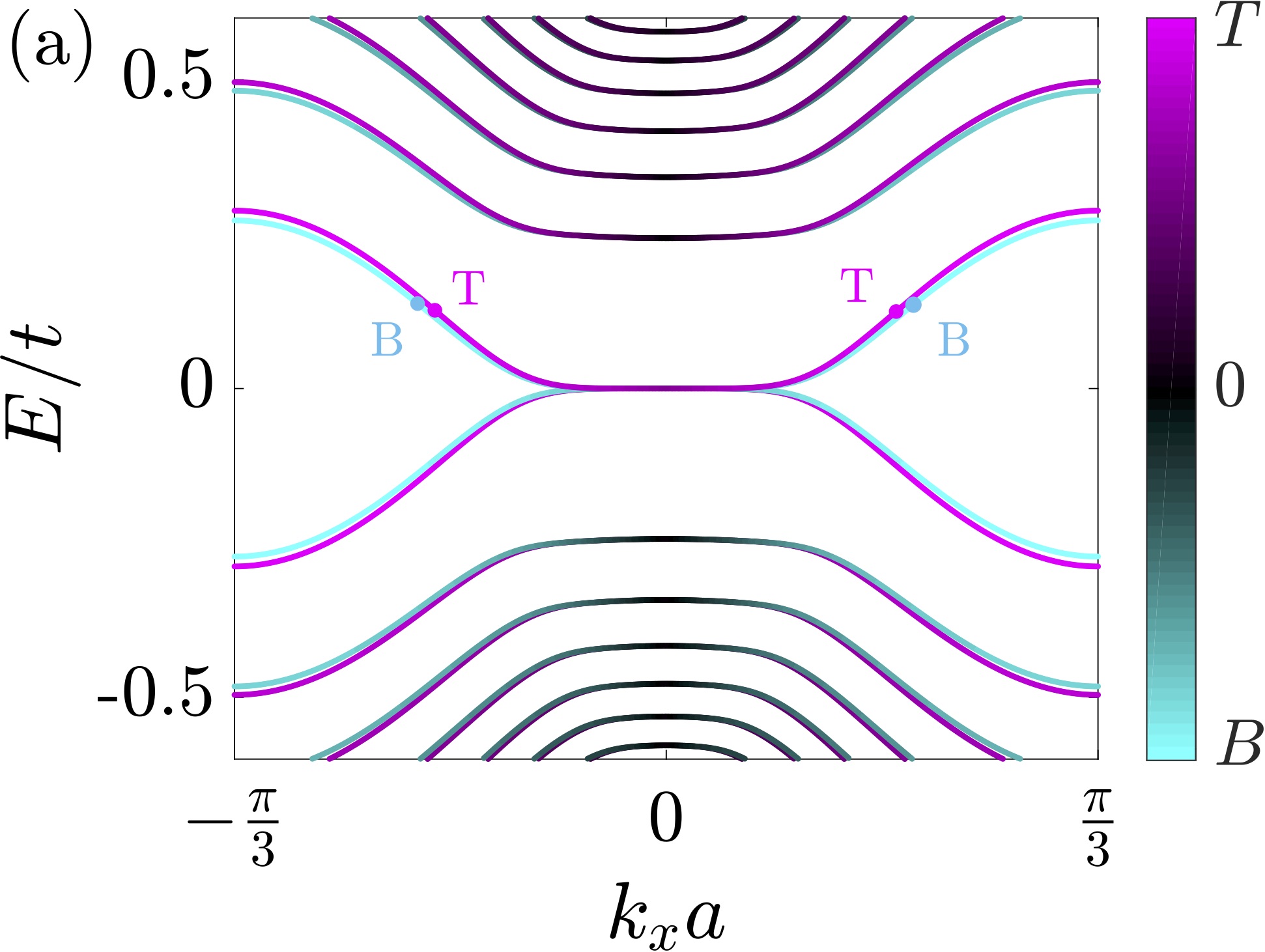}
\includegraphics[width=0.23\textwidth]{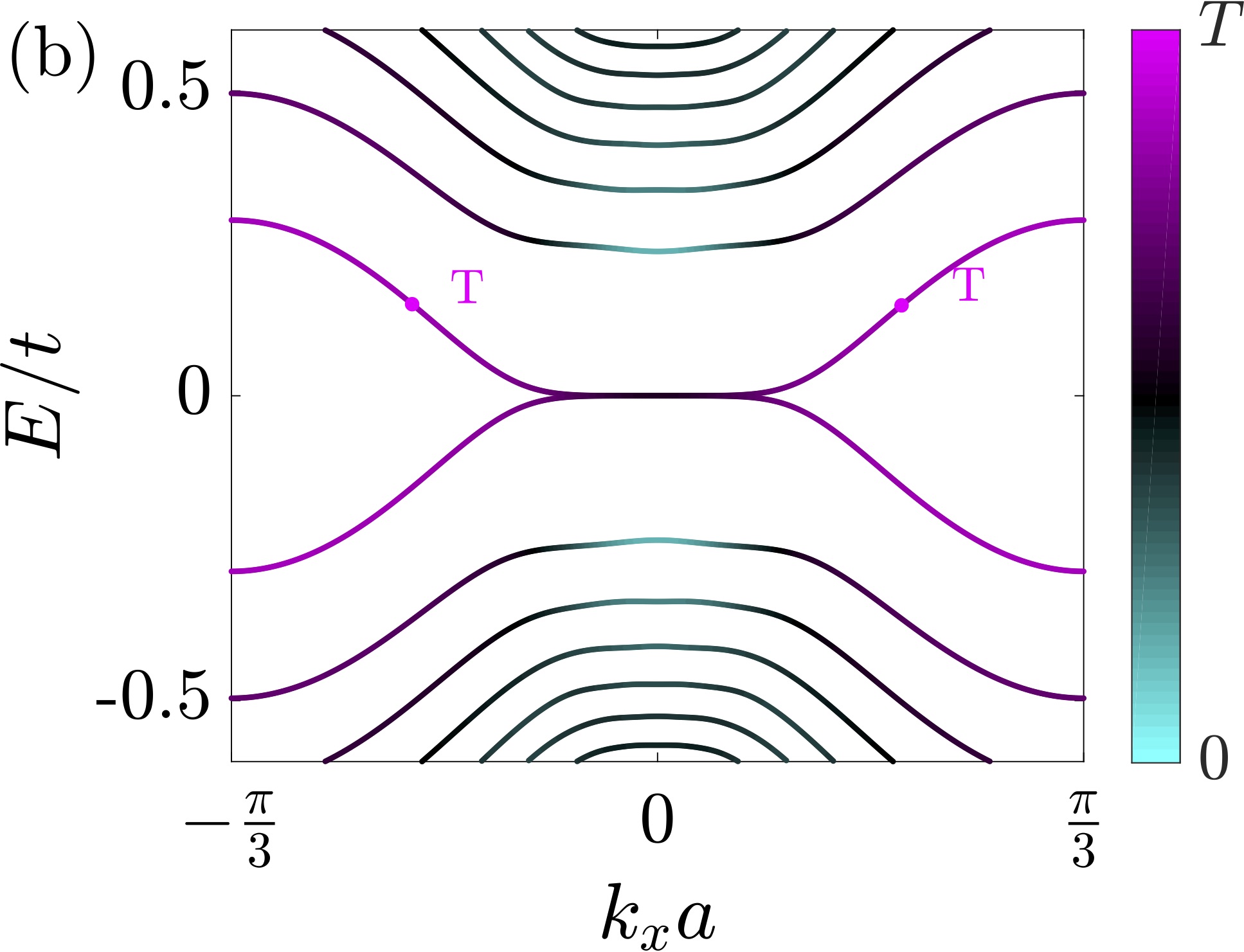}
\includegraphics[width=0.225\textwidth]{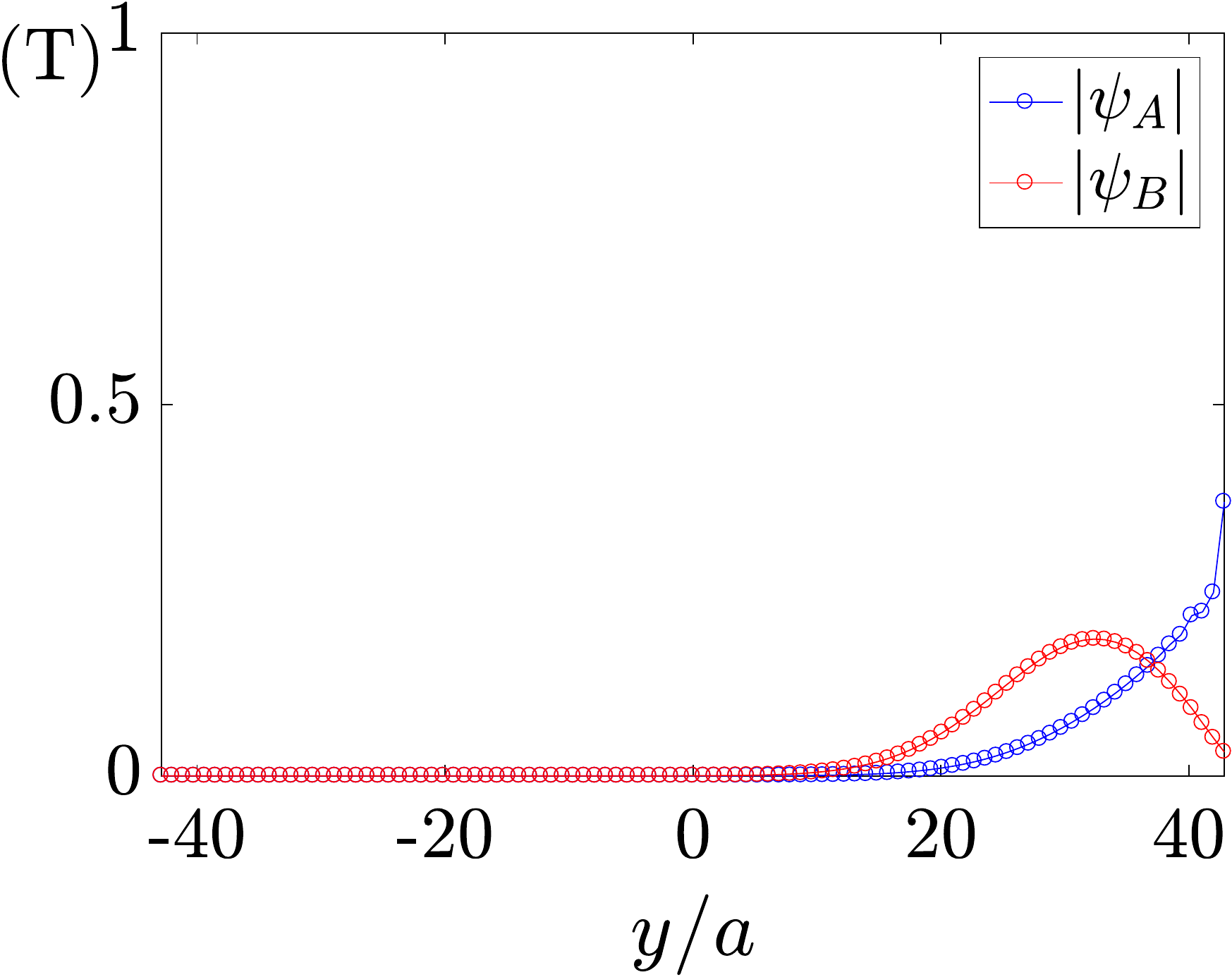}
\includegraphics[width=0.225\textwidth]{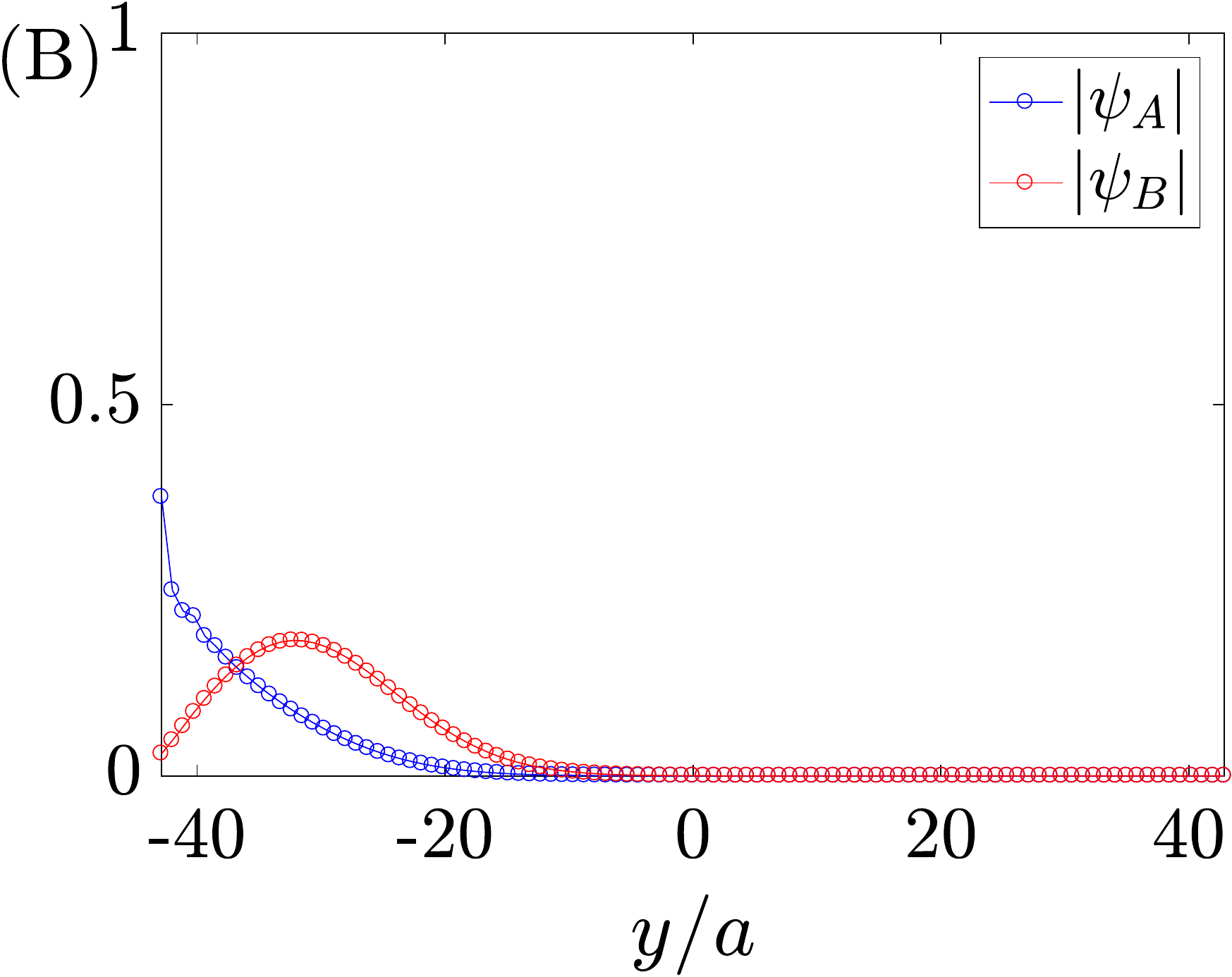}
\caption{Energy dispersion and modulus of the numerical wavefunctions for a uni-axially strained ribbon along $y$ with periodic boundary conditions along $x$.
The strain strength is $\tau=0.02$, corresponding to a pseudo-magnetic field of $\mathcal{B}=120$~T, for parameters of solid-state graphene.
In panel~(a) $N_y=99$ along the vertical direction and $t_{2,3}=t$ is in the center of the ribbon, thus both the top and the bottom armchair edge can sustain propagating edge states. In panel~(b) $N_y=49$ along the vertical direction and the hoppings are equal at the bottom edge, where no propagating edge state is supported. 
Each state is colored according to the mean $\langle y \rangle$ position of its spatial wavefunction, where cyan stands for states localized on the bottom edge, magenta stands for states localized on the top edge and black is for bulk states.
The spatial wavefunction of the states associated with the cyan and magenta dots are plotted in panel~(B)~and~(T), corresponding to the eigenstate localized on the bottom or on the top edge, for parameters in panel~(a). }
\label{fig:Spectra_all_terminations_armchair}
\end{figure}

In Fig.~\ref{fig:Spectra_all_terminations_armchair} we show the low-energy dispersion of a ribbon with $\tau=0.02$.
Figure~\ref{fig:Spectra_all_terminations_armchair}(a) is obtained for a ribbon of $N_y=99$ unit cells along the vertical direction where the hoppings are equal at the center of the ribbon.
In this case, as we have discussed in the previous Section, both the top and the bottom armchair edge can sustain propagating edge states.
The energy dispersion shows the quantized pseudo-Landau levels around the two Dirac points, both located at $k_xa=0$, with a pair of propagating edge states localized at the top and bottom edges, as indicated with respectively the T and B label
\footnote{The small energy splitting between the top and bottom edge states in Fig.~\ref{fig:Spectra_all_terminations_armchair}(a) is due to the breaking of the reflection symmetry in the $y$ direction, which comes from the choice of $y=0$ in implementing the strain in Eq.~\ref{uniaxialY}.}.
As done in Fig.~\ref{fig:Spectra_all_terminations}, we have calculated the mean $\langle y\rangle$ position of the spatial wavefunction, which is reported as a color scale in the energy dispersion. States that are localized on the top (bottom) edge are colored in magenta (cyan), while bulk states are represented with darker colors. The eigenstates corresponding to the states indicated with the magenta and cyan dots are shown in Figs.~\ref{fig:Spectra_all_terminations_armchair}(T)-(B) respectively. 
Figure~\ref{fig:Spectra_all_terminations_armchair}(b) shows the energy dispersion of a ribbon of $N_y=49$ unit cells along the vertical direction where the hoppings are equal at the bottom end of the ribbon. This configuration has been previously discussed in \cite{Atteia} and propagating edge states were found only at the top edge. 
A similar configuration was also discussed in \cite{Ghaemi}, where a semi-infinite system was considered instead of a finite system, showing that propagating edge states can not exist on the bottom edge.
This result for the armchair edge is well explained by our criterion. In this case, the bottom armchair edge does not have a zero-energy edge state in the local picture. At the top end, instead, the strain is such that in the local picture, the zero-energy edge state is present. These predictions are confirmed in Fig.~\ref{fig:Spectra_all_terminations_armchair}(b) where we see a propagating state in the dispersion localized only at the top edge.

\section{Propagating edge states of a system in the presence of a real magnetic field.}
\label{sec:magnetic_dispersion}

We now present the case of a honeycomb lattice in the presence of a real magnetic field, and clarify the differences with the case of a strained system.
In particular, we show that our criterion for the existence of the propagating edge states can also be applied to the case of the chiral edge states of a system in the presence of a real magnetic field.

The effect of a real magnetic field $\mathcal{B}_r$ can be expressed, within a Landau gauge with the vector potential oriented along $y$, as a complex hopping in the chirally-symmetric tight-binding Hamiltonian: 
\begin{align}
\mathcal{H} = -\sum_{\mathbf{r}} t \Big( &\hat{a}^{\dagger}_{\mathbf{r} - \mathbf{R}_1} \hat{b}_{\mathbf{r}}+ \hat{a}^{\dagger}_{\mathbf{r} - \mathbf{R}_2} \hat{b}_{\mathbf{r}} e^{-i \pi \varphi \mathbf{r}\cdot \mathbf{R}_1} + \notag \\& \hat{a}^{\dagger}_{\mathbf{r} - \mathbf{R}_3} \hat{b}_{\mathbf{r}}e^{i \pi \varphi \mathbf{r}\cdot \mathbf{R}_1} +\text{H.c.}\Big),
\label{magnetic_ham}
\end{align}
where the magnetic flux per plaquette is $e \mathcal{B}_r S/\hbar = 2\pi\varphi$, and $S=3\sqrt{3}a^2/2$ is the area of the hexagonal plaquette.
It is straightforward to diagonalize the Hamiltonian in Eq.~\eqref{magnetic_ham} by applying periodic boundary conditions along $y$ to obtain the Landau levels and the resulting wavefunctions that depend on the valley index $\xi$ \cite{Goerbig}. 
A very important consequence of such valley index dependence is that the $0$-th Landau level wavefuntions from different valleys are localized on different sublattices. 
For a positive magnetic field, the $0$-th Landau level wavefunction of the $K$($K'$) valley is localized on the $B$($A$)-sublattice.
The $0$-th Landau level wavefunctions can therefore always give rise to propagating edge states, regardless of the type of site on the edge. Moreover, these edge states are chirally propagating, because the real magnetic field breaks the time-reversal symmetry.

\begin{figure}[t]
\centering
\includegraphics[width=0.47\textwidth]{colorbar.jpg}
\includegraphics[width=0.23\textwidth]{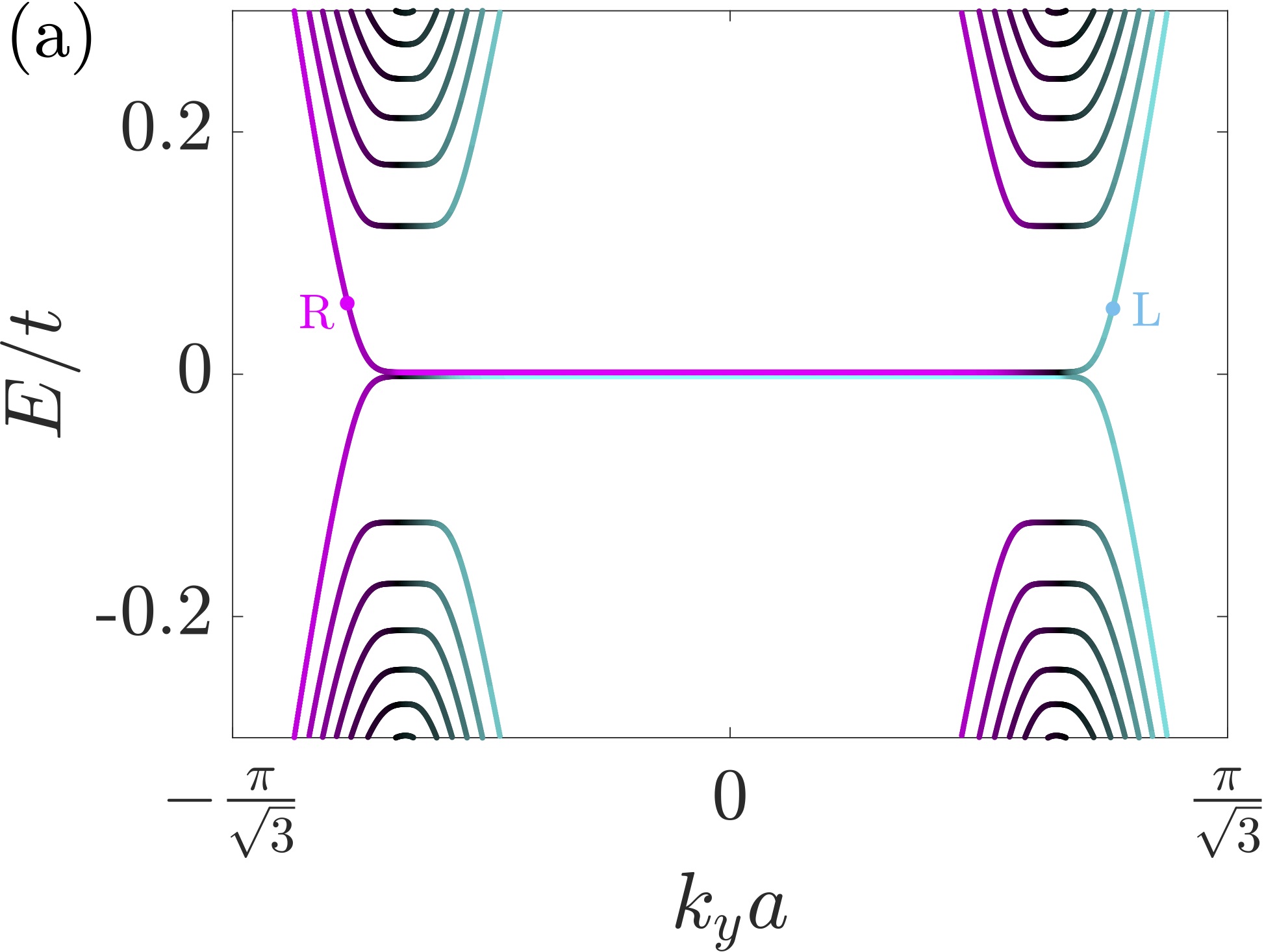}
\includegraphics[width=0.23\textwidth]{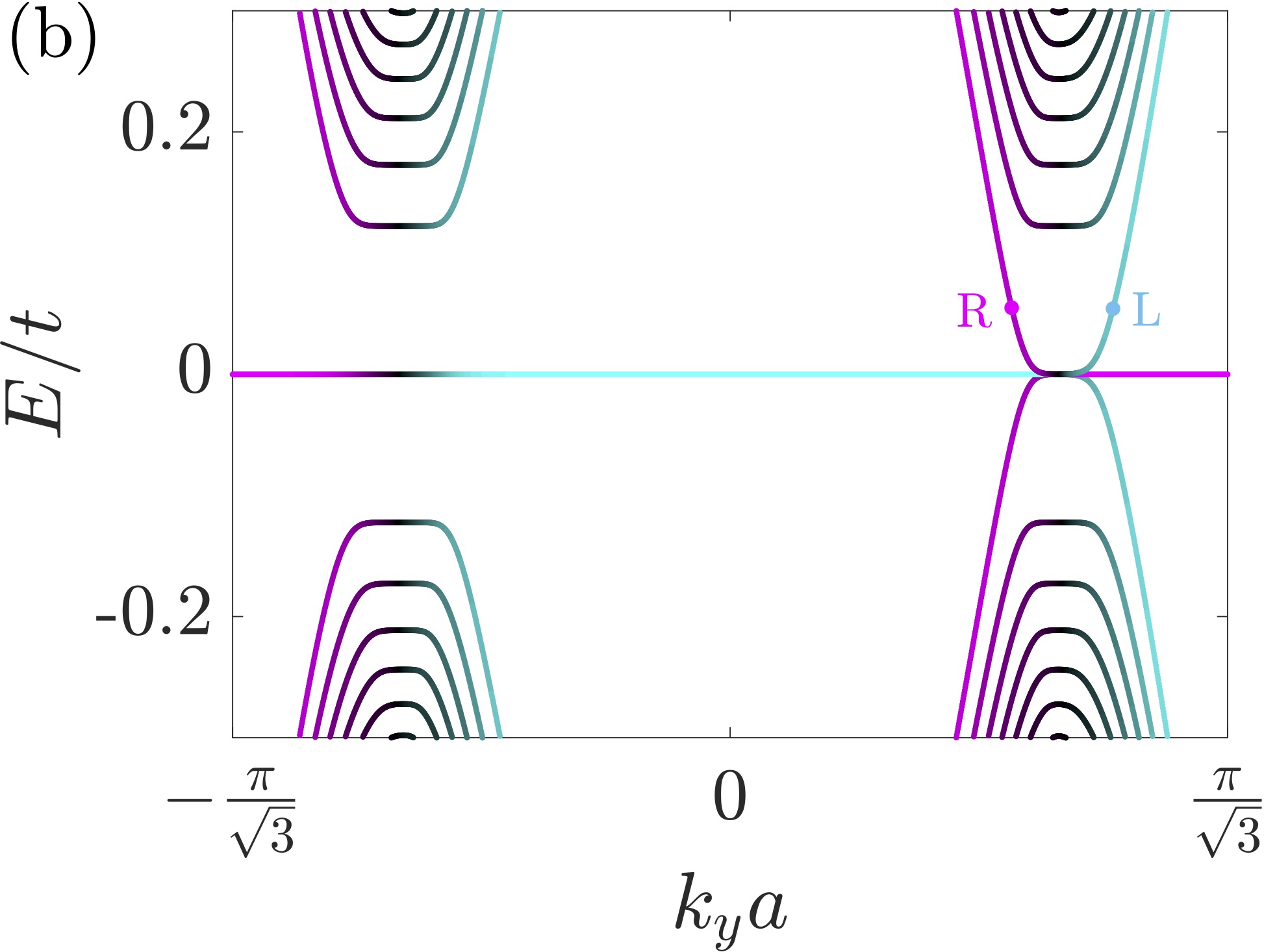}
\includegraphics[width=0.23\textwidth]{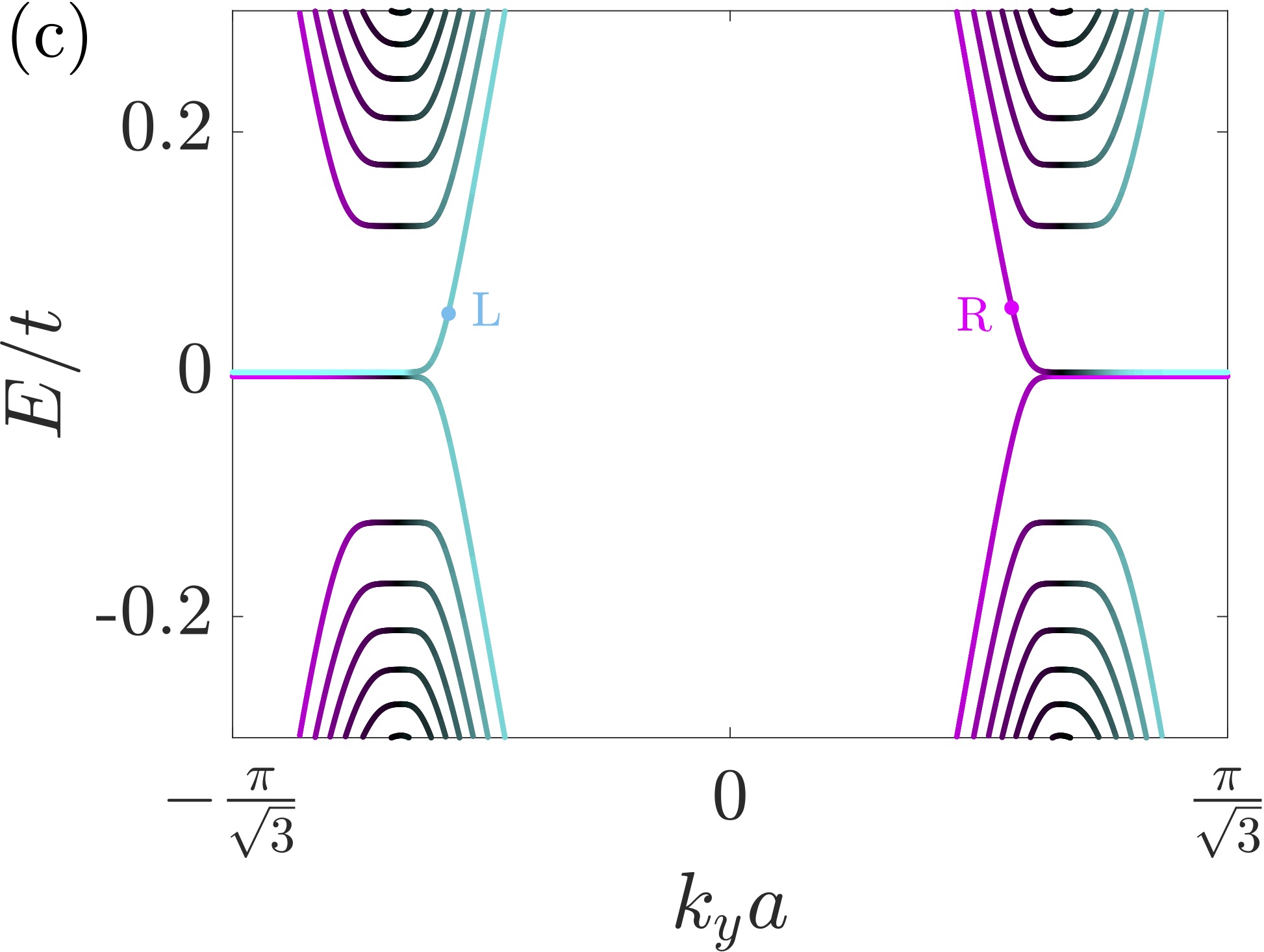}
\includegraphics[width=0.23\textwidth]{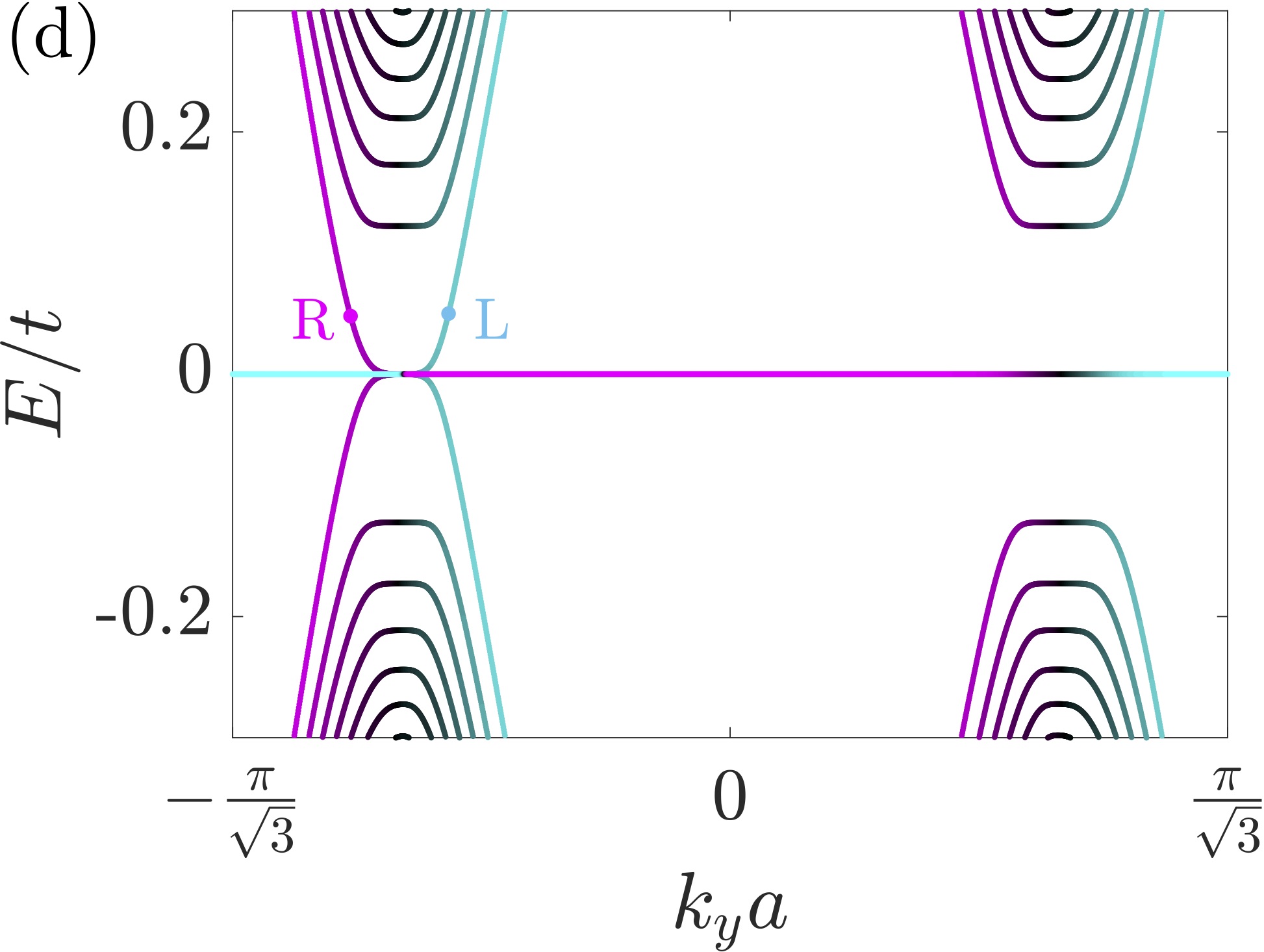}
\caption{Energy dispersion in the presence of a real magnetic field for a ribbon of $N_x=99$ along the armchair direction and periodic boundary conditions along $y$, with different terminations on the left and right edges. The magnetic flux per plaquette has strength $2\pi \varphi = 0.0087$. The intensity of the associated real magnetic field, $\mathcal{B}_r=30$~T for parameters of solid-state graphene, is the same as the pseudo-magnetic field generated by the strain in Fig.~\ref{fig:Spectra_all_terminations}. 
Panel~(a) is for bearded terminations on both ends, panel~(b) is for a bearded termination on the left and a zigzag termination on the right, panel~(c) is for zigzag terminations on both ends and panel~(d) is for a zigzag termination on the left and a bearded termination on the right.
The states are colored according to the mean position of their wavefunction.}
\label{fig:Spectra_all_terminations_real_field}
\end{figure}

Our criterion developed in Section~\ref{sec:criterion} can also be applied to the case of the magnetic field to predict which valley hosts the propagating edge states. 
In Fig.~\ref{fig:Spectra_all_terminations_real_field} we show the energy dispersion calculated for a ribbon of $N_x=99$ unit cells along the armchair direction and periodic boundary conditions along the $y$-direction, with a flux per plaquette $2\pi \varphi=0.0087$. 
Figures~\ref{fig:Spectra_all_terminations_real_field}(a)-(d) are respectively for: bearded terminations on both ends; a bearded termination on the left and a zigzag termination on the right; zigzag terminations on both ends; and a zigzag termination on the left and a bearded termination on the right. 
As previously done in Figs.~\ref{fig:Spectra_all_terminations}-\ref{fig:Spectra_all_terminations_armchair}, each state is colored according to the mean $\langle x \rangle$ position of its spatial wavefunction, as indicated by the color bar.
We notice that, differently from the case of a strained system, the quantized relativistic Landau levels are exactly flat, and in the first gap there are chirally propagating edge states associated to the $0$-th Landau level, existing on both the left and right ends. 
In particular, we see that in Fig.~\ref{fig:Spectra_all_terminations_real_field}(a) for the bearded-bearded case, the propagating edge state is left-localized on the $A$-sublattice with a positive group velocity in the $K$ valley, and right-localized on the $B$-sublattice with a negative group velocity in the $K'$ valley.
In Fig.~\ref{fig:Spectra_all_terminations_real_field}(b), the lattice always terminates with an $A$-site, as it has bearded edge on the left and zigzag edge on the right ends. According to our criterion, the $0$-th Landau level wavefunction localized on the $B$-sublattice is the only one that can give rise to a propagating edge state, hence such state is present only for the $K$ valley and not for the $K'$ valley. Similar arguments apply to Figs.~\ref{fig:Spectra_all_terminations_real_field}(c) and~\ref{fig:Spectra_all_terminations_real_field}(d), clearly demonstrating the validity of our criterion also in the case of a honeycomb lattice in the presence of a real magnetic field.

\begin{figure}[t]
\centering
\includegraphics[width=0.23\textwidth]{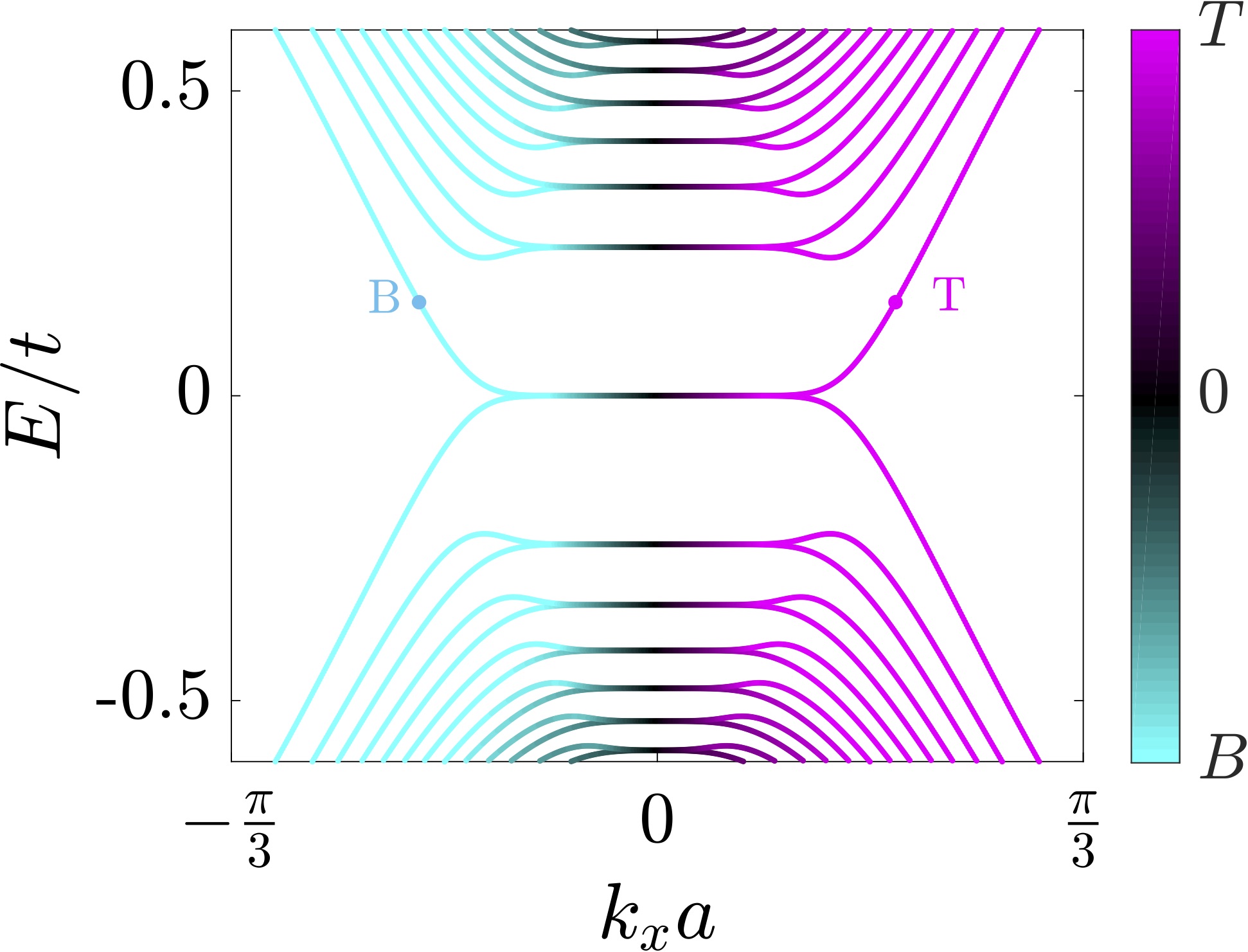}
\caption{Energy dispersion in the presence of a real magnetic field for a ribbon of $N_y=99$ along the vertical direction and periodic boundary conditions along the $x$ direction. The strength of the magnetic flux per plaquette is $2\pi \varphi = 0.0346$. The intensity of the associated real magnetic field, $\mathcal{B}_r=120$~T for parameters of solid-state graphene, is the same as the pseudo-magnetic field generated by the strain in Fig.~\ref{fig:Spectra_all_terminations_armchair}. 
The states are colored according to the mean position of their wavefunction.}
\label{fig:Spectra_all_terminations_real_field_armchair}
\end{figure}

We now discuss the chirally propagating edge states of the armchair edge. 
The tight-binding Hamiltonian within a Landau gauge with the vector potential oriented along $x$ is:
\begin{align}
\mathcal{H} = -\sum_{\mathbf{r}} t \Big( &\hat{a}^{\dagger}_{\mathbf{r} - \mathbf{R}_1} \hat{b}_{\mathbf{r}}e^{-i 2\pi \varphi y}+ \hat{a}^{\dagger}_{\mathbf{r} - \mathbf{R}_2} \hat{b}_{\mathbf{r}}+ \hat{a}^{\dagger}_{\mathbf{r} - \mathbf{R}_3} \hat{b}_{\mathbf{r}} +\text{H.c.}\Big).
\label{magnetic_ham2}
\end{align}
The energy dispersion in Fig.~\ref{fig:Spectra_all_terminations_real_field_armchair} is obtained by diagonalizing the Hamiltonian in Eq.~\eqref{magnetic_ham2} for a ribbon of $N_y=99$ unit cells along the vertical direction and periodic boundary conditions along $x$, where the magnetic flux per plaquette is $2\pi \varphi = 0.0346$. 
In this case, the Landau levels located at $k_xa=0$ are doubly degenerate because they belong to both $K$ and $K'$ valleys. Therefore, the $B$-localized $0$-th Landau level of the $K$ valley can always mix with the $A$-localized $0$-th Landau level of the $K'$ valley and form a propagating edge state. 

We can conclude that, in the presence of a real magnetic field, all terminations can host chirally-propagating edge states of the $0$-th Landau level. On the contrary, for a pseudo-magnetic field stemming from strain, the helically-propagating edge states of the $0$-th Landau level depends on the type of termination and the type of strain.

\section{Propagating edge states from the steady-state of a coherently driven-dissipative system} 
\label{sec:steadystate}

We now demonstrate how our findings can be observed in a driven-dissipative strained system, such as artificial photonic graphene made, for example, of coupled cavity arrays \cite{Jacqmin} or microwave resonators \cite{Bellec2013a, Bellec2013b, Bellec2014}. As discussed in \cite{Salerno, Bellec2014}, general spectroscopic techniques can be used to extract the main properties of the pseudo-Landau levels, such as their wavefunction. We now apply these techniques to probe the edge physics and to verify our general criterion. 

We consider a finite lattice, with $N_x$ and $N_y$ unit cells along the horizontal and vertical directions. 
The system is coherently driven by a monochromatic field at frequency $\omega_0$, while the dissipation is assumed to be uniform for all lattice sites and to have a rate of $\gamma$. 
Due to continuous pump and loss, the fields in the lattice sites $A$ and $B$ reach a steady-state, described as $a_{nm}(T)= a_{nm} e^{-i \omega_0 T}$ and $b_{nm}(T)=b_{nm} e^{-i \omega_0 T}$ at time $T$.
The steady-state amplitudes $a_{nm}$ and $b_{nm}$ are obtained by solving a system of linear Heisenberg equations \cite{Carusotto}:
$$\left[\hbar(\omega_0+i\gamma) \mathbb{I} - \mathcal{H}\right] \Psi = f,$$
where $\Psi$ is the vector with all the amplitudes $a_{nm}$ and $b_{nm}$ for all sites, $\mathcal{H}$ is the real space tight-binding Hamiltonian matrix associated with Eq.~\eqref{strained_ham}, $\mathbb{I}$ is the identity matrix, and $f$ is a vector describing the amplitude of the pump on each site.
In the following, we assume that only one site is pumped, therefore both valleys $K$ and $K'$ are simultaneously excited.

\subsection{Uni-axial strain along x}
\begin{figure}[t]
\centering
\includegraphics[width=0.15\textwidth]{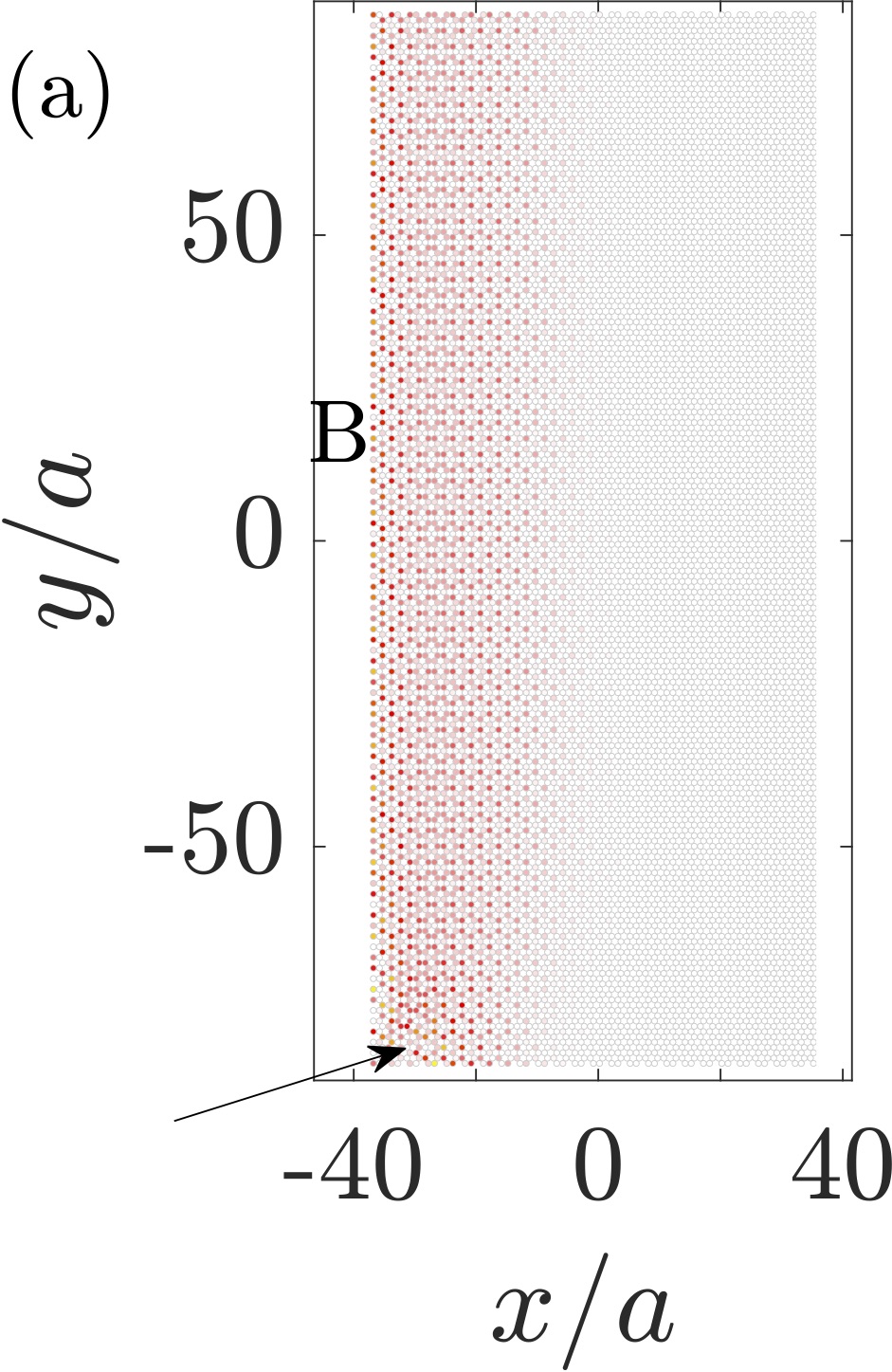}
\includegraphics[width=0.15\textwidth]{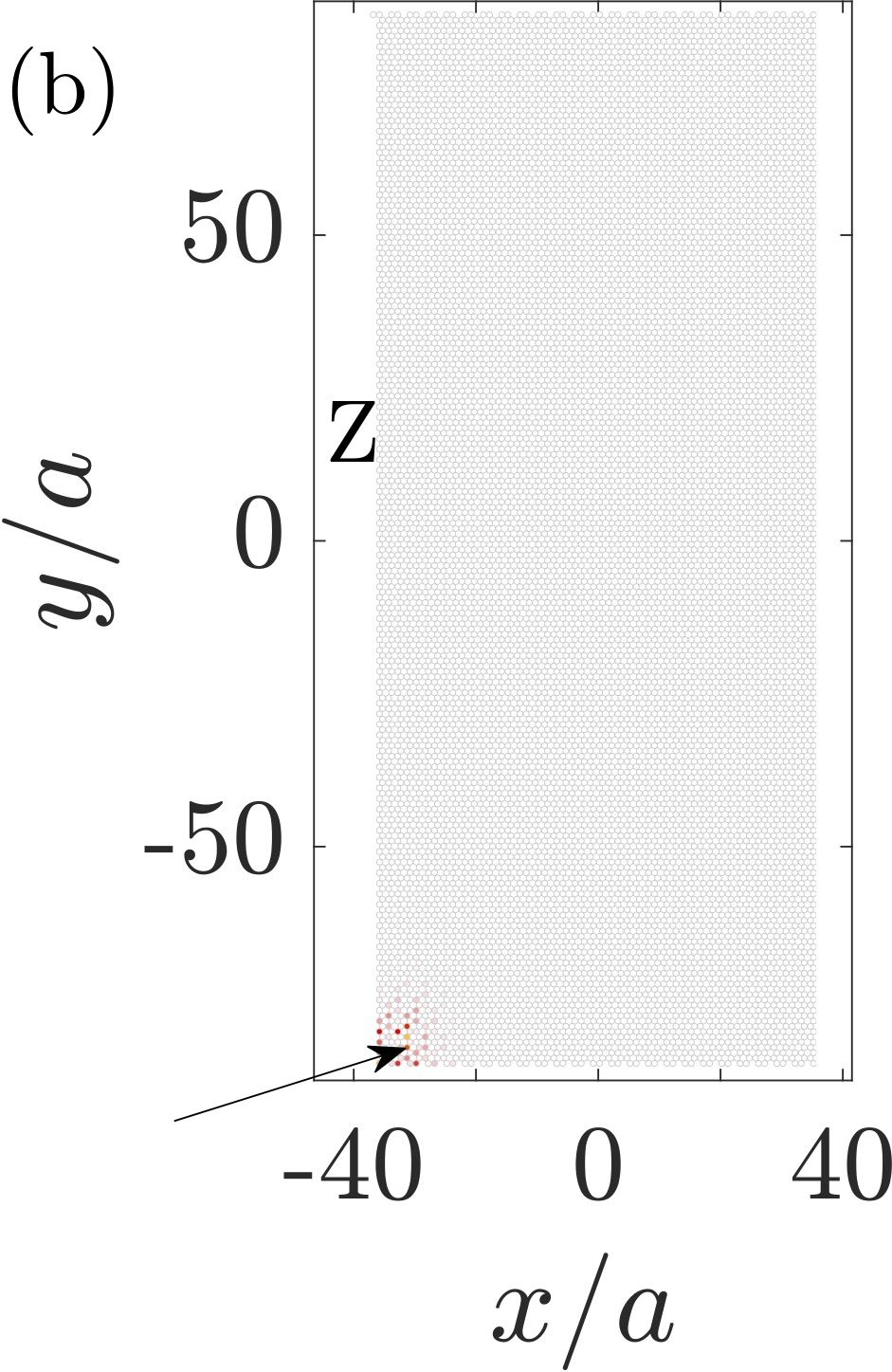}
\includegraphics[width=0.15\textwidth]{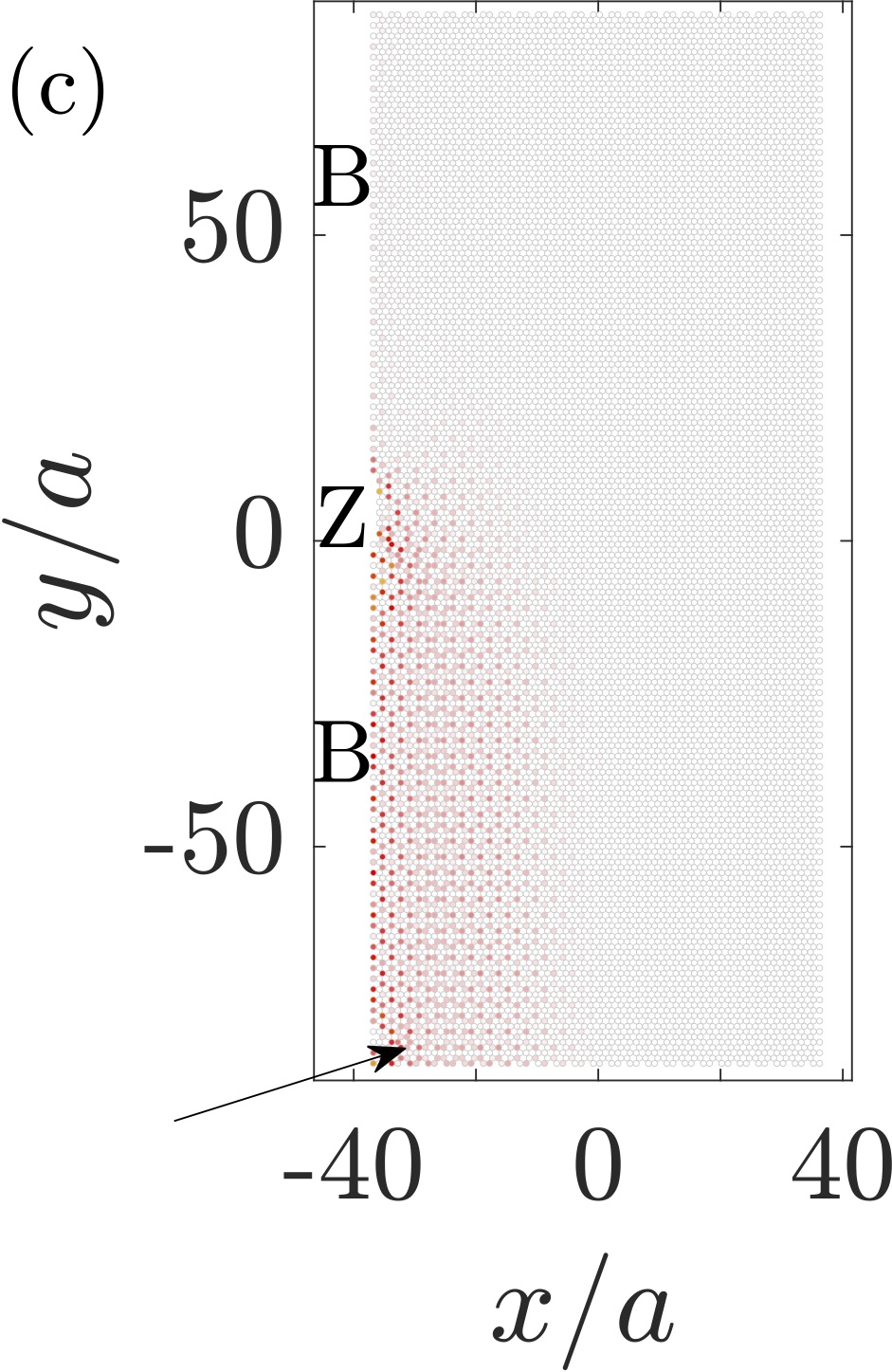}
\caption{Panel (a)-(c): spatial amplitudes in the steady-state showing the propagation of the edge state in the uni-axially-strained system of $N_x=49$ and $N_y=199$ unit cells in the horizontal and vertical direction respectively.
Parameters are $\tau=0.02$, $\hbar\gamma/t=0.0025$, $\hbar\omega_0/t=(E_{n=0}+E_{n=1})/2t=0.0707$.
The pumped site is indicated by the arrow. In panel (a), the left edge is bearded, and so supports a $0$-th Landau level propagating edge states.
In panel (b), the left edge is changed to zigzag, and the propagation of the edge state is clearly suppressed.
In panel (c), the left bearded edge has six defect sites around $y=0$, and the propagation of the edge state is strongly suppressed.
}
\label{fig:uni-axial_edges}
\end{figure}

In Figs.~\ref{fig:uni-axial_edges}(a)-(c), we show the steady-state amplitudes for a system with bearded and zigzag edges at the left end, under the uni-axial strain along $x$ in Eq.~\eqref{uniaxialX}.
The pumped site, indicated by the arrow, is on the $A$-sublattice at the bottom left corner of the system and the pumping frequency is in between the lowest $n=0$ and the first $n=1$ pseudo-Landau level.
We assume a positive strain, such that only the left edge supports the propagating edge state of the $0$-th pseudo-Landau level for the bearded termination, as previously shown in the energy dispersion of Fig.~\ref{fig:Spectra_all_terminations}(a) and now clearly visible from the steady-state shown in Fig.~\ref{fig:uni-axial_edges}(a). 
When the left termination is changed to zigzag in Fig.~\ref{fig:uni-axial_edges}(b), the propagating edge state is suppressed on the left edge, as we have seen in the energy dispersion of Fig.~\ref{fig:Spectra_all_terminations}(c).
In Fig.~\ref{fig:uni-axial_edges}(c) we have removed 6 sites of type $A$ at $y=0$, and we see that the propagation of the edge state is strongly suppressed. When hitting the defect, the edge state scatter to the other valley and propagate backwards; no propagation into the bulk is observed.
Figure~\ref{fig:uni-axial_edges}(a) also shows that the edge states propagate until it is reflected at the corners of the system by the top and bottom armchair edges, which, for this particular type of strain, do not have (within a local picture) any zero-energy edge states to mix with the $0$-th pseudo-Landau level. 

\subsection{Uni-axial strain along y}

\begin{figure}[t]
\centering
\includegraphics[width=0.23\textwidth]{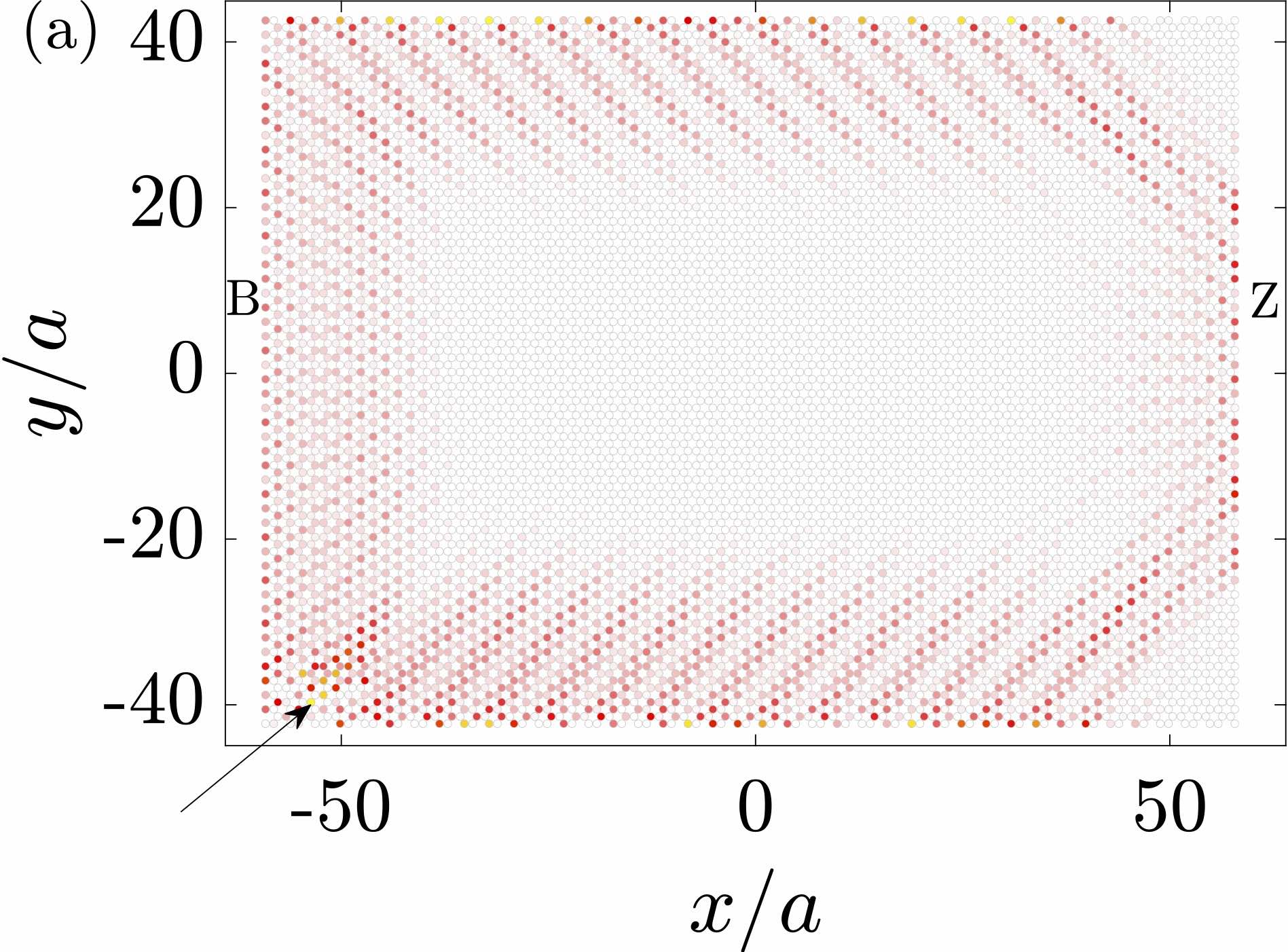}
\includegraphics[width=0.23\textwidth]{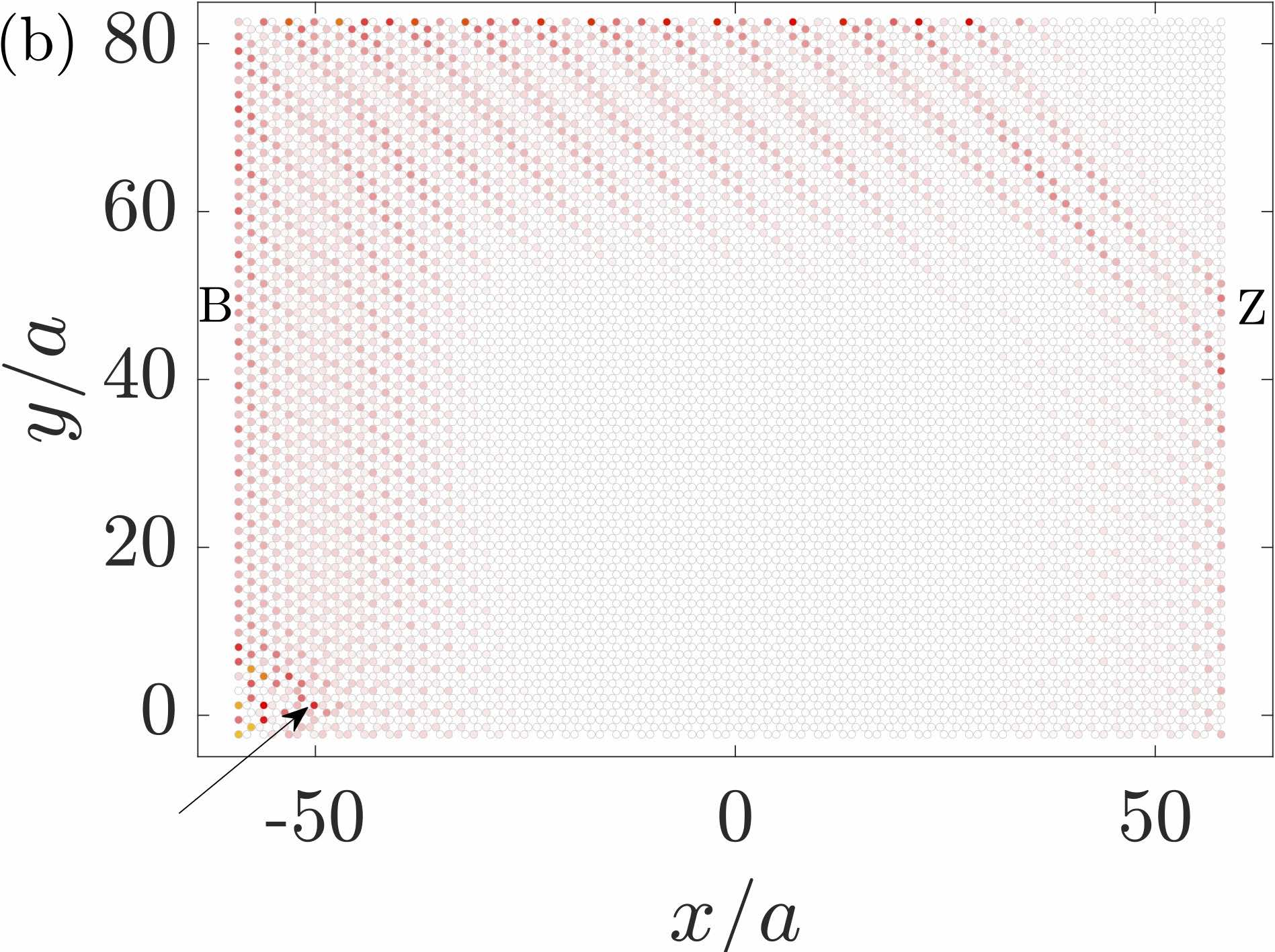}
\caption{Panel (a) and (b): spatial amplitudes in the steady-state showing the propagation of the edge state in the armchair-strained system. 
The maximum numbers of unit cells along the horizontal and vertical directions are $N_x=79$ and $N_y=99$. 
The pumped site is indicated by the arrow. In panel (a), all edges support helically-propagating edge states. Parameters are $\tau=0.02$, $\hbar\gamma/t=0.0025$, $\hbar\omega_0/t=(E_{n=0}+E_{n=1})/4t=0.1225$.
In panel (b) the strain is such that the propagation of the edge state is forbidden on the bottom edge. 
Parameters are $\tau=0.01$, $\hbar\gamma/t=0.0025$, $\hbar\omega_0/t=(E_{n=0}+E_{n=1})/4t=0.0866$.
}
\label{fig:armchair}
\end{figure}

In Figs.~\ref{fig:armchair}(a) and~\ref{fig:armchair}(b) we show the steady-state amplitudes for a system with the uni-axial strain along $y$ given in Eq.~\eqref{uniaxialY}.
The pumped site, indicated by the arrow, is on the $A$-sublattice at the bottom left corner of the system and the pumping frequency is in between the lowest $n=0$ and the first $n=1$ pseudo-Landau level. 
In Fig.~\ref{fig:armchair}(a), the strain strength is $\tau=0.02$ and $t_2=t_3=t$ for $y=0$, and the vertical edges are both terminating with $A$-sites. 
As correctly predicted by our criterion and shown in the energy dispersion of Fig.~\ref{fig:Spectra_all_terminations_armchair}, we see that all edges, and in particular both the top and the bottom armchair edges, sustain propagating edge states.

In Fig.~\ref{fig:armchair}(b) the strain starts from the bottom end of the ribbon, as $t_2=t_3=t$ for $y=-L_y$, and $\tau=0.01$.
In this case, as we have shown in Fig.~\ref{fig:Spectra_all_terminations_armchair}(b), the bottom armchair edge does not have a zero-energy edge state in the local picture, therefore our criterion predicts that the propagating edge states are not supported on the bottom end, which is confirmed by the steady-state calculation.

In Fig.~\ref{fig:armchair} we notice a fringe pattern, which is due to the interference between counter propagating edge states of the two $K$ and $K'$ valleys, and are enhanced by the reflection of the states at each corner. 
We also notice that the corners connecting the armchair with the zigzag edge are avoided. We believe this is due to a subtle interplay between the zigzag edge state and the armchair edge state.

\subsection{Trigonal strain}

We now numerically demonstrate that our criterion for the existence of propagating edge states still applies in the case of trigonal strain. 
As mentioned earlier, for the trigonal strain the $0$-th pseudo-Landau level wavefunction is again localized only on $B$($A$)-sublattice for a positive $\tau>0$ (negative $\tau<0$) for both valleys. Our criterion predicts that the $0$-th pseudo-Landau level possesses helical edge states when there are zero-energy edge states on the $A$($B$)-sublattice within the local picture.

\begin{figure}[t]
\centering
\includegraphics[width=0.23\textwidth]{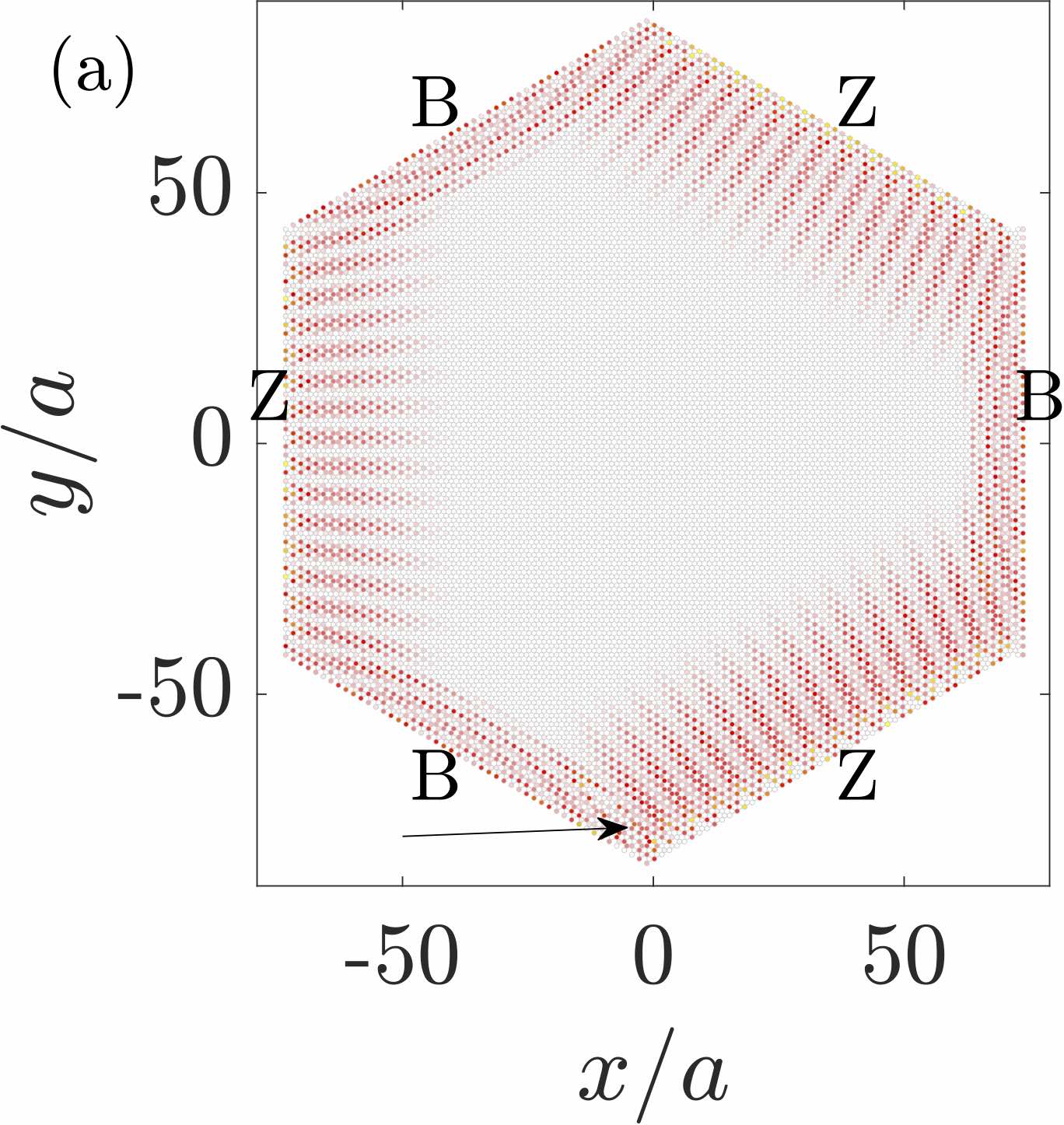}
\includegraphics[width=0.23\textwidth]{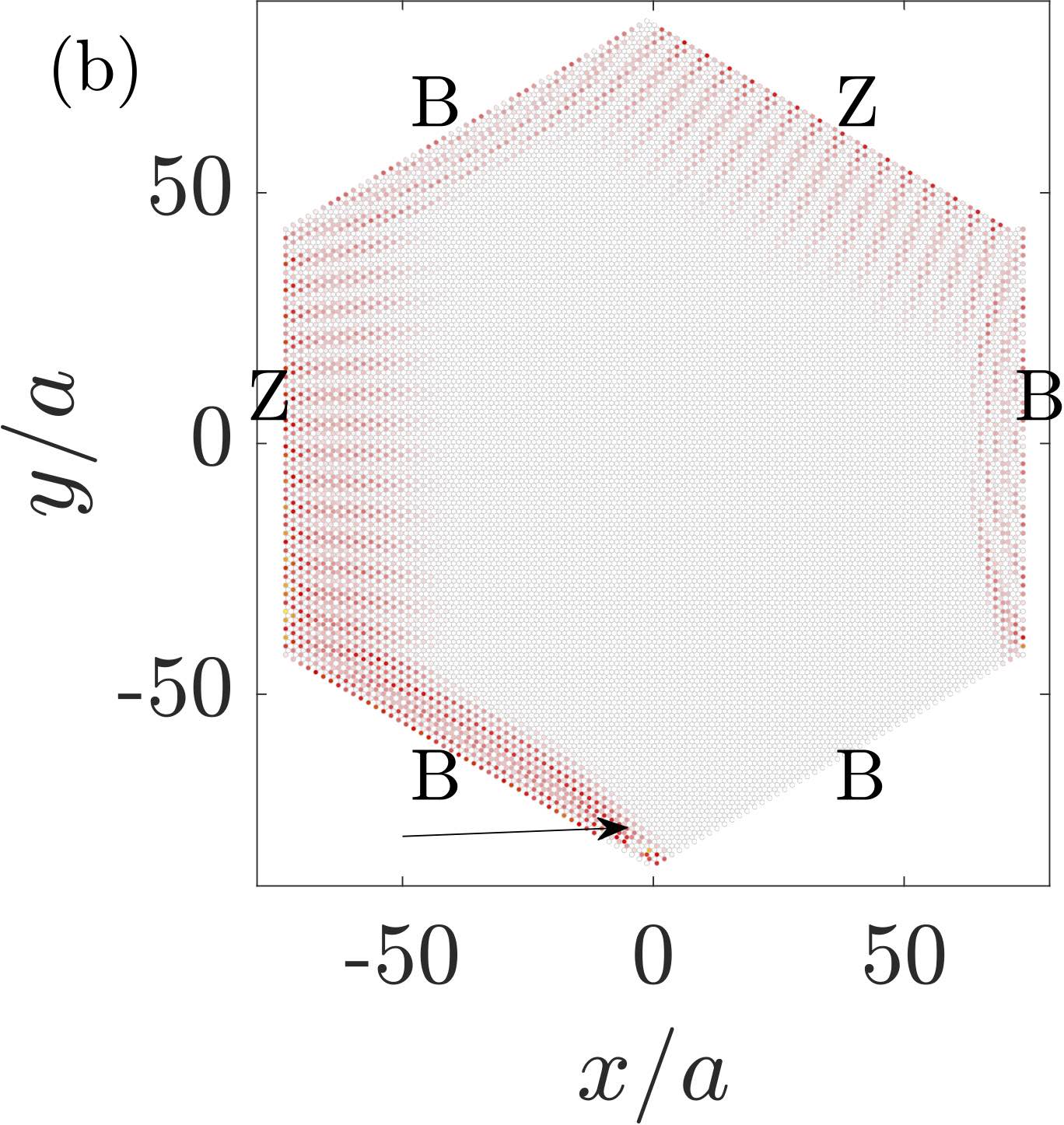}\\
\includegraphics[width=0.23\textwidth]{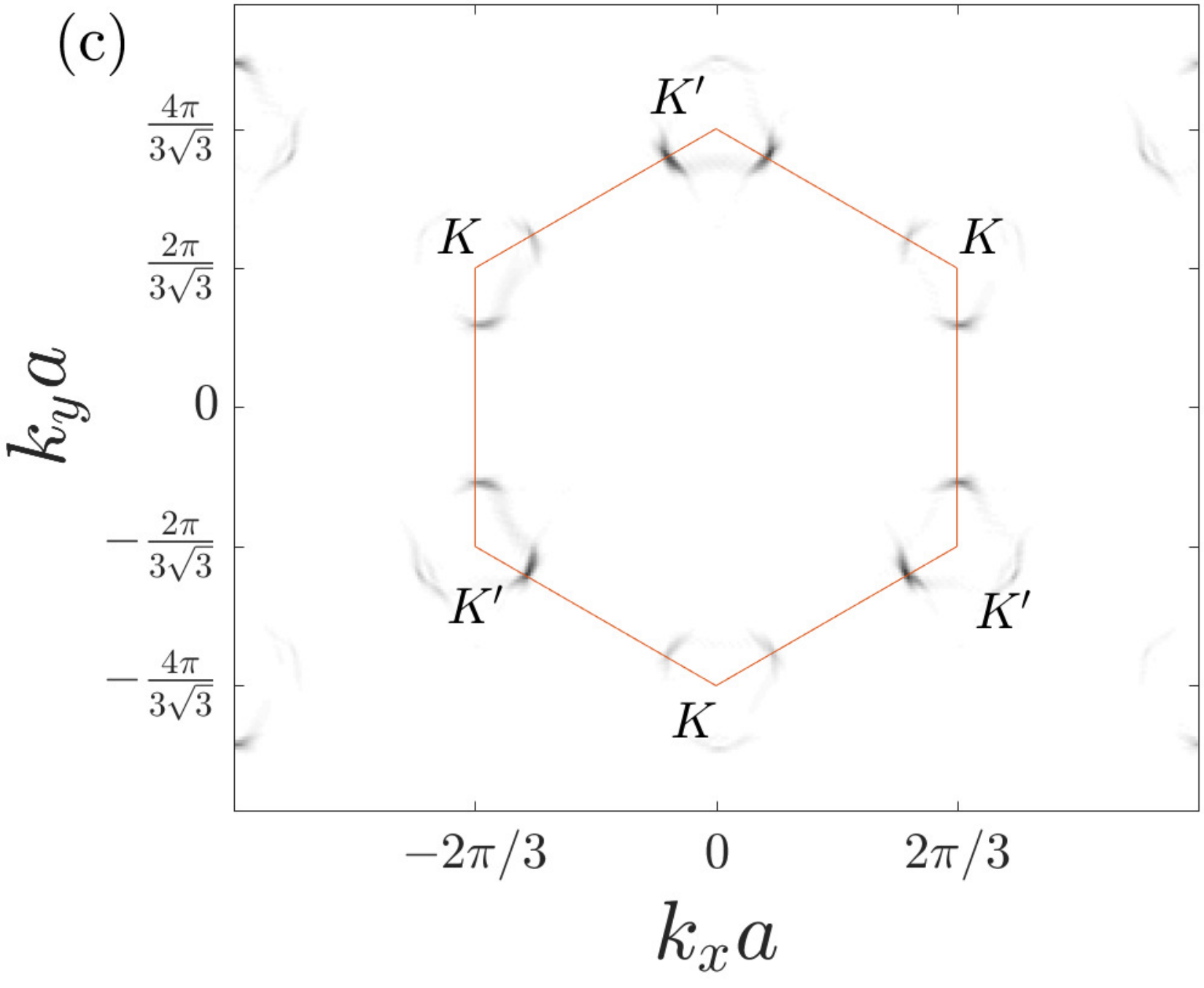}$\,$
\includegraphics[width=0.23\textwidth]{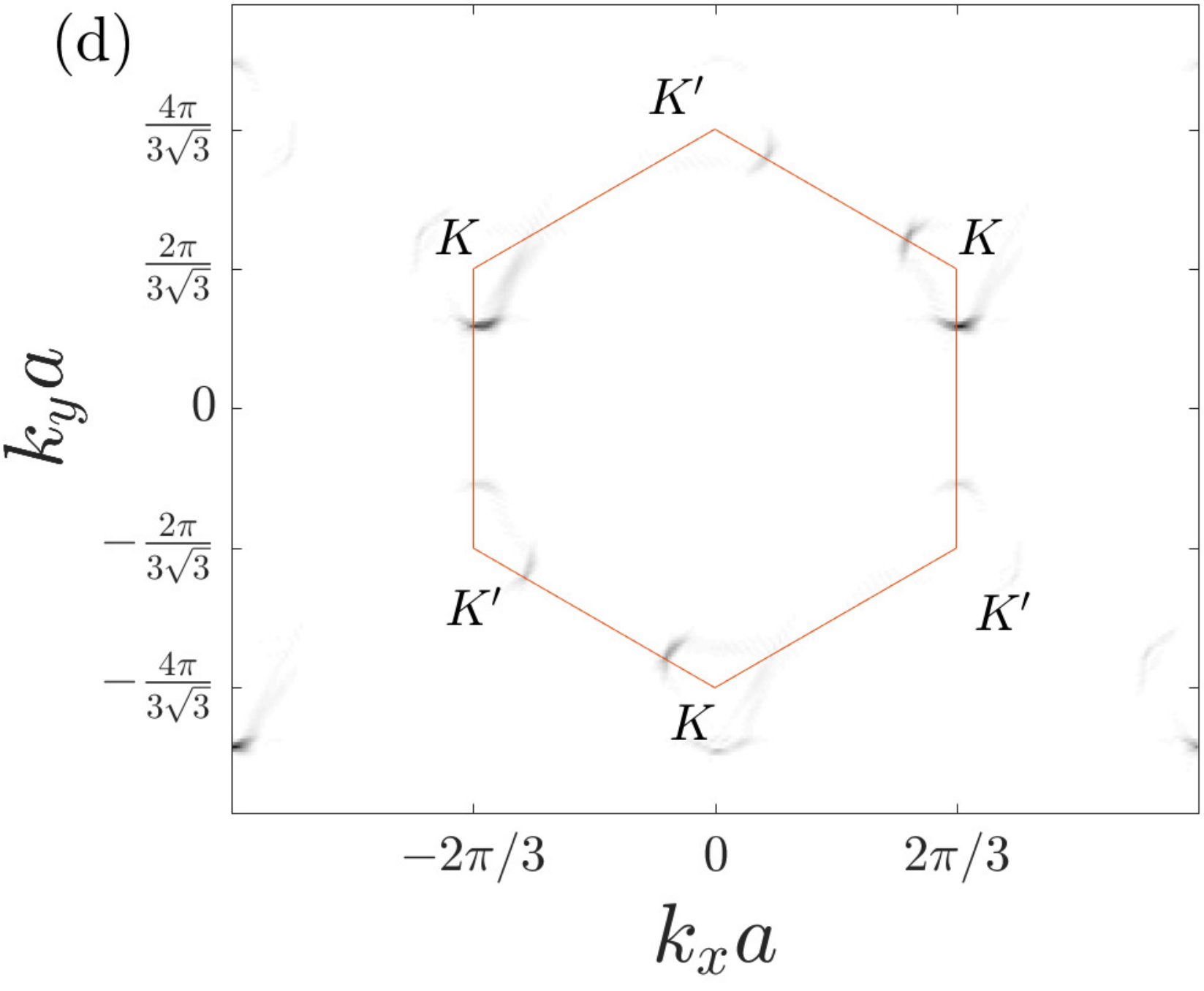}
\caption{Panel (a) and (b): spatial amplitudes in the steady-state showing the propagation of the edge state in the trigonally-strained hexagonally shaped system. Edges are labelled by $Z$ and $B$ to indicate if they are respectively zigzag or bearded terminations.
The maximum numbers of unit cells along the horizontal and vertical directions are $N_x=99$ and $N_y=199$ respectively. 
The pumped site is indicated by the arrow. In panel (a), all edges support helically-propagating edge states. 
In panel (b) the bottom right edge is changed from a zigzag to bearded, and the propagation of the edge state is clearly blocked on that edge. 
Parameters are $\tau=-0.02$, $\hbar\gamma/t=0.0025$, $\hbar\omega_0/t=(E_{n=0}+E_{n=1})/4t=0.061$.
Panels (c) and (d) show the Fourier transform of panels (a) and (b) respectively.
We see that all valleys are excited in panel (c), that corresponds to the case of both helical edge states going around the system in panel (a).
In panel (d) one valley is predominantly excited, corresponding to the case of mostly the clock-wise edge state in panel (b).
}
\label{fig:valleyfilter}
\end{figure}

In Figs.~\ref{fig:valleyfilter}(a)-(b) we show the steady-state amplitudes for a trigonally strained system with $\tau=-0.02$, terminated with bearded or zigzag edges respectively on the ends labelled by B or Z.
The pumped site is located on a $B$-site close to the bottom corner, as indicated by the arrow.
The pumping frequency is in the first gap at $\hbar\omega_0/t=(E_{n=0}+E_{n=1})/4t=0.061$, and the loss rate is small enough such that the particles can travel around the whole system. 
In Fig.~\ref{fig:valleyfilter}(a) the system is shaped such that all the edges are terminating on a $B$-sublattice, either with zigzag or bearded edges.
In agreement with our criterion, we see that a pair of edge states helically propagate around the entire system in both directions.
On the contrary, in Fig.~\ref{fig:valleyfilter}(b) the bottom right edge terminates on the $A$-sublattice and we see that propagation along that edge is forbidden, and so only the edge state propagating along the bottom left edge is excited.
Each valley $K$ or $K'$ is associated with the direction of propagation of the helical edge states being clock-wise or counter-clock-wise respectively. 
We plot the spatial Fourier transform of the steady-state amplitudes of Figs.~\ref{fig:valleyfilter}(a)-(b) in Figs.~\ref{fig:valleyfilter}(c)-(d) and show that unidirectional propagation is associated with valley filtering.
In Fig.~\ref{fig:valleyfilter}(a), where the edge states are propagating in both directions starting from the pumped site, we see in the corresponding Fourier transform in Fig.~\ref{fig:valleyfilter}(c) that regions around both $K$ and $K'$ points in the Brillouin zone are excited. 
In Fig.~\ref{fig:valleyfilter}(b), the edge state mainly propagates in the clock-wise direction and the Fourier transform in Fig.~\ref{fig:valleyfilter}(d) shows that the $K'$ points have been filtered out.
The residual excitation that is present around the $K'$ points is due to the reflection at the corner with the edge that does not support the propagating edge state. This is the same reflection process mentioned previously for the uni-axial strain.

\begin{figure}[t]
\centering
\includegraphics[width=0.33\textwidth]{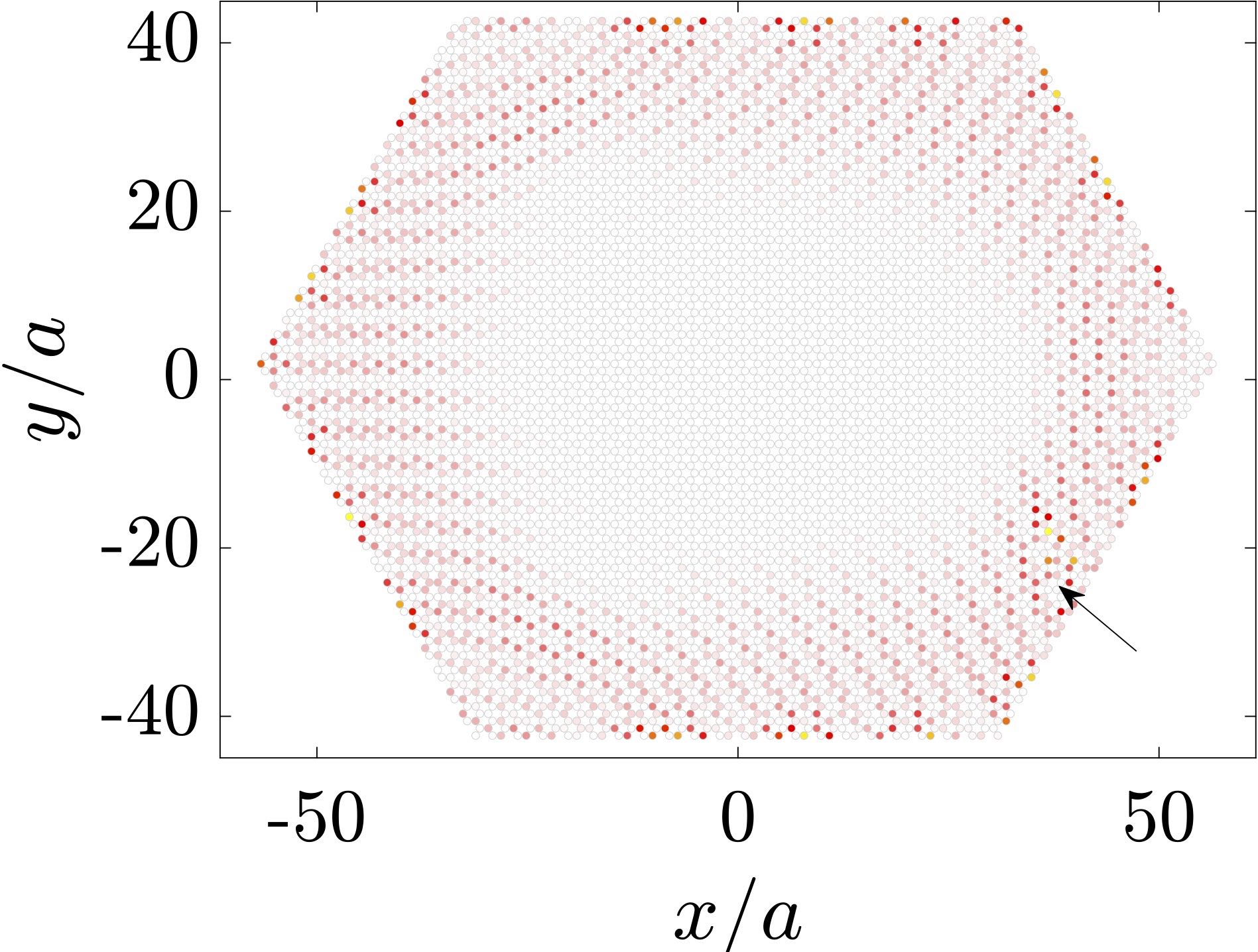}
\caption{Spatial amplitudes in the steady-state showing the propagation of the edge state in the trigonally-strained hexagonally shaped system. All edges are armchair.
The maximum numbers of unit cells along the horizontal and vertical directions are $N_x=79$ and $N_y=99$ respectively. 
The pumped site is indicated by the arrow. Parameters are the same as in Fig.~\ref{fig:valleyfilter}.
}
\label{fig:allArmchair}
\end{figure}

Finally, in Fig.~\ref{fig:allArmchair}, we show a hexagonally-shaped system terminated with armchair edges, which are all able to support the propagating edge states. This is because, according to our criterion, the strain is such that, in the local picture, all edges support zero-energy edge states on the $B$-sublattice and hence can mix with the $0$-th pseudo-Landau level, which is localized on the $A$-sublattice. The pumped site, indicated by the arrow, is a site of type $A$, and the parameters are the same as in Fig.~\ref{fig:valleyfilter}. 
As expected, we see that the edge states propagate around the entire system.

\section{Propagating edge states with next-nearest-neighbour hoppings} 
\label{sec:NNN}

In this final section, it is worth assessing the effect of next-nearest-neighbour (NNN) hoppings. As these terms break chiral symmetry, which is at the heart of our argument, our criterion is not valid any more, and so new features are expected to arise.

We define the vectors that connect NNN sites as $\mathbf{D}_1=\mathbf{R}_3-\mathbf{R}_2$, $\mathbf{D}_2=\mathbf{R}_2-\mathbf{R}_1$ and $\mathbf{D}_3=\mathbf{R}_3-\mathbf{R}_1$.
The tight-binding Hamiltonian with NNN hoppings is:
\begin{equation}
\begin{split}
\mathcal{H}=-\sum_{\mathbf{r},j} \Big[ &t_j(\mathbf{r}) \hat{a}_{\mathbf{r}-\mathbf{R}_j}^\dagger \hat{b}_{\mathbf{r}}+t_j'(\mathbf{r}) \hat{a}_{\mathbf{r}-\mathbf{D}_j}^\dagger \hat{a}_{\mathbf{r}}\\& +t_j'(\mathbf{r}) \hat{b}_{\mathbf{r}-\mathbf{D}_j}^\dagger \hat{b}_{\mathbf{r}}+ \text{H.c.}\Big]
\end{split}
\label{HNNN}
\end{equation}

To a first approximation, we consider the NNN hopping to be constant over the lattice $t'_j(\mathbf{r}) \rightarrow t'$. This approximation is sufficient when the strain and the bare NNN hopping are both small. For a more refined study, one could instead assume that the NNN hoppings have the same spatial dependence as the strained nearest-neighbour hoppings, as done in \cite{Salerno} for the uni-axial strain along $x$, which however does not qualitatively change the main features of the results.

\subsection{Energy dispersions}

\begin{figure}[t]
\centering
\includegraphics[width=0.23\textwidth]{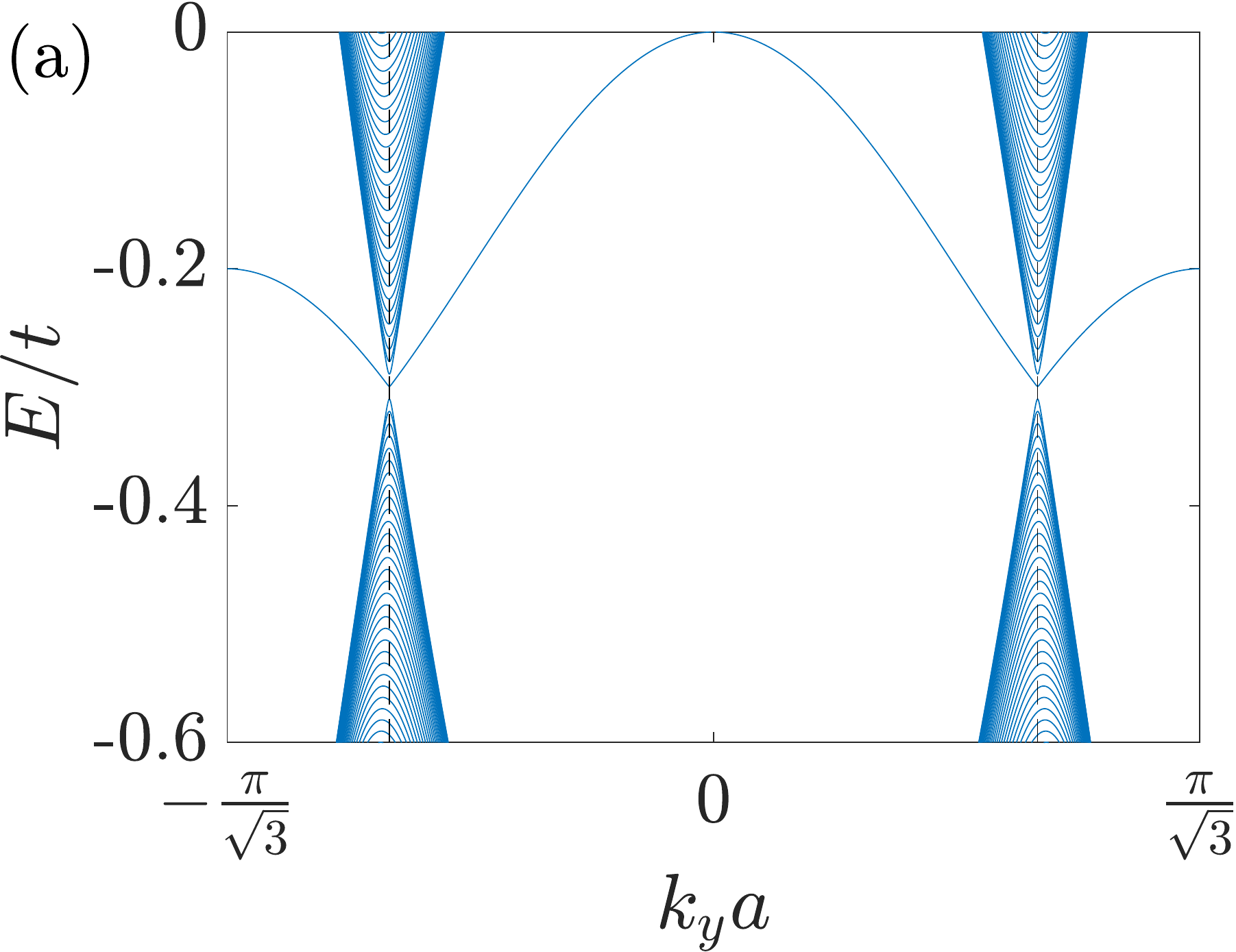}
\includegraphics[width=0.23\textwidth]{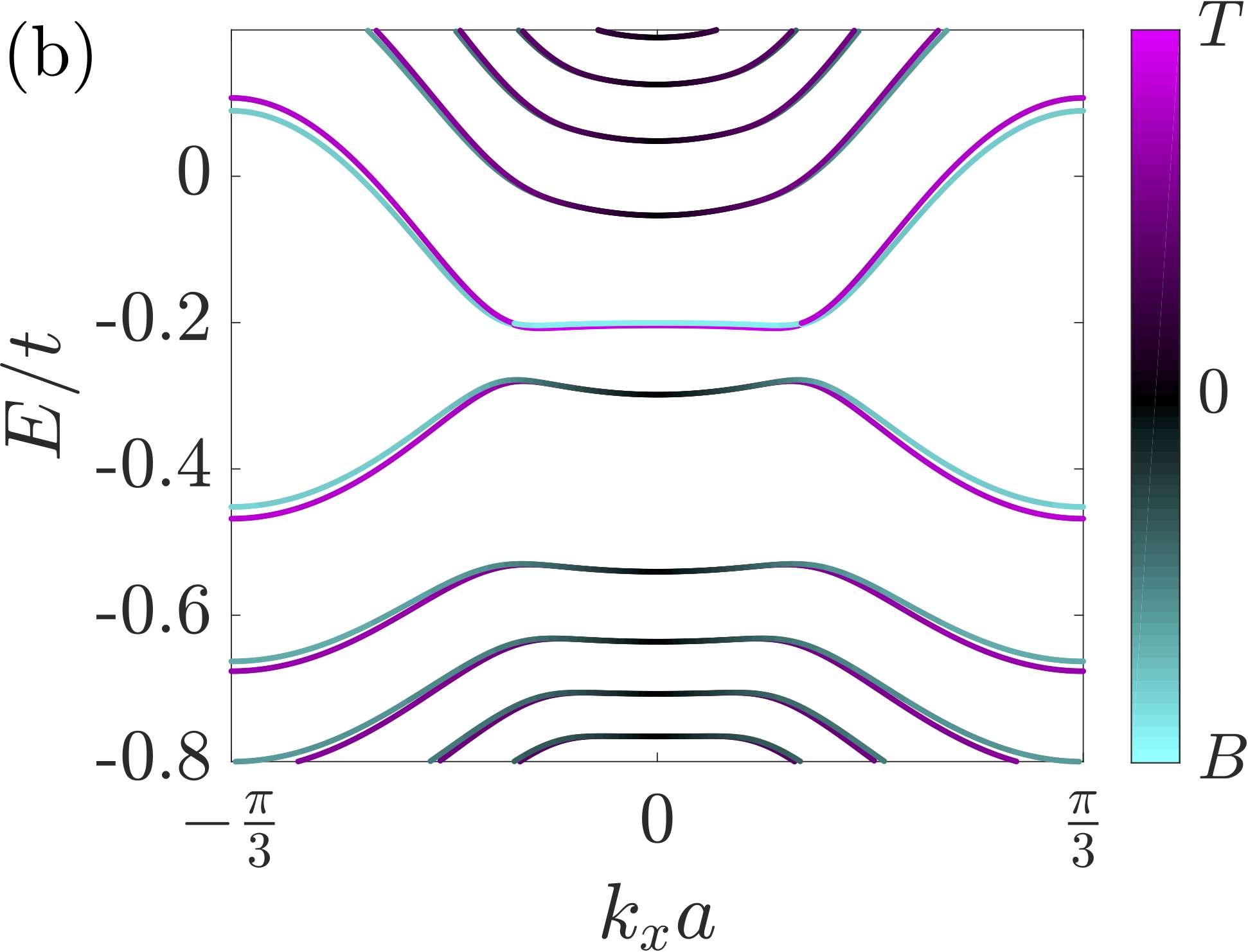}
\includegraphics[width=0.23\textwidth]{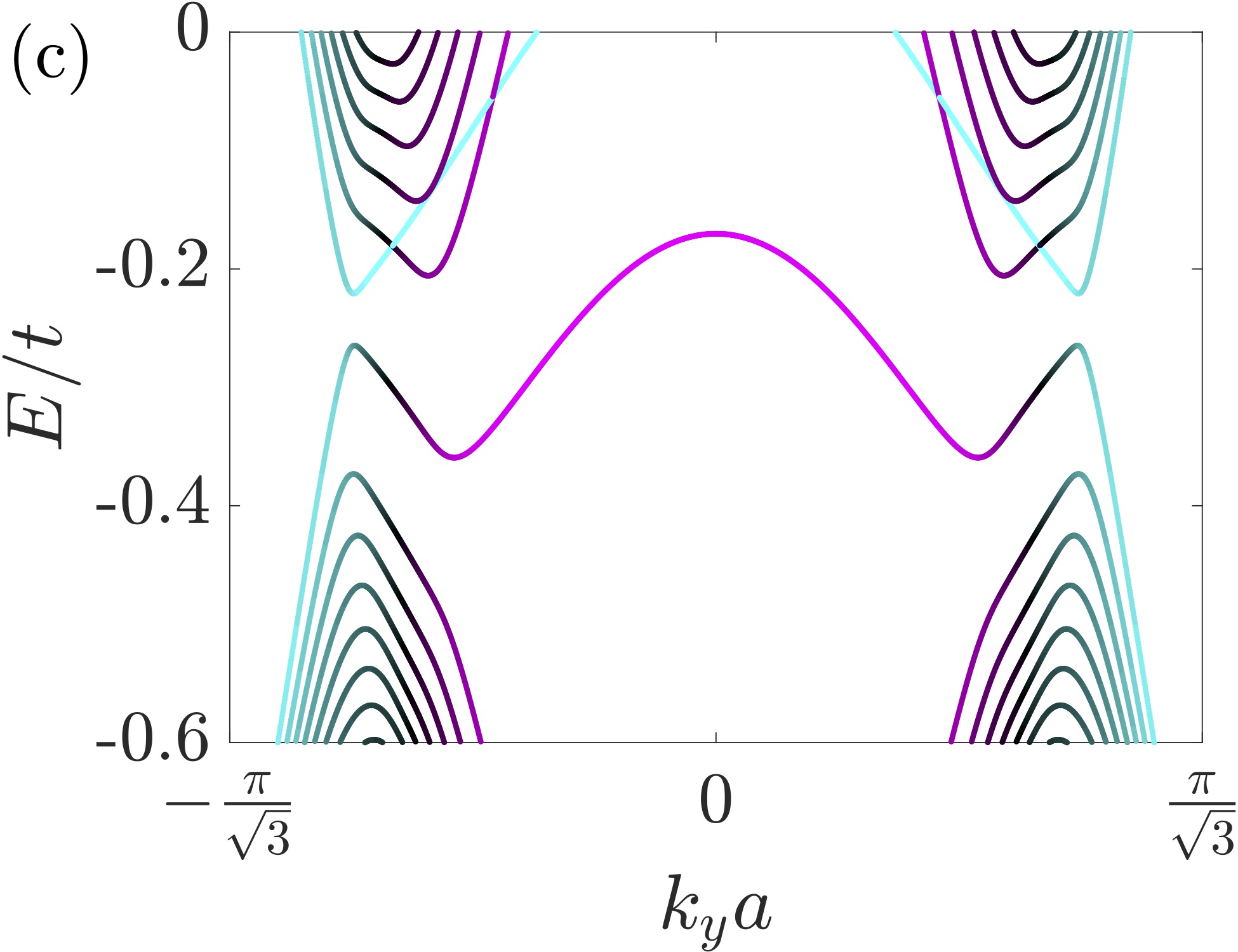}
\includegraphics[width=0.23\textwidth]{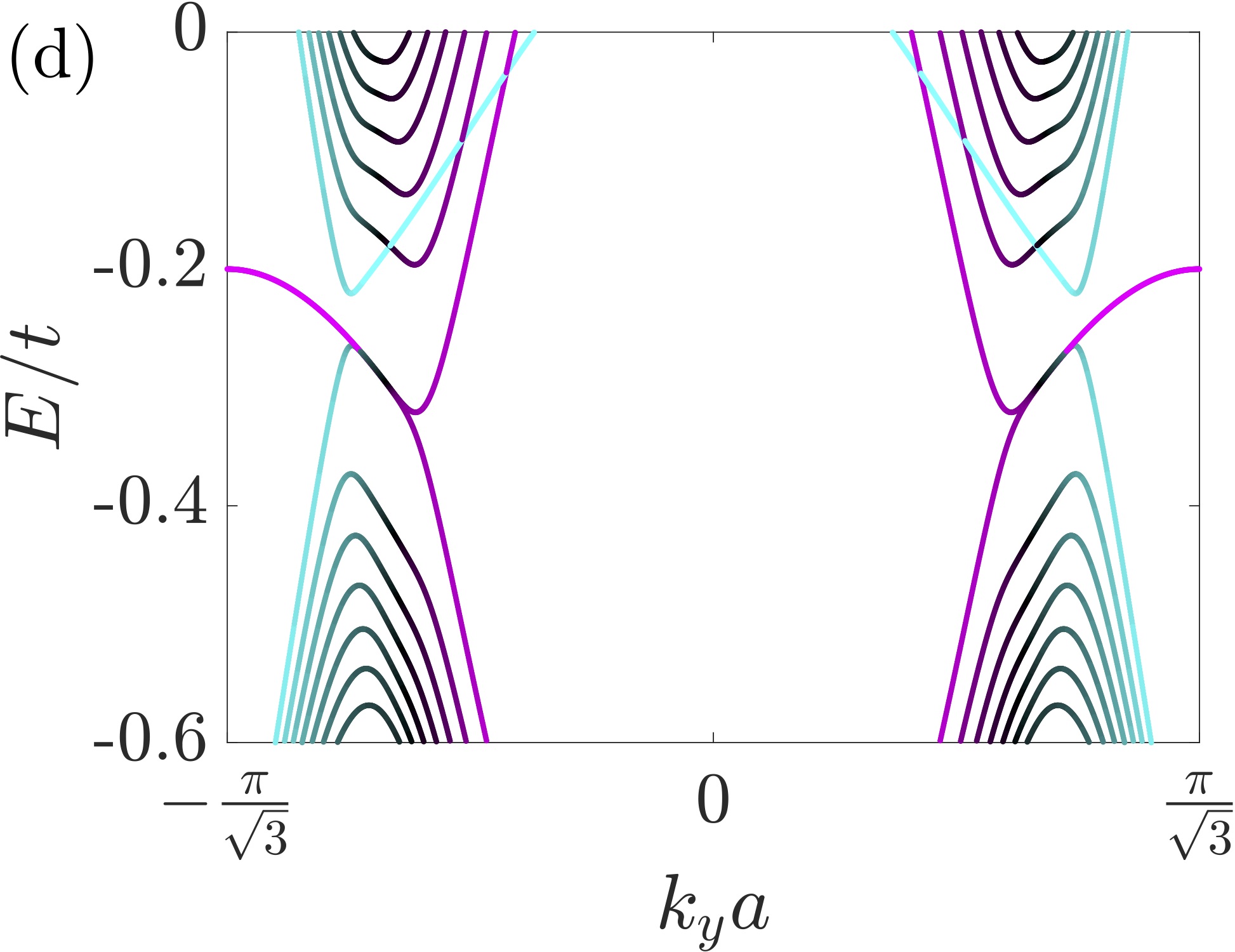}
\includegraphics[width=0.23\textwidth]{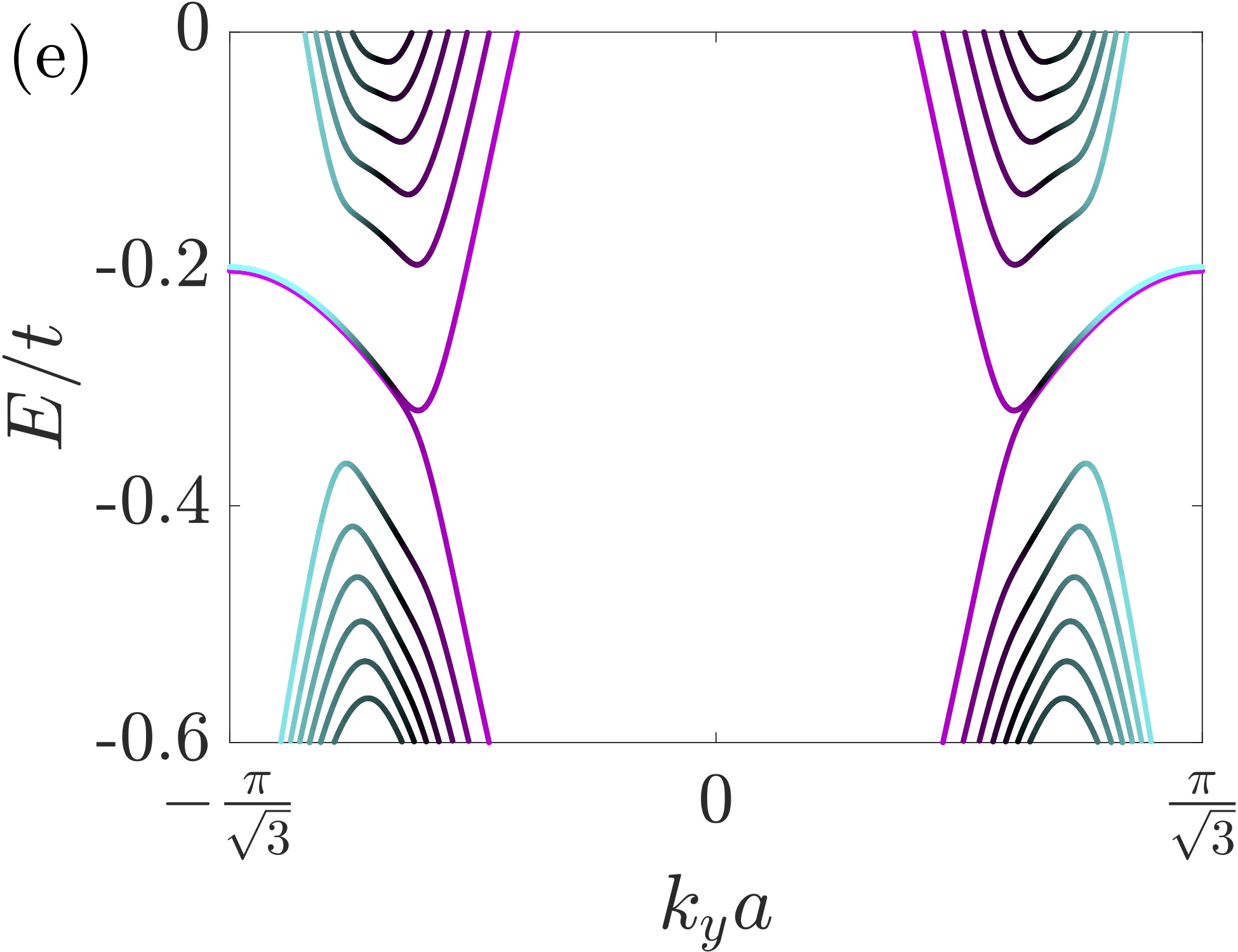}
\includegraphics[width=0.23\textwidth]{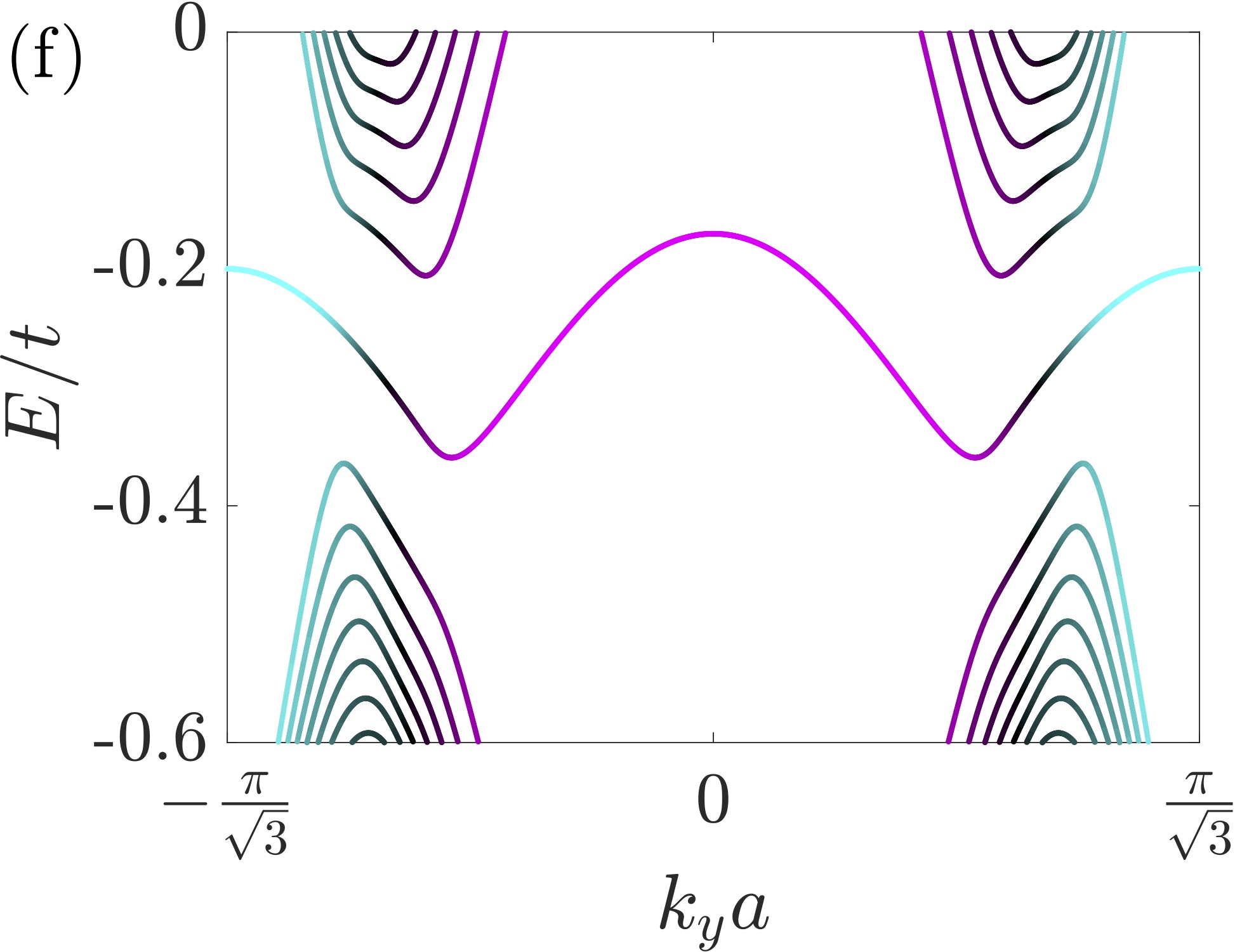}
\includegraphics[width=0.47\textwidth]{colorbar.jpg}
\caption{
Panel~(a): Energy dispersion for a unstrained system $\tau=0$, with $N_y=99$ along the vertical direction and $t'=0.1 t$, with a bearded termination on the left and a zigzag termination on the right.
Panel~(b): Energy dispersion for a uni-axially strained ribbon along $y$ with periodic boundary conditions along $x$ and armchair terminations at the top and bottom, with a next-nearest-neighbour hopping strength of $t'=0.1 t$. The strain strength is $\tau=0.02$, with $N_y=99$ along the vertical direction and $t_{2,3}=t$ is in the center of the ribbon. 
Panel~(c)-(f): Energy dispersions for a uni-axially strained ribbon along $x$, with $N_x=99$ along the armchair direction and periodic boundary conditions along $y$, for different terminations with a constant next-nearest-neighbour hopping strength of $t'=0.1 t$, and a strain strength of $\tau=0.015$.
Panel~(c) is for bearded terminations on both ends, panel (d) is for a bearded termination on the left and a zigzag termination on the right, panel~(e) is for zigzag terminations on both ends and panel~(f) is for a zigzag termination on the left and a bearded termination on the right.
In panels~(b)-(f), the states are colored according to the mean position of their wavefunction, as indicated by the colorbar.}
\label{fig:Spectra_all_terminationsNNN}
\end{figure}

We repeat the numerical calculation of Sec.~\ref{sec:uniaxial_dispersion}, including a NNN hopping strength of $t'=0.1 t$. This value is consistent with estimates for NNN hopping in solid-state graphene \cite{CastroNeto} and in artificial graphene \cite{Jacqmin, Bellec2013a}. 
In this way, we obtain the low-energy dispersion presented in Fig.~\ref{fig:Spectra_all_terminationsNNN} for various edge terminations and with the two types of uni-axial strain (along $x$ and along $y$).

In Fig.~\ref{fig:Spectra_all_terminationsNNN}~(a) we show the energy dispersion for a unstrained system $\tau=0$ with periodic boundary conditions along $y$ and $t'=0.1 t$, with a bearded termination on the left and a zigzag termination on the right.
Because of the breaking of the chiral symmetry due to the NNN hoppings, the edge states are no longer pinned at zero energy as in Fig.~\ref{fig:Spectra_all_terminations}, but they are now dispersive. 

In Figs.~\ref{fig:Spectra_all_terminationsNNN}~(b)-(f) we show the energy dispersion for the two different uni-axial strains in the presence of a NNN hopping of strength $t'=0.1t$. In all panels, we identify the Landau levels appearing in the vicinity of the $K$ and $K'$ points.
Figure~\ref{fig:Spectra_all_terminationsNNN}~(b) is the energy dispersion of a ribbon with uni-axial strain along $y$ and non-zero NNN hoppings, corresponding to the NNN generalization of the spectra in Fig.~\ref{fig:Spectra_all_terminations_armchair}~(a). In this case we see that the energy dispersion is almost unaffected by the NNN hoppings, apart from the opening of a gap that removes the degeneracy between the $n=0$ Landau levels and the non-propagating edge states of the strained armchair termination. Also the localization of the wavefunction, which is indicated by the color bar, is unaffected by the NNN hoppings.

In Figs.~\ref{fig:Spectra_all_terminationsNNN}~(c)-(f) we show the energy dispersions of a ribbon with uni-axial strain along $x$ and non-zero NNN hoppings, corresponding to the NNN generalization of the spectra in Fig.s~\ref{fig:Spectra_all_terminations}(a)-(d).
As already noted in \cite{Salerno}, Landau levels in panels~\ref{fig:Spectra_all_terminationsNNN}~(c)-(f) are more tilted than the ones in Fig.~\ref{fig:Spectra_all_terminations} without NNN hoppings.

The mean localization of the wavefunction in each level, which is shown by the color scale, is similar to the one shown in Figs.~\ref{fig:Spectra_all_terminations}(a)-(d) for $t'=0$.
We still see that the presence of propagating edge states of the $n=0$ Landau level still depends on the type of terminations of the ribbon. This is particularly clear for the energies between the lower bands and the shifted energy of the Dirac point $-3t'$ (lower LL0 gap).
As compared to the edge states that become propagating as a consequence of the NNN hoppings and shown in \ref{fig:Spectra_all_terminationsNNN}~(a), some of the propagating edge states associated with $n=0$ Landau level may have a steeper dispersion.

\subsection{Steady-state of the coherently driven system}

\begin{figure}[t]
\centering
\includegraphics[width=0.23\textwidth]{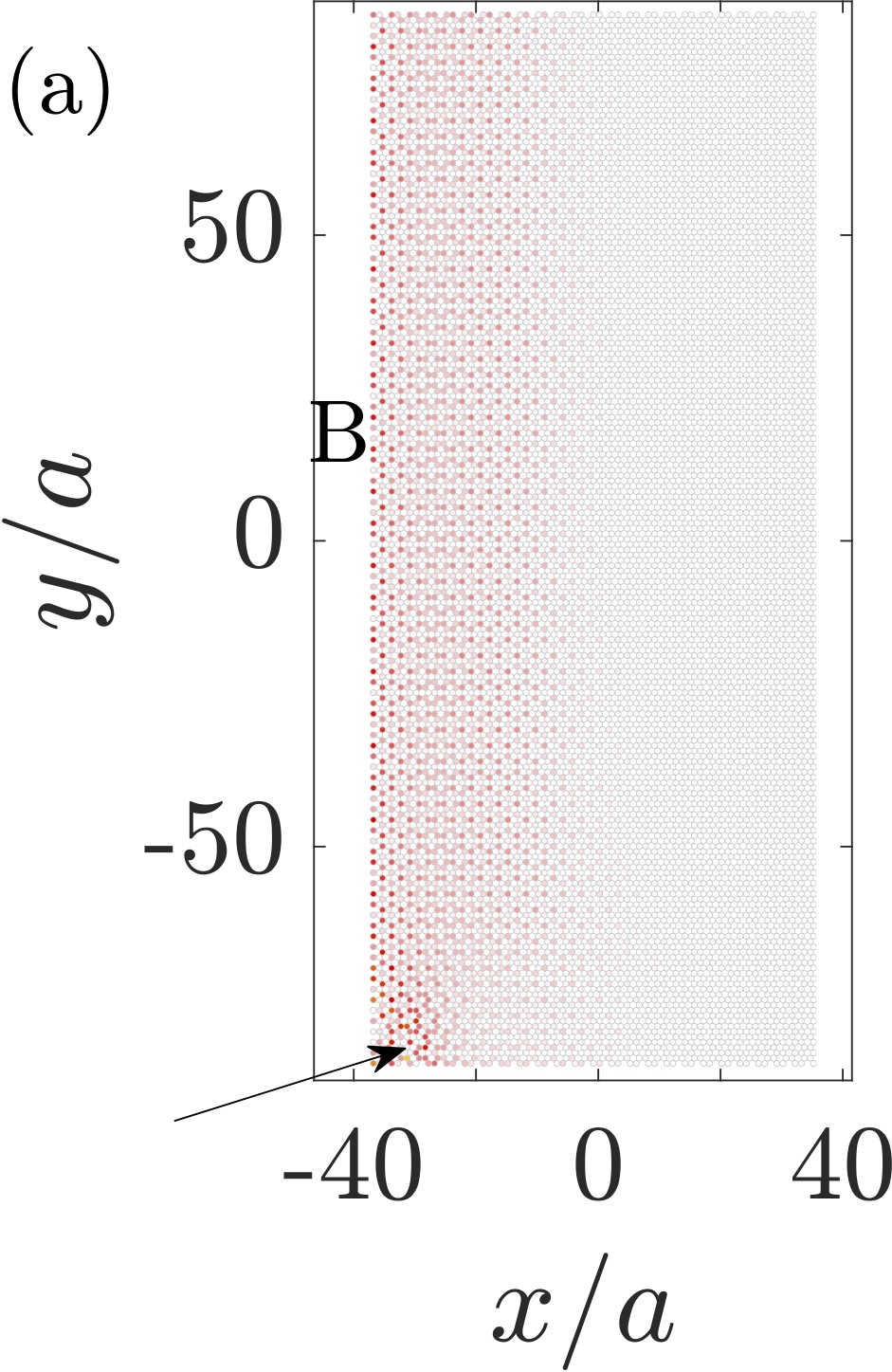}
\includegraphics[width=0.23\textwidth]{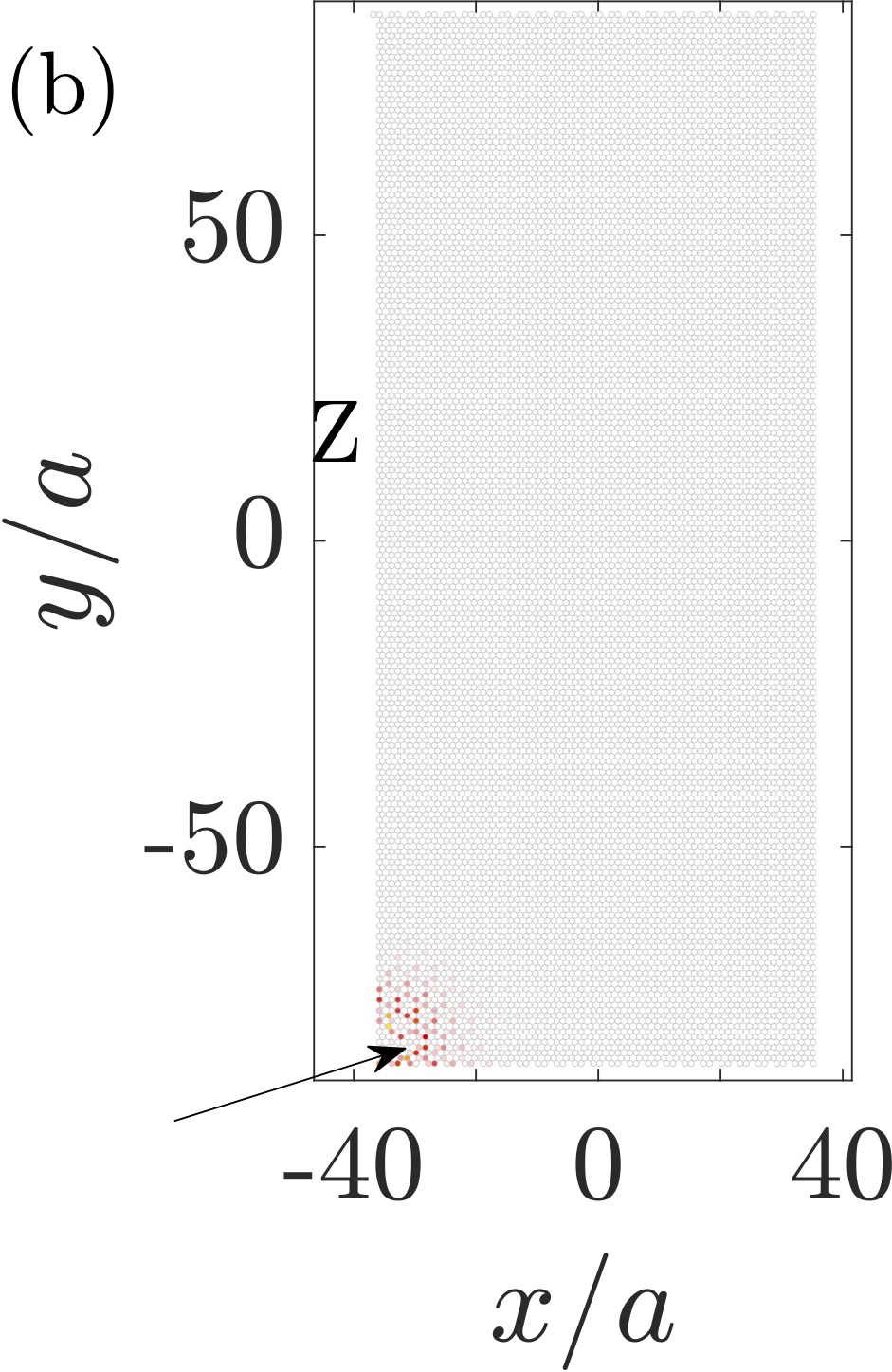}\\
\includegraphics[width=0.23\textwidth]{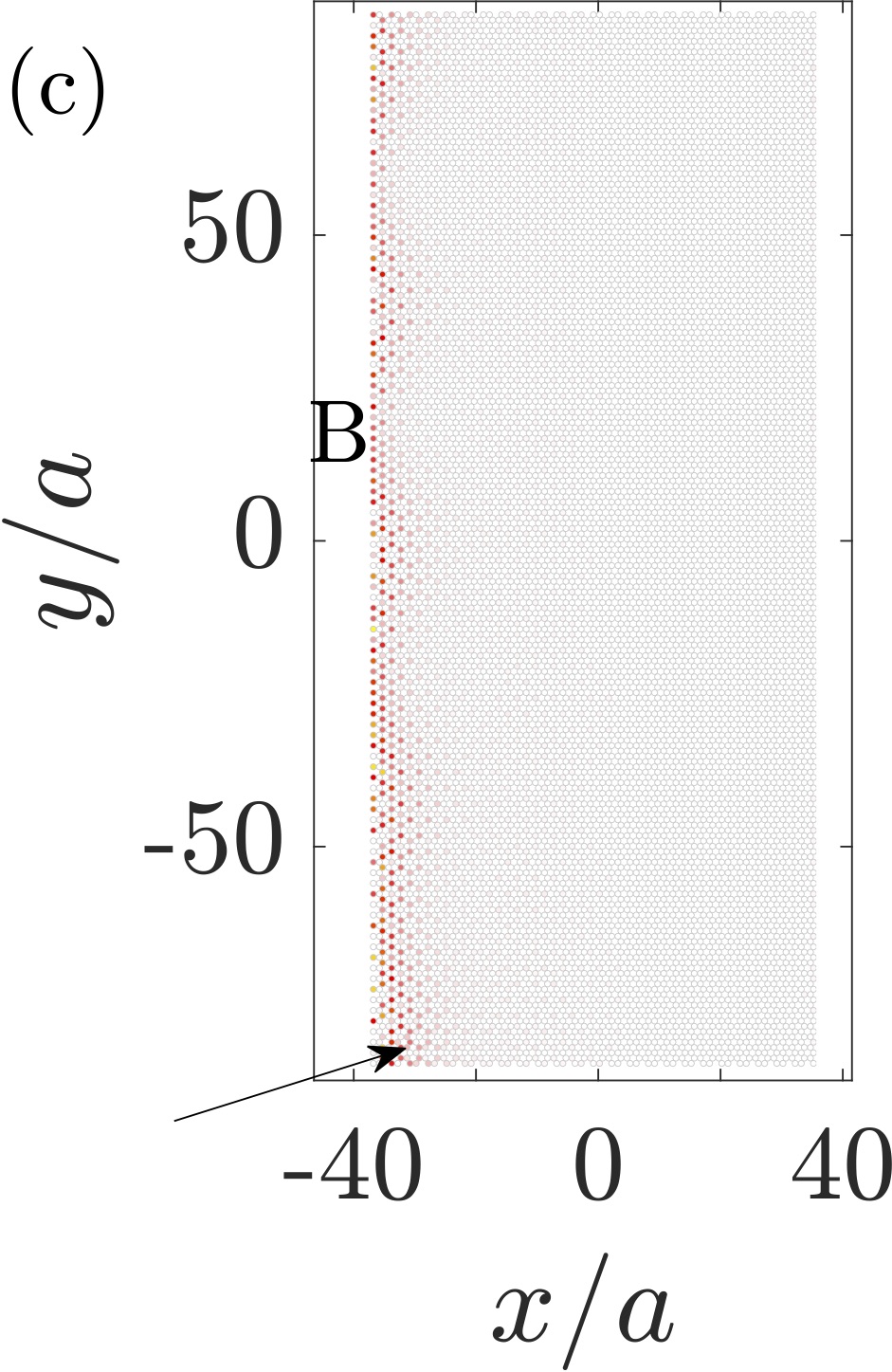}
\includegraphics[width=0.23\textwidth]{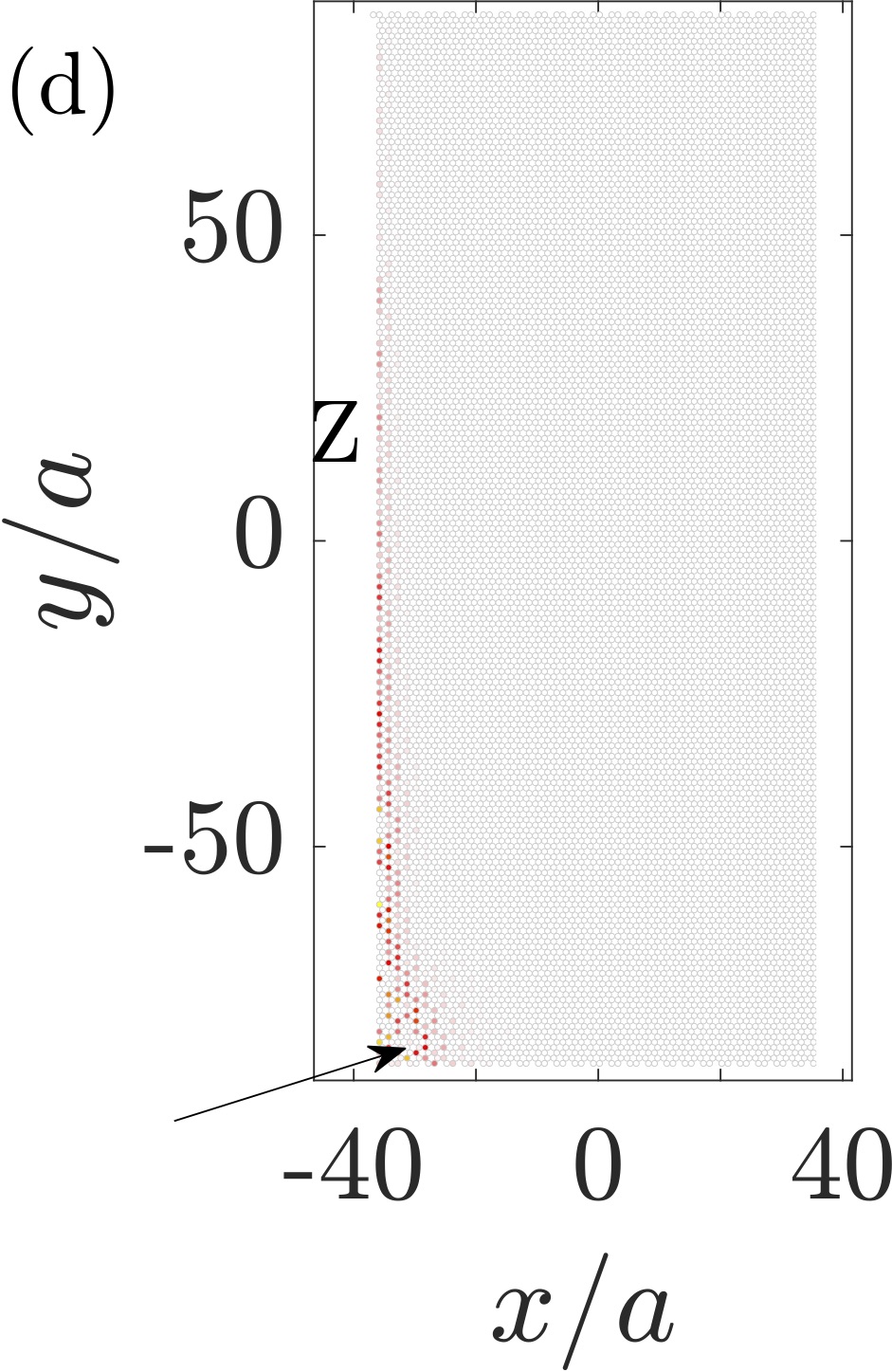}
\caption{
Panel (a)-(d): spatial amplitudes in the steady-state showing the propagation of the edge state in the uni-axially-strained system of $N_x=49$ and $N_y=199$ unit cells in the horizontal and vertical direction respectively.
Parameters are $t'=0.1 t$, $\tau=0.02$, $\hbar\gamma/t=0.0025$. Panels (a)-(b) are for a pump frequency in the lower LL0 gap $\hbar\omega_0/t=-(E_{n=0}+E_{n=1})/2t=-0.36$, while panels (c)-(d) are for pump frequency in the upper LL0 gap $\hbar\omega_0/t=(E_{n=0}+E_{n=1})/2t=-0.24$.
The pumped site is indicated by the arrow. In panel (a), the left edge is bearded, and so supports a $0$-th Landau level propagating edge state.
In panel (b), the left edge is changed to zigzag, and the propagation of the edge state is clearly suppressed.
Panel (c)-(d) show the same configuration as in panel (a)-(b) respectively. We see the mixing of the propagating edge state associated with pseudo-Landau levels with previously non-propagating edge states, which are localized on a shorter length. In particular, now we see that the zigzag termination in the upper LL0 gap has an edge state that propagates along the edge. }
\label{fig:NNNuni-axial_edges}
\end{figure}

We now repeat the calculations done in Sec.~\ref{sec:steadystate} for the uni-axial strain along $x$ to show how the results for a coherently driven system are affected by the NNN hoppings. 
In Fig.~\ref{fig:NNNuni-axial_edges} we present the steady-state amplitude for the same configuration as in Fig.~\ref{fig:uni-axial_edges}, including a NNN hopping strength of $t'=0.1 t$. The energy dispersion of these configurations is shown in Fig.~\ref{fig:Spectra_all_terminationsNNN}~(c) and \ref{fig:Spectra_all_terminationsNNN}~(e).

The driving frequency is between the lower bands and the shifted energy of the Dirac point $-3t'$ (lower LL0 gap) for Figs.~\ref{fig:NNNuni-axial_edges}~(a)-(b), while it is between the upper bands and the shifted energy of the Dirac point $-3t'$ (upper LL0 gap) for Figs.~\ref{fig:NNNuni-axial_edges}~(c)-(d).
While for $t'=0$ the system is symmetric with respect to the energy of the $n=0$ Landau level, here we see that the pattern of the steady-state amplitudes depends on the chosen gap.
Fig.~\ref{fig:NNNuni-axial_edges}~(a) shows the propagating edge state supported by the bearded edge, while Fig.~\ref{fig:NNNuni-axial_edges}~(b) shows the absence of a propagating edge state. For this specific pump frequency, the results are then very similar to the ones obtained in the absence of NNN hoppings and shown in Fig.~\ref{fig:uni-axial_edges}~(a)-(b).

When pumping in in the lower LL0 gap, instead, we see in Fig.~\ref{fig:NNNuni-axial_edges}~(c) that the propagating edge state supported by the bearded edge is much more localized than the one shown in  Fig.~\ref{fig:NNNuni-axial_edges}~(a). This is because the pump excites also the previously non-propagating edge state localized on the left edge. 
The same argument is valid in Fig.~\ref{fig:NNNuni-axial_edges}~(d), where now we see a propagating edge state where before it was non-propagating as in Fig.~\ref{fig:uni-axial_edges}(b).
The lower group velocity is reflected in the edge state having a shorter propagation distance than in Figs.~\ref{fig:uni-axial_edges}(a) and \ref{fig:uni-axial_edges}(c).

\section{Conclusions}
\label{sec:conclusions}

To conclude, we have given a general criterion for the existence of helically-propagating edge states associated with pseudo-Landau levels in strained honeycomb lattices with only nearest-neighbour hoppings.
Our criterion is based on the chiral symmetry of the tight-binding Hamiltonian and on the fact that the wavefunction of the $0$-th Landau level is localized only on one sublattice.
The criterion can be applied to any type of edges and to various strains, showing that the existence of propagating edge states in strained honeycomb lattices is termination-dependent as well as strain-dependent.
We have numerically verified our criterion by calculating the energy dispersion of uni-axially strained systems. 
We have also shown, with a view to artificial graphene, how the helically-propagating edge states appear in the steady state of a driven-dissipative system for uni-axial as well as trigonal strains, for all the three types of edges.
We have seen that, by suitably engineering the edges, the system can act as a valley filter for honeycomb lattices.
Finally, we have commented on the effects of NNN hoppings, which break the chiral symmetry, such that our criterion only holds approximately.

\begin{acknowledgments}
We are grateful to M. Bellec and F. Mortessagne for useful discussions. 
We also thank S. Ryu and S. Gopalakrishnan for helpful exchanges on their paper.
This work was funded by the EU--FET Proactive grant AQuS, Project No. 640800, and by the Autonomous Province of Trento, partially through the project SiQuro. 
H.M.P. was also supported by the EC through the H2020 Marie Sk\l{}odowska--Curie Action, Individual Fellowship Grant No: 656093 ``SynOptic''.
\end{acknowledgments}


\begin{thebibliography}{}

\bibitem{Girvin}S. M. Girvin, \textit{The quantum Hall effect: novel excitations and broken symmetries}, (Springer 1999).
\bibitem{Yoshioka} D. Yoshioka, {\it The Quantum Hall Effect}, (Springer 2002).

\bibitem{Kane} C. L. Kane and E. J. Mele, \textit{Phys. Rev. Lett.} \textbf{78}, 1932 (1997).
\bibitem{Vozmediano} M. A. H Vozmediano, M. I. Katsnelson, and F. Guinea, \textit{Phys. Rep.} \textbf{496}, 109 (2010).
\bibitem{CastroNeto} A. H. Castro Neto, F. Guinea, N. M. R. Peres, K. S. Novoselov, and A. K. Geim, \textit{Rev. Mod. Phys.} {\bf 81}, 109 (2009).
\bibitem{Goerbig} M. O. Goerbig, \textit{Rev. Mod. Phys.} \textbf{83}, 1193 (2011).
\bibitem{Guinea} F. Guinea, M. I. Katsnelson, and A. K. Geim, \textit{Nat. Phys.} \textbf{6}, 30 (2010).
\bibitem{GuineaPRB} F. Guinea, A. K. Geim, M. I. Katsnelson, and K. S. Novoselov, \textit{Phys. Rev. B} \textbf{81}, 035408 (2010).
\bibitem{deJuan} F. de~Juan, M. Sturla, and M. A. H. Vozmediano, \textit{Phys. Rev. Lett.} \textbf{108}, 227205 (2012).
\bibitem{Levy} N. Levy, S. A. Burke, K. L. Meaker, M. Panlasigui, A. Zettl, F. Guinea, A. H. Castro~Neto, and M. F. Crommie, \textit{Science} \textbf{329}, 544 (2010).

\bibitem{Low} T. Low and F. Guinea, \textit{Nano Lett.} \textbf{10}, 3551 (2010).
\bibitem{Chang} Y. Chang, T. Albash and S. Haas, \textit{Phys. Rev. B} \textbf{86} 125402 (2012).
\bibitem{Ghaemi} P. Ghaemi, S. Gopalakrishnan, and S. Ryu, \textit{Phys. Rev. B} \textbf{87}, 155422 (2013).
\bibitem{Brendel} C. Brendel, V. Peano, O. Painter, and F. Marquardt, arXiv:1607.04321.
\bibitem{Atteia} J. Atteia, Stage de Master 2, Universit\'{e} de Bordeaux (2014) \href{http://www.ens-lyon.fr/DSM/SDMsite/M2/stages_M2/Atteia2014.pdf}{http://www.ens-lyon.fr/DSM/SDMsite/M2/stages\_M2/Atteia2014.pdf}.

\bibitem{Bellec2014} M. Bellec, U. Kuhl, G. Montambaux, and F. Mortessagne, \textit{New J. Phys.} \textbf{16}, 113023 (2014).
\bibitem{Jacqmin} T. Jacqmin, I. Carusotto, I. Sagnes, M. Abbarchi, D. D. Solnyshkov, G. Malpuech, E. Galopin, A. Lemai\^{t}re, J. Bloch, and A. Amo, \textit{Phys. Rev. Lett.} \textbf{112}, 116402 (2014).
\bibitem{Bellec2013a} M. Bellec, U. Kuhl, G. Montambaux, and F. Mortessagne, \textit{Phys. Rev. B} \textbf{88} 115437 (2013).
\bibitem{Bellec2013b} M. Bellec, U. Kuhl, G. Montambaux, and F. Mortessagne, \textit{Phys. Rev. Lett.} \textbf{110}, 033902 (2013).
\bibitem{Rechtsman} M. C. Rechtsman, J. M. Zeuner, A. T\"{u}nnermann, S. Nolte, M. Segev, and A. Szameit, \textit{Nat. Phot.} \textbf{7}, 153 (2013)

\bibitem{Salerno} G. Salerno, T. Ozawa, H. M. Price, and I. Carusotto,  \textit{2d Mat.} \textbf{2}, 034015 (2015).

\bibitem{Delplace} P. Delplace, D. Ullmo, and G. Montambaux, \textit{Phys. Rev. B} \textbf{84}, 195452 (2011).

\bibitem{Kohmoto} M. Kohmoto and Y. Hasegawa, \textit{Phys. Rev. B} \textbf{76}, 205402 (2007).

\bibitem{Poli} C. Poli, J. Arkinstall, and H. Schomerus, \textit{Phys. Rev. B} \textbf{90}, 155418 (2014).

\bibitem{Carusotto} I. Carusotto and C. Ciuti, \textit{Rev. Mod. Phys.} {\bf 85}, 299 (2013).

\end{thebibliography}
\end{document}